\documentclass[sigconf]{acmart}

\AtBeginDocument{%
  \providecommand\BibTeX{{%
    \normalfont B\kern-0.5em{\scshape i\kern-0.25em b}\kern-0.8em\TeX}}}

\copyrightyear{2026}
\acmYear{2026}
\setcopyright{cc}
\setcctype{by}
\acmConference[CHI '26]{Proceedings of the 2026 CHI Conference on Human Factors in Computing Systems}{April 13--17, 2026}{Barcelona, Spain}
\acmBooktitle{Proceedings of the 2026 CHI Conference on Human Factors in Computing Systems (CHI '26), April 13--17, 2026, Barcelona, Spain}
\acmPrice{}
\acmDOI{10.1145/3772318.3790802}
\acmISBN{979-8-4007-2278-3/2026/04}

\acmSubmissionID{4827}

\usepackage{acmart-taps}
\usepackage{cprotect}
\usepackage{algorithm}
\usepackage{algpseudocode}
\usepackage{listings}
\usepackage{fancyvrb}
\usepackage{multirow}
\usepackage{ragged2e}
\usepackage{seqsplit}
\usepackage{array}
\usepackage{svg}
\usepackage{graphicx}
\usepackage[utf8]{inputenc}
\usepackage{comment} 
\usepackage{xspace}
\usepackage{soul}

\usepackage{multicol}
\usepackage{comment}
\usepackage{makecell}
\usepackage[shortlabels]{enumitem}
\setlist[enumerate]{leftmargin=*, labelsep=0.5em} 

\usepackage{booktabs}
\usepackage{siunitx}
\usepackage{multirow}
\usepackage{amsfonts}
\usepackage{amsmath}
\usepackage{xspace}
\usepackage{booktabs}
\usepackage{graphicx}
\usepackage{comment}
\usepackage{float}
\usepackage{listings}
\usepackage{changepage}

\usepackage{tikz}
\usepackage{circledsteps}
\usepackage{textcomp}
\DeclareTextSymbolDefault{\ohorn}{T5}
\DeclareTextSymbolDefault{\uhorn}{T5}

\usepackage{bm}
\usepackage{fix-cm}
\usepackage{amsthm}
\usepackage{subcaption}
\usepackage{xcolor,colortbl}
\usepackage{framed}

\usepackage[textwidth=2.5cm]{todonotes}

\newcolumntype{P}[1]{>{\raggedright\arraybackslash}p{#1}}

\lstdefinestyle{rstyle}{
    language=R,
    basicstyle=\smaller\ttfamily,
    keywordstyle=\color{blue},
    stringstyle=\color{green!60!black},
    commentstyle=\color{gray},
    numbers=none,
    numberstyle=\tiny,
    numbersep=5pt,
    frame=none,
    breaklines=true,
    breakatwhitespace=true,
    showstringspaces=false,
    captionpos=b
}

\DeclareRobustCommand{\rr}[1]{{#1}}

\DeclareRobustCommand{\rr}[1]{#1}

 \newcommand{\xhdr}[1]{\smallskip\noindent{{\bf #1.}}}

\newcommand{\syncsense}{\textsc{SyncSense}\xspace}

\newcommand{\chatTimeline}{{\textsf{Outline}}\xspace}
\newcommand{\chatContents}{{\textsf{Detail Explorer}}\xspace}
\newcommand{\origChat}{{\textsf{Annotated Conversation}}\xspace}
\newcommand{\authoring}{{\textsf{Authoring}}\xspace}

\definecolor{toturnBorder}{HTML}{213652} 
\definecolor{yellowbg}{RGB}{255,242,204}
\definecolor{yellowtxt}{RGB}{0, 0, 0}
\definecolor{greentxt}{HTML}{237804} 
\definecolor{greenbg}{HTML}{f6ffed}
\definecolor{graybg}{HTML}{F3F3F3}
\definecolor{orangebg}{RGB}{255,200,170}
\definecolor{orangetxt}{RGB}{117,56,0}

\newcommand{\componentArtifact}{\includegraphics[height=0.8em]{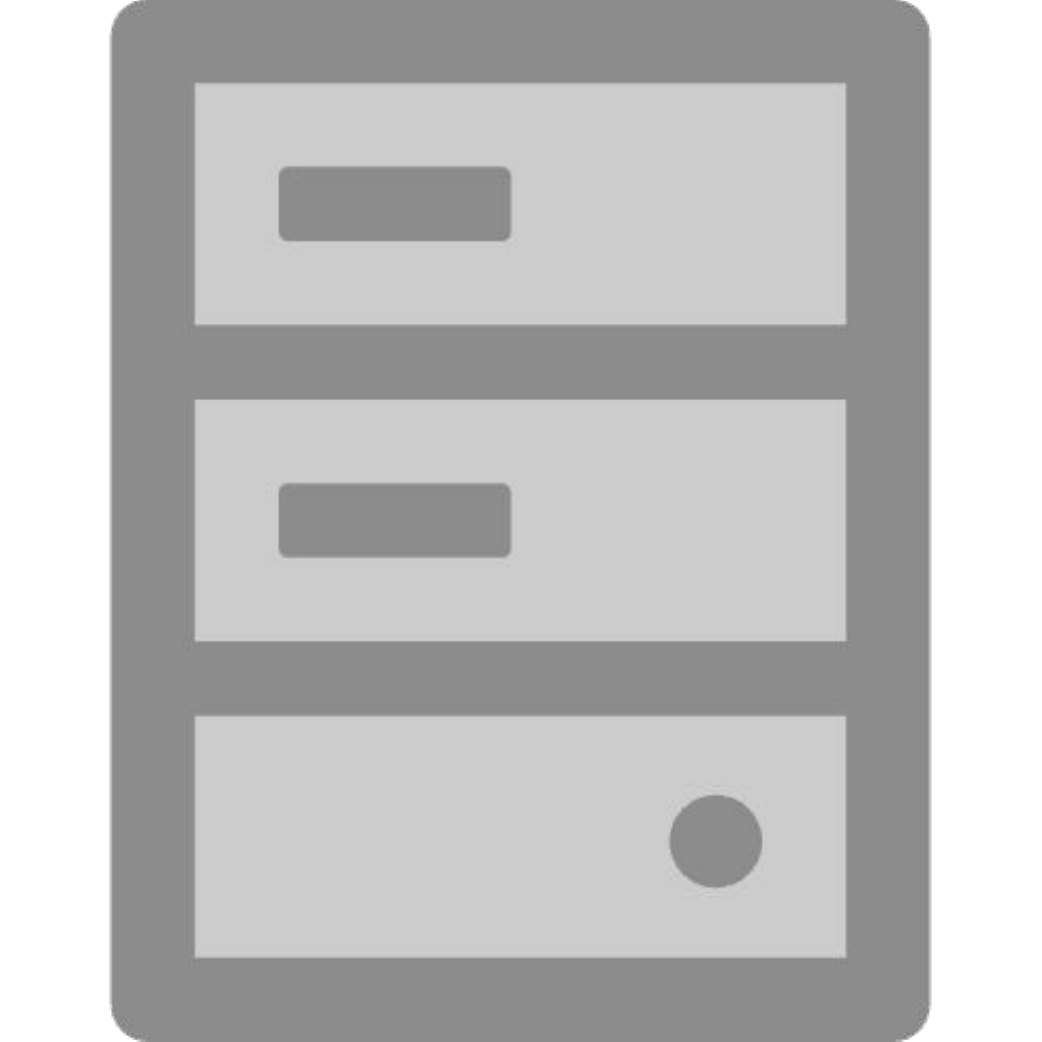}~Analysis Artifacts\xspace}
\newcommand{\componentInsight}{\includegraphics[height=0.8em]{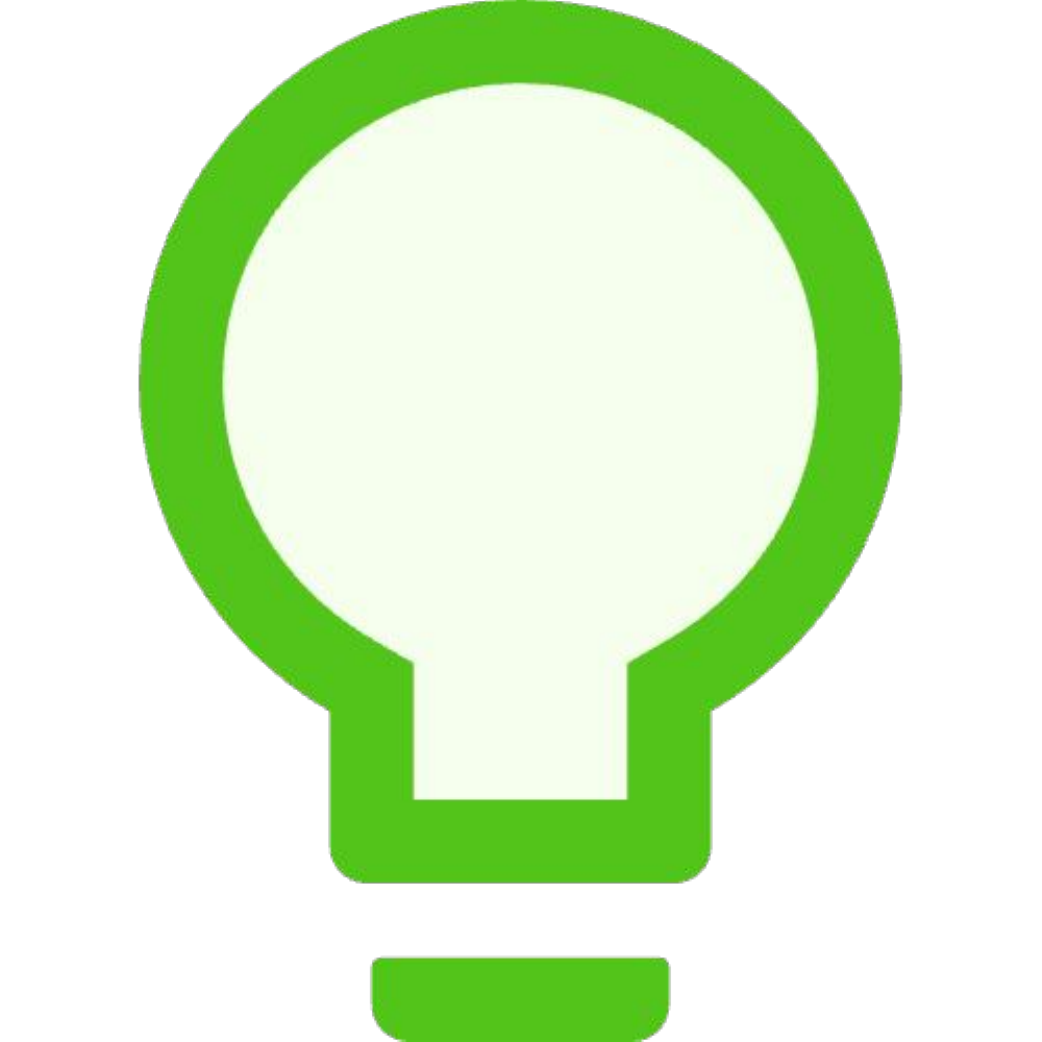}~Analysis Insights\xspace}
\newcommand{\componentSpeechAct}{\includegraphics[height=0.8em]{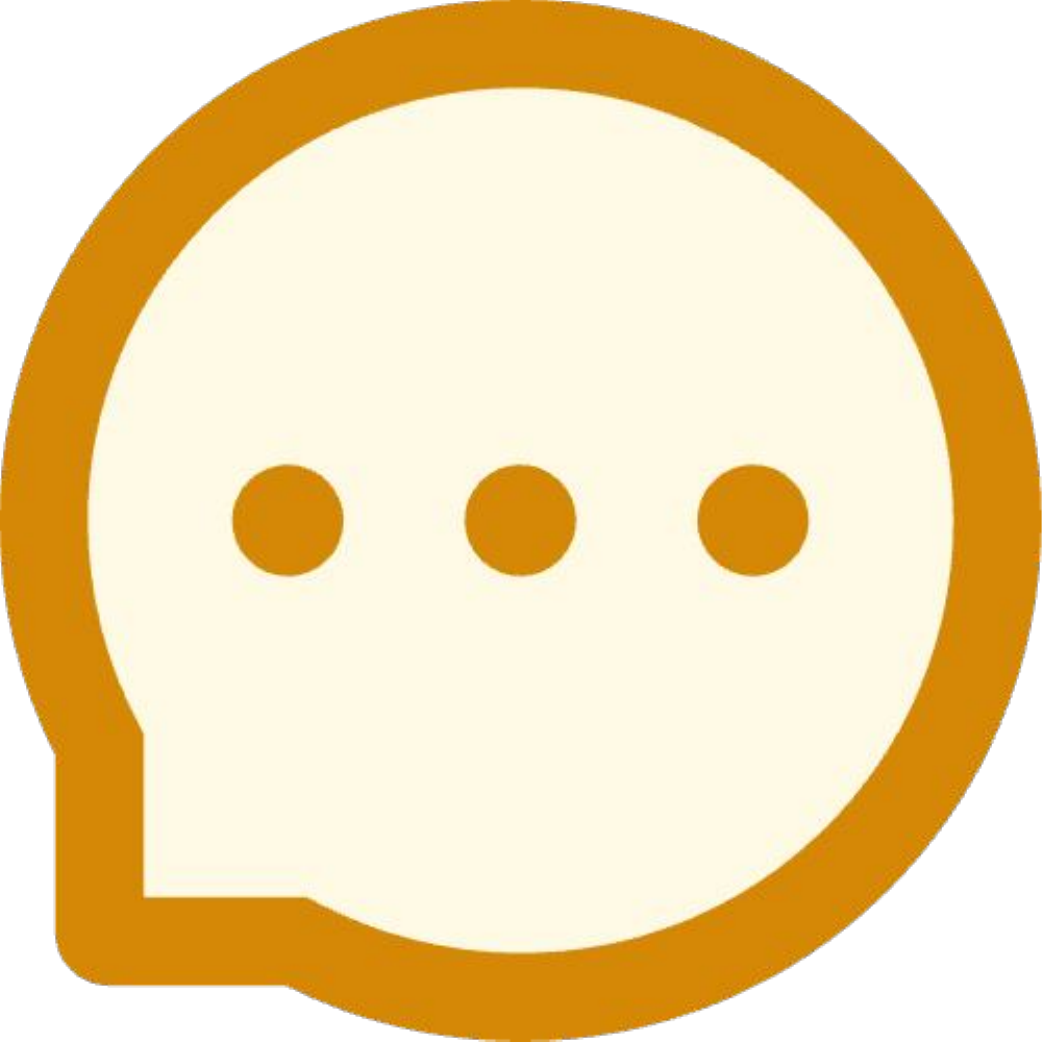}~Speech Acts\xspace}

\newcommand{\componentThreads}{\includegraphics[height=0.8em]{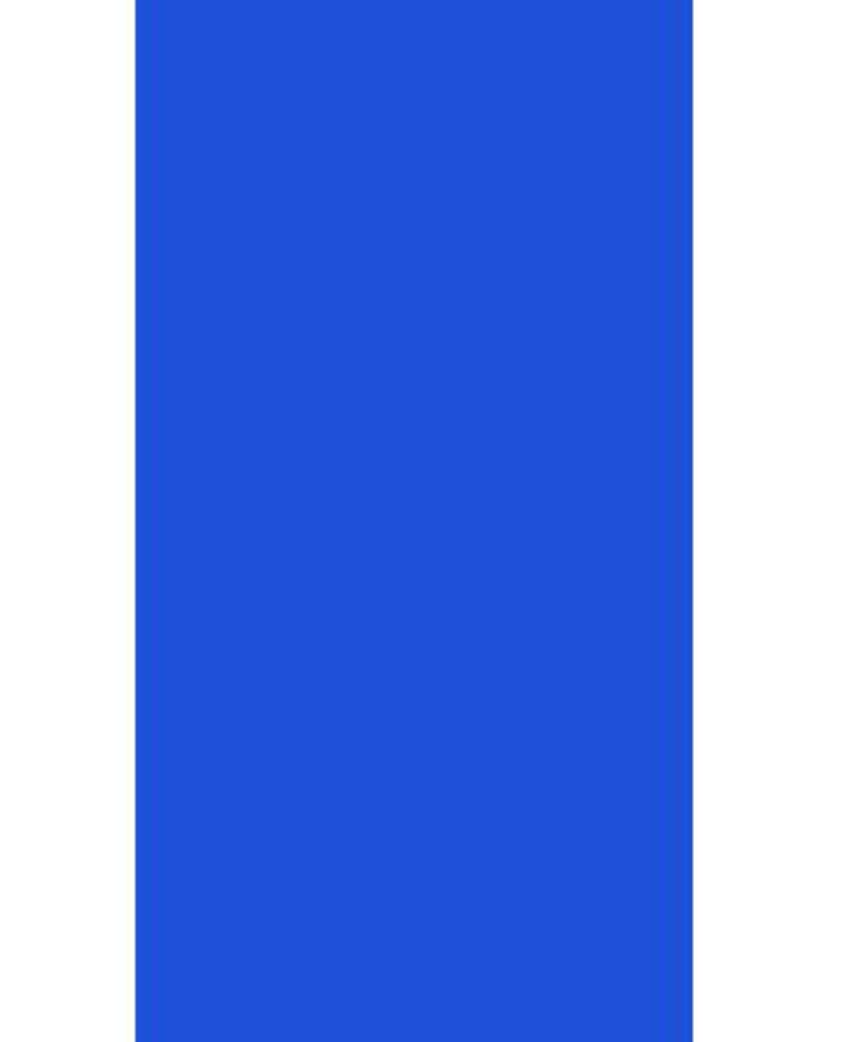}~Threads\xspace}

\newcommand{\componentArtifactShort}{\includegraphics[height=0.8em]{Figures/icons/component-artifact.pdf}~artifacts\xspace}
\newcommand{\componentInsightShort}{\includegraphics[height=0.8em]{Figures/icons/component-insight.pdf}~insights\xspace}
\newcommand{\componentSpeechActShort}{\includegraphics[height=0.8em]{Figures/icons/component-speechAct.pdf}~speech acts\xspace}
\newcommand{\componentThreadsShort}{\includegraphics[height=0.8em]{Figures/icons/component-threads.pdf}~threads\xspace}

\newcommand{\insightTrend}{%
  \protect\hspace{0pt}%
  \begingroup
    \setlength{\fboxsep}{2.5pt}%
    \colorbox{greenbg}{%
      \textcolor{greentxt}{%
        \includegraphics[height=0.85em]{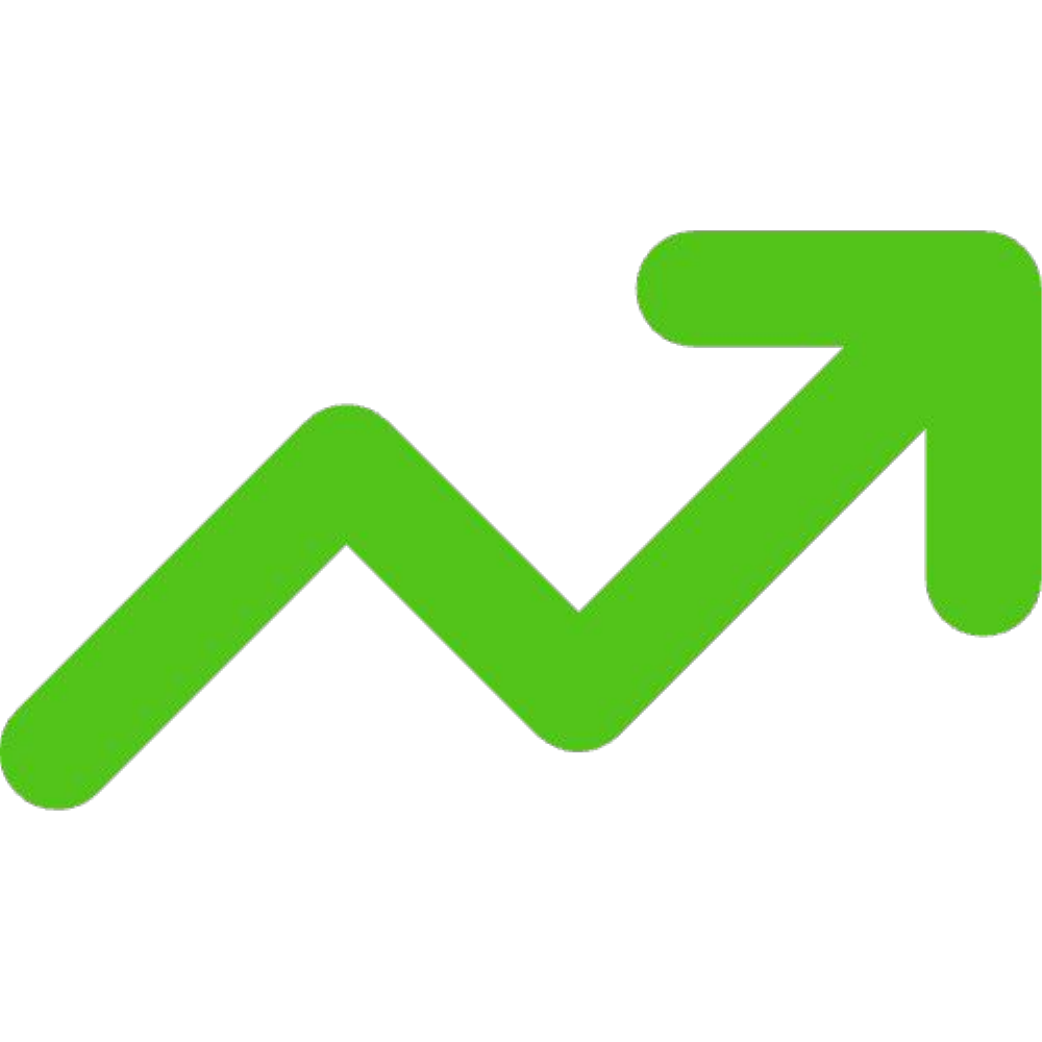}~Trend}}\endgroup\xspace}

\newcommand{\insightRank}{%
  \begingroup
    \setlength{\fboxsep}{2.5pt}%
    \colorbox{greenbg}{%
      \textcolor{greentxt}{%
        \includegraphics[height=0.85em]{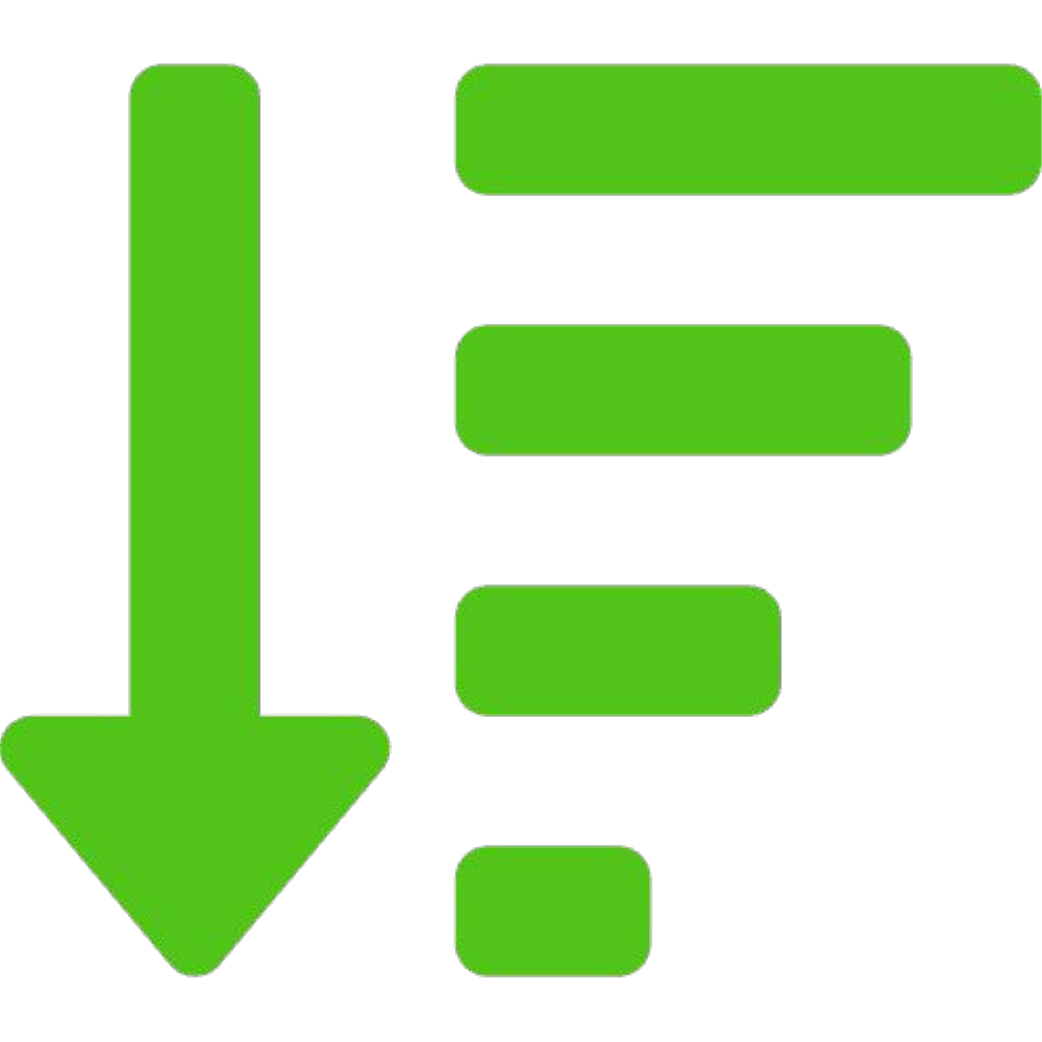}~Rank}}\endgroup\xspace}

\newcommand{\insightProportion}{%
  \begingroup
    \setlength{\fboxsep}{2.5pt}%
    \colorbox{greenbg}{%
      \textcolor{greentxt}{%
        \includegraphics[height=0.85em]{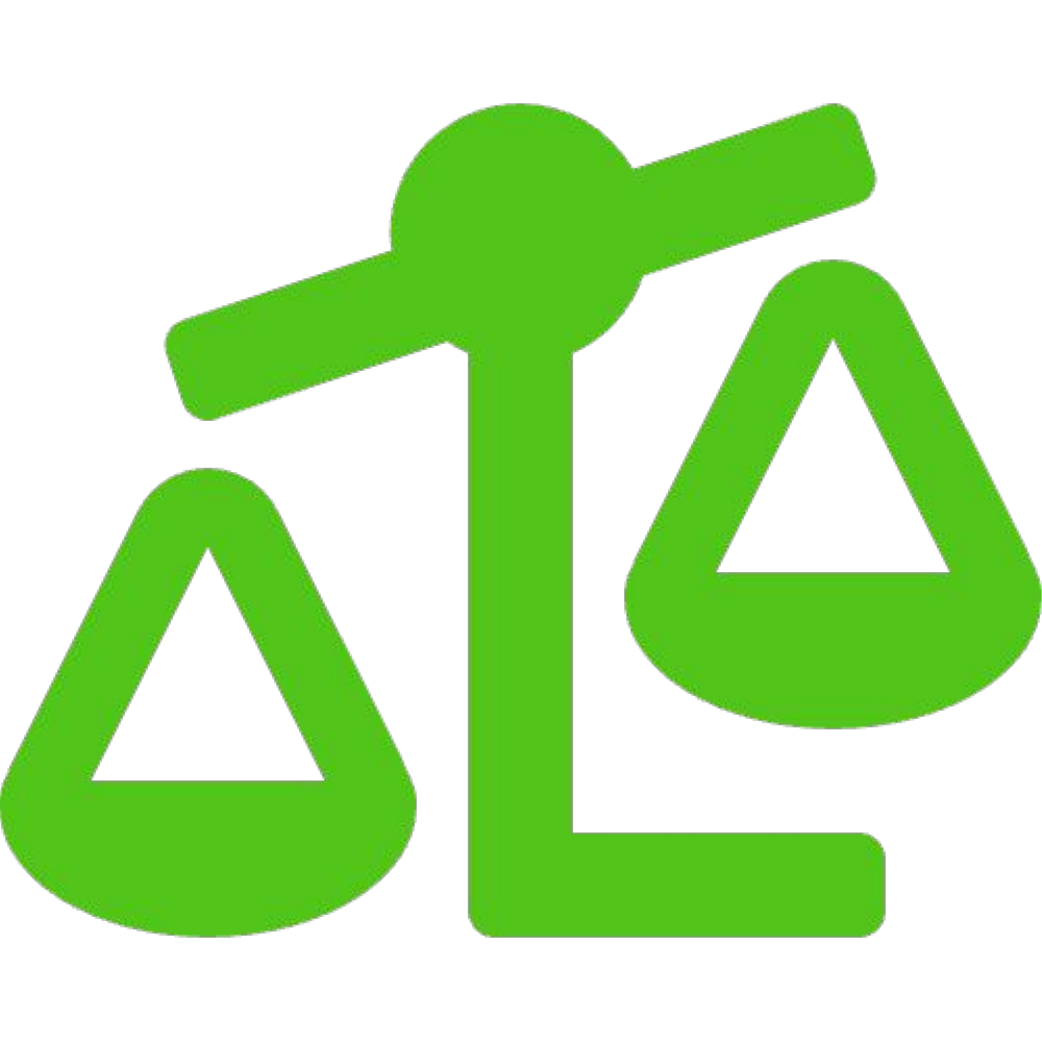}~Proportion}}\endgroup\xspace}

\newcommand{\insightOutlier}{%
  \begingroup
    \setlength{\fboxsep}{2.5pt}%
    \colorbox{greenbg}{%
      \textcolor{greentxt}{%
        \includegraphics[height=0.85em]{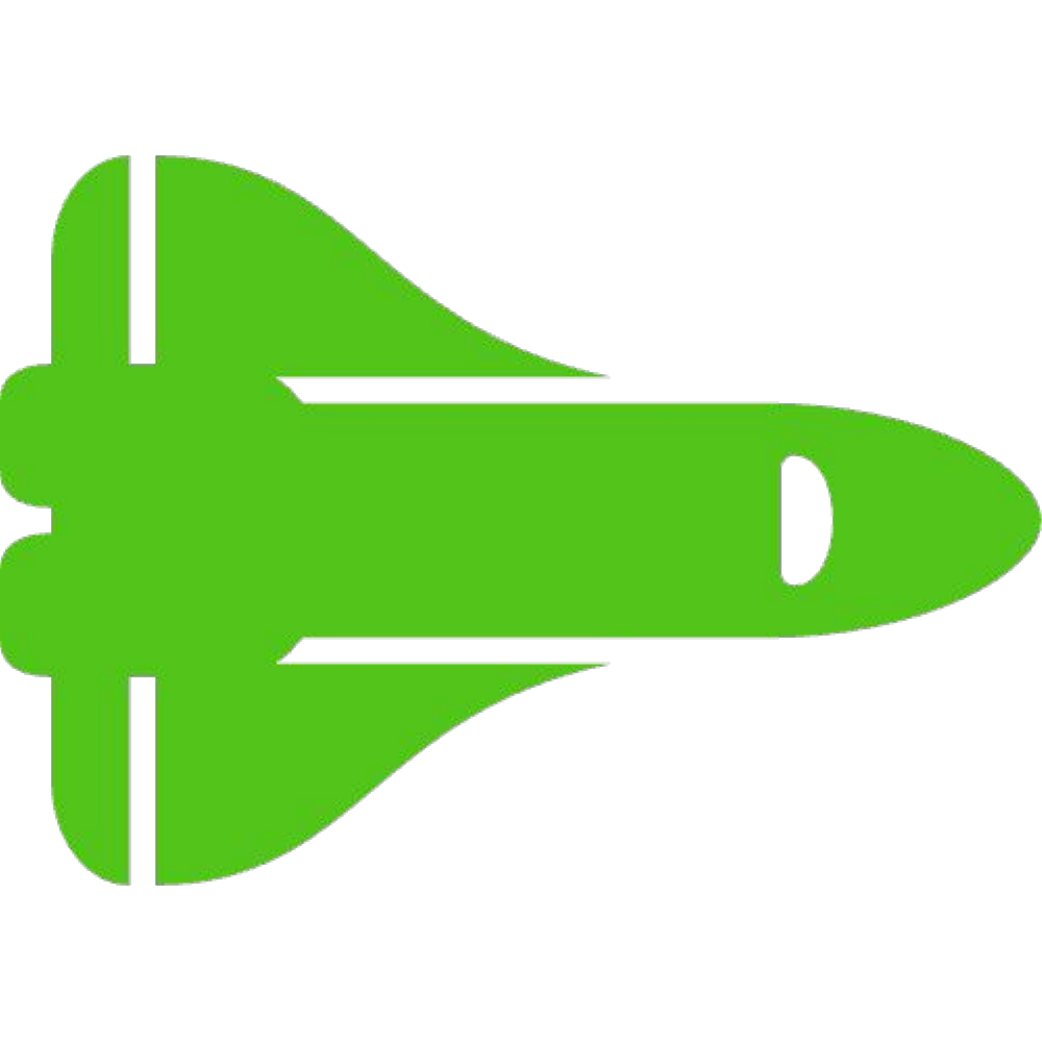}~Outlier}}\endgroup\xspace}

\newcommand{\insightExtreme}{%
  \begingroup
    \setlength{\fboxsep}{2.5pt}%
    \colorbox{greenbg}{%
      \textcolor{greentxt}{%
        \includegraphics[height=0.85em]{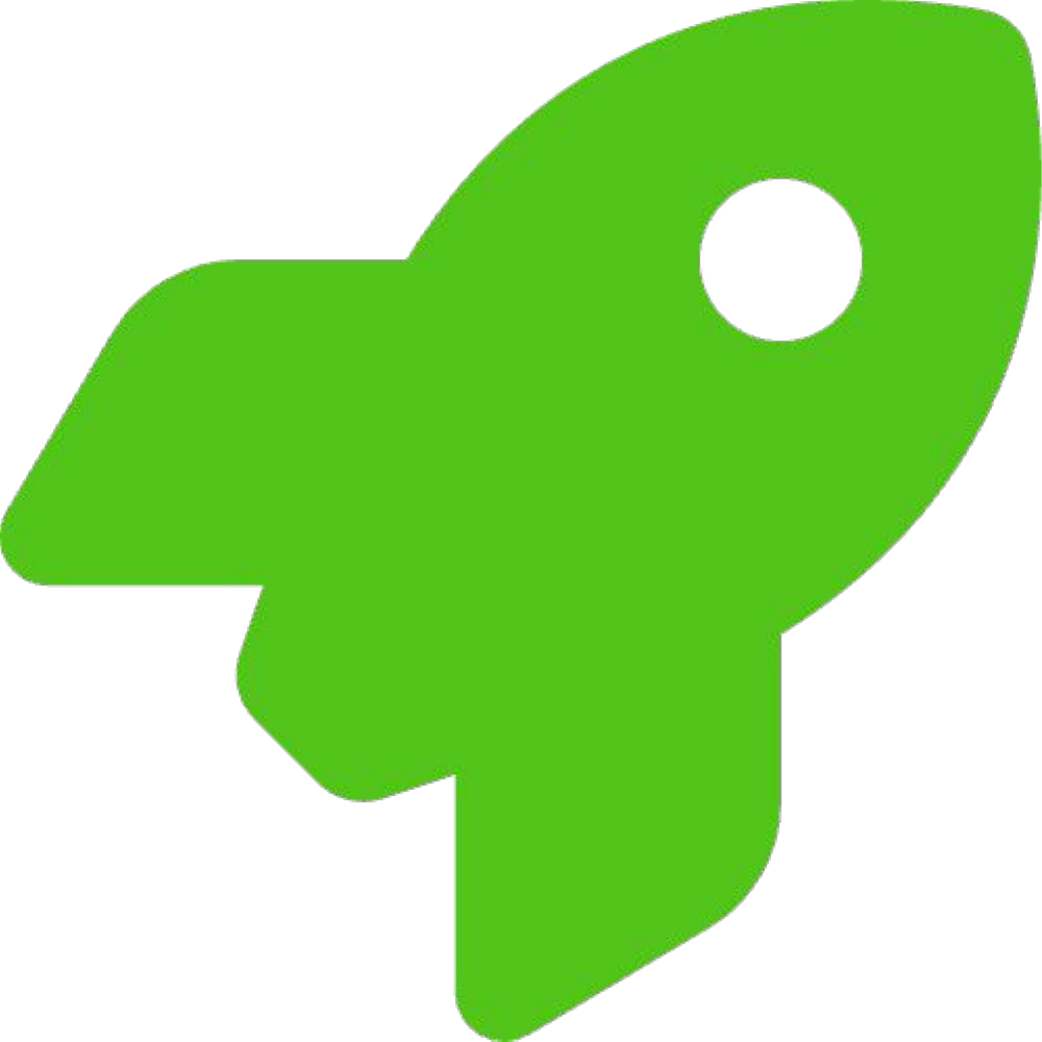}~Extreme}}\endgroup\xspace}

\newcommand{\insightDistribution}{%
  \begingroup
    \setlength{\fboxsep}{2.5pt}%
    \colorbox{greenbg}{%
      \textcolor{greentxt}{%
        \includegraphics[height=0.85em]{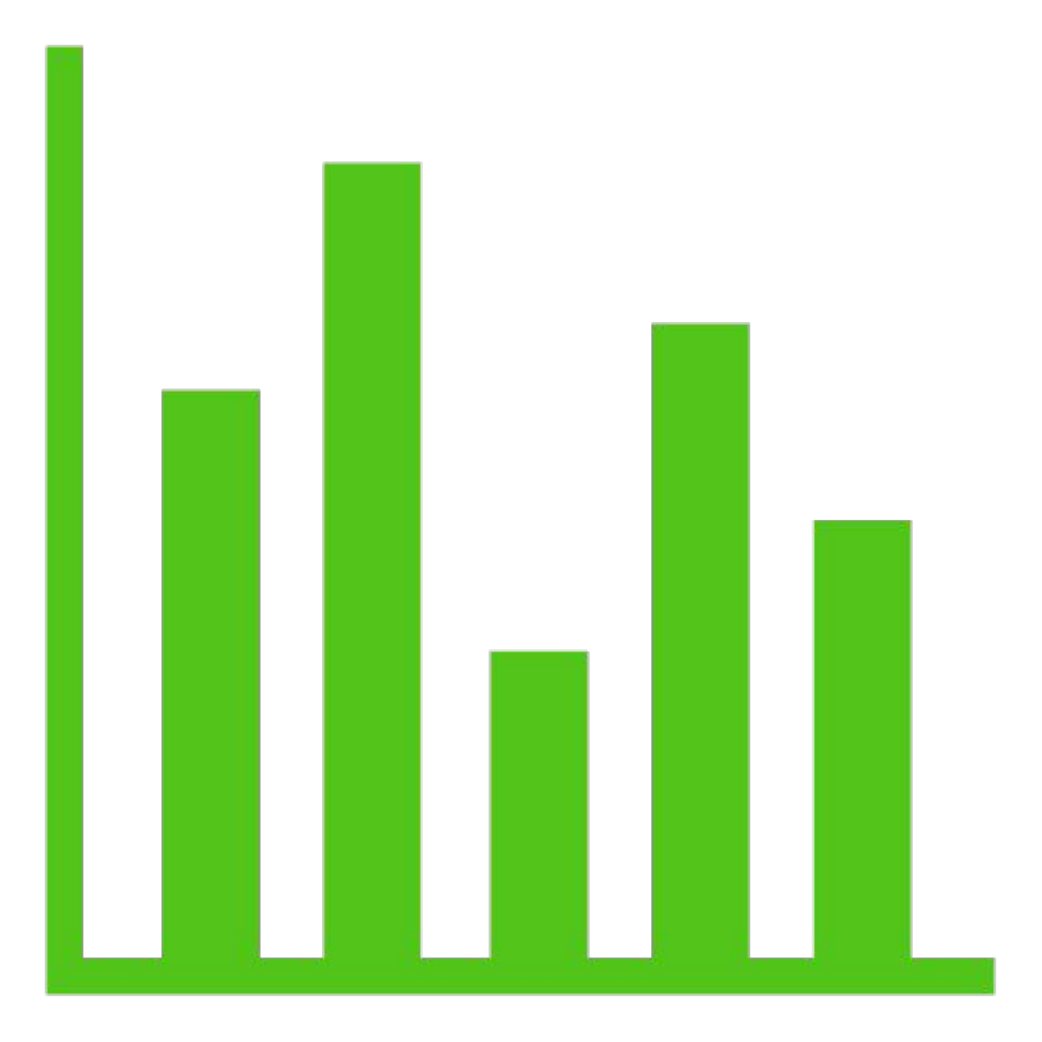}~Distribution}}\endgroup\xspace}

\newcommand{\insightDifference}{%
  \begingroup
    \setlength{\fboxsep}{2.5pt}%
    \colorbox{greenbg}{%
      \textcolor{greentxt}{%
        \includegraphics[height=0.85em]{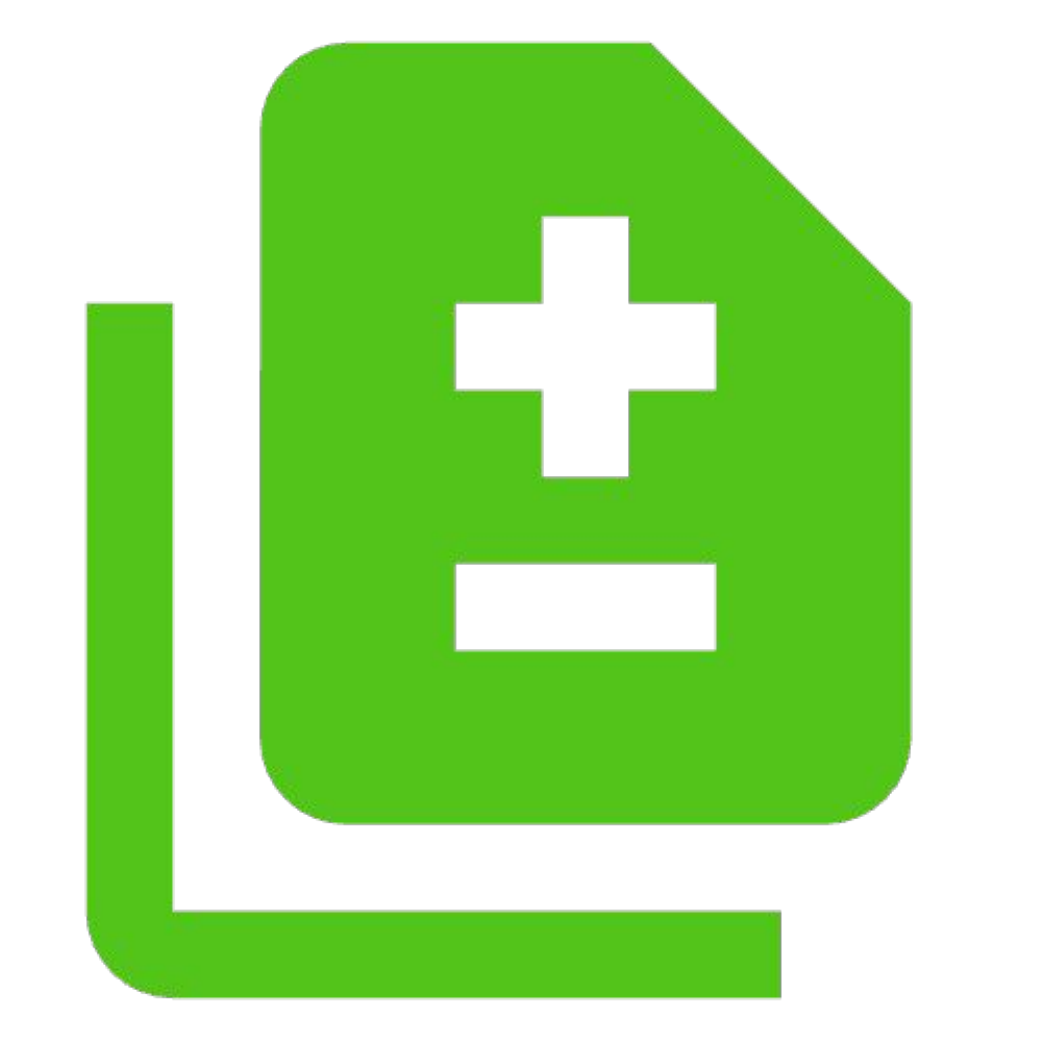}~Difference}}\endgroup\xspace}

\newcommand{\insightCategorization}{%
  \begingroup
    \setlength{\fboxsep}{2.5pt}%
    \colorbox{greenbg}{%
      \textcolor{greentxt}{%
        \includegraphics[height=0.85em]{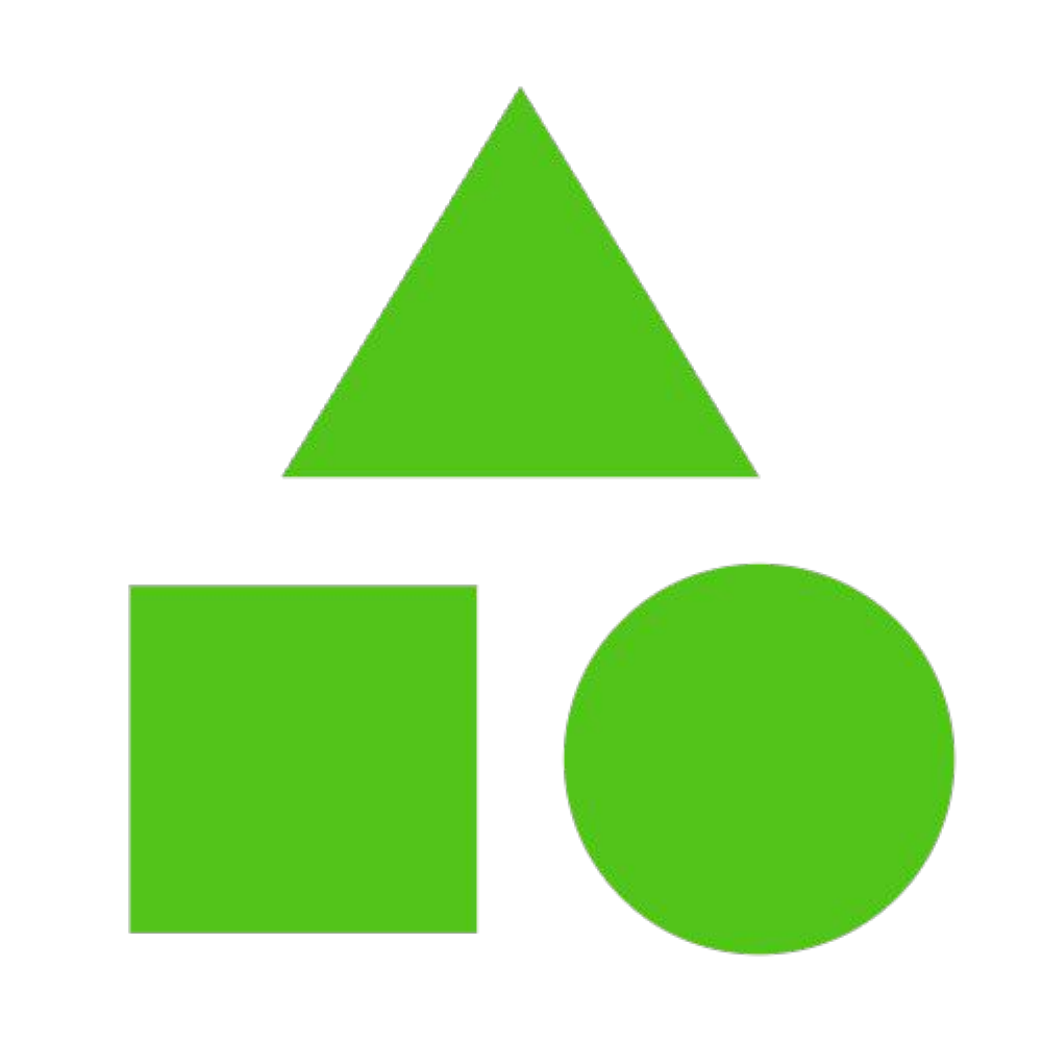}~Categorization}}\endgroup\xspace}
        
\newcommand{\insightAssociation}{%
  \begingroup
    \setlength{\fboxsep}{2.5pt}%
    \colorbox{greenbg}{%
      \textcolor{greentxt}{%
        \includegraphics[height=0.85em]{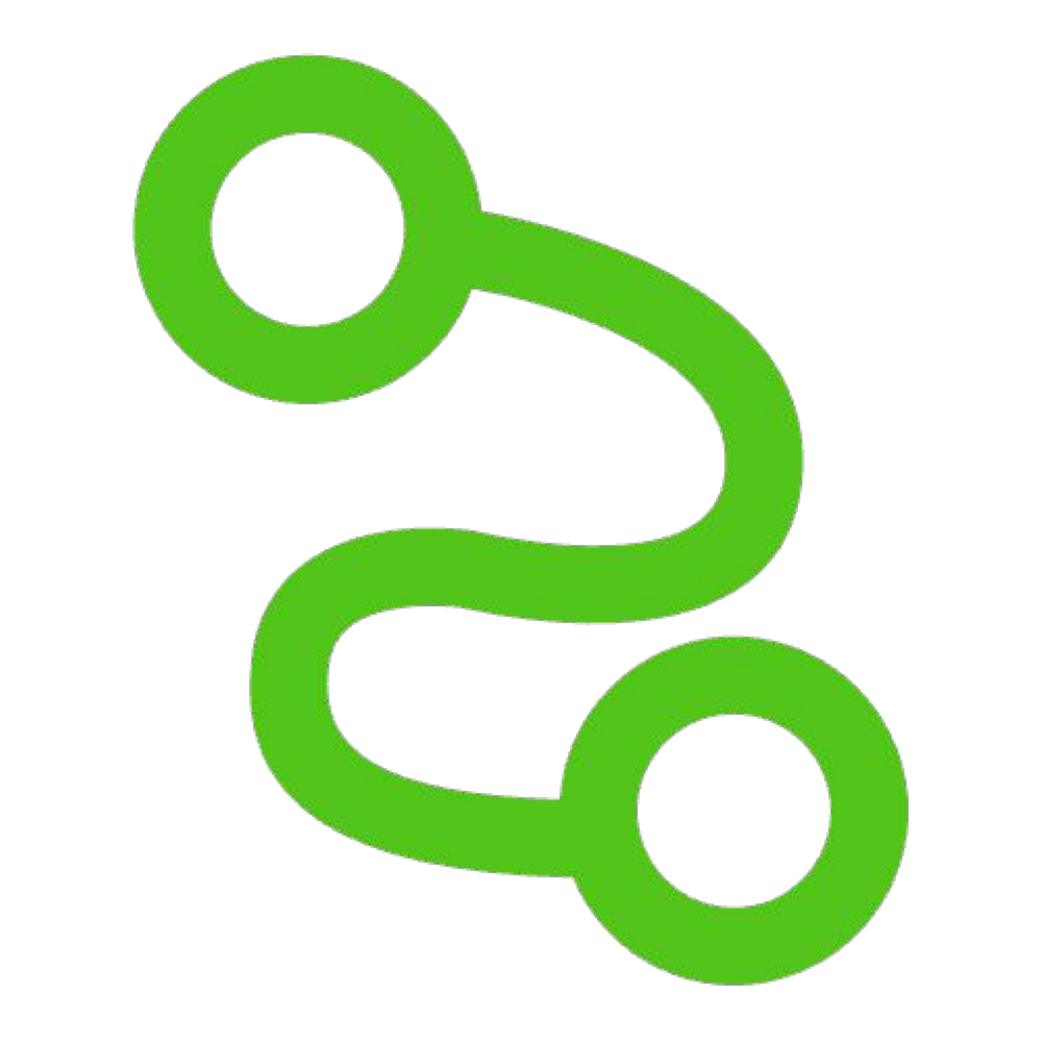}~Association}}\endgroup\xspace}
        
\newcommand{\insightAggregation}{%
  \begingroup
    \setlength{\fboxsep}{2.5pt}%
    \colorbox{greenbg}{%
      \textcolor{greentxt}{%
        \includegraphics[height=0.85em]{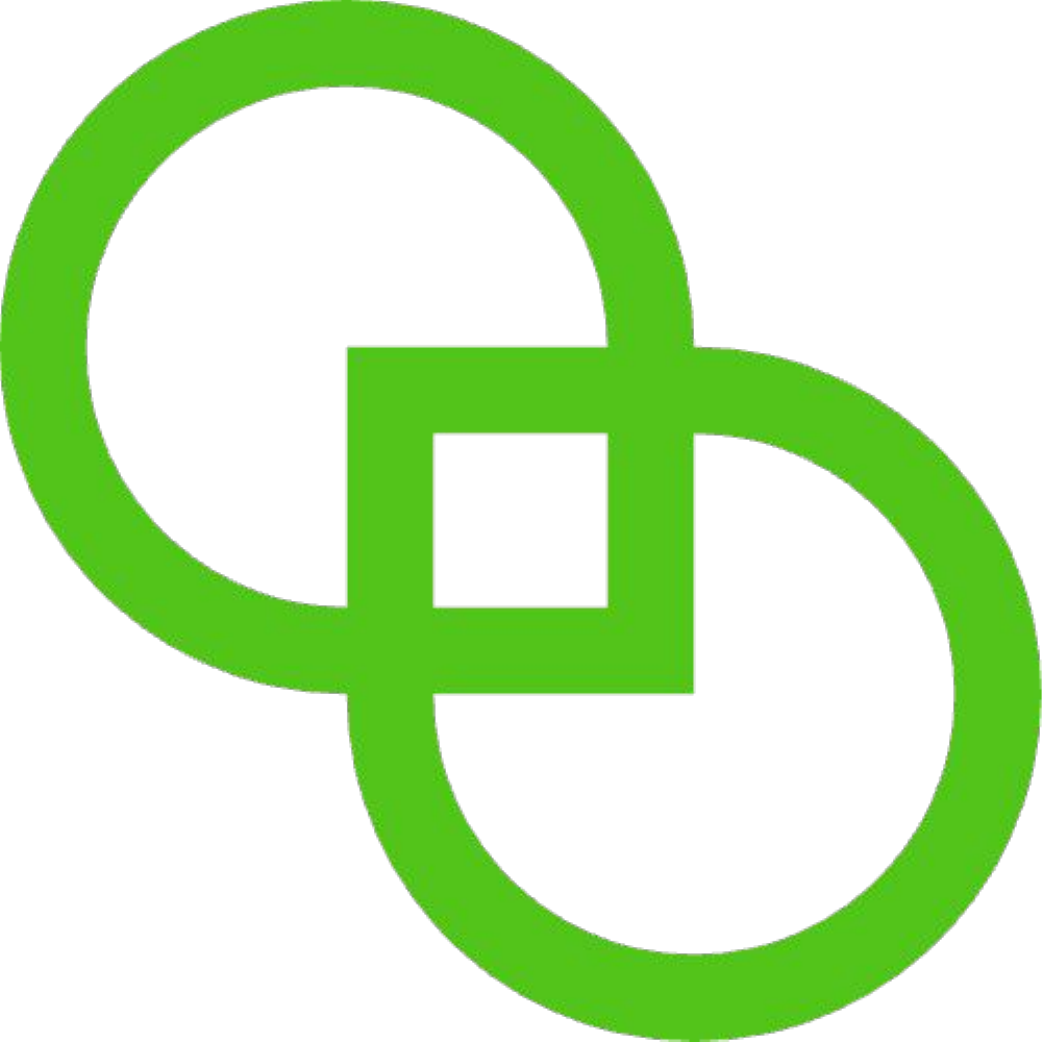}~Aggregation}}\endgroup\xspace}

\newcommand{\artifactViz}{%
  \begingroup
    \setlength{\fboxsep}{2.5pt}%
    \colorbox{graybg}{%
      \includegraphics[height=0.8em]{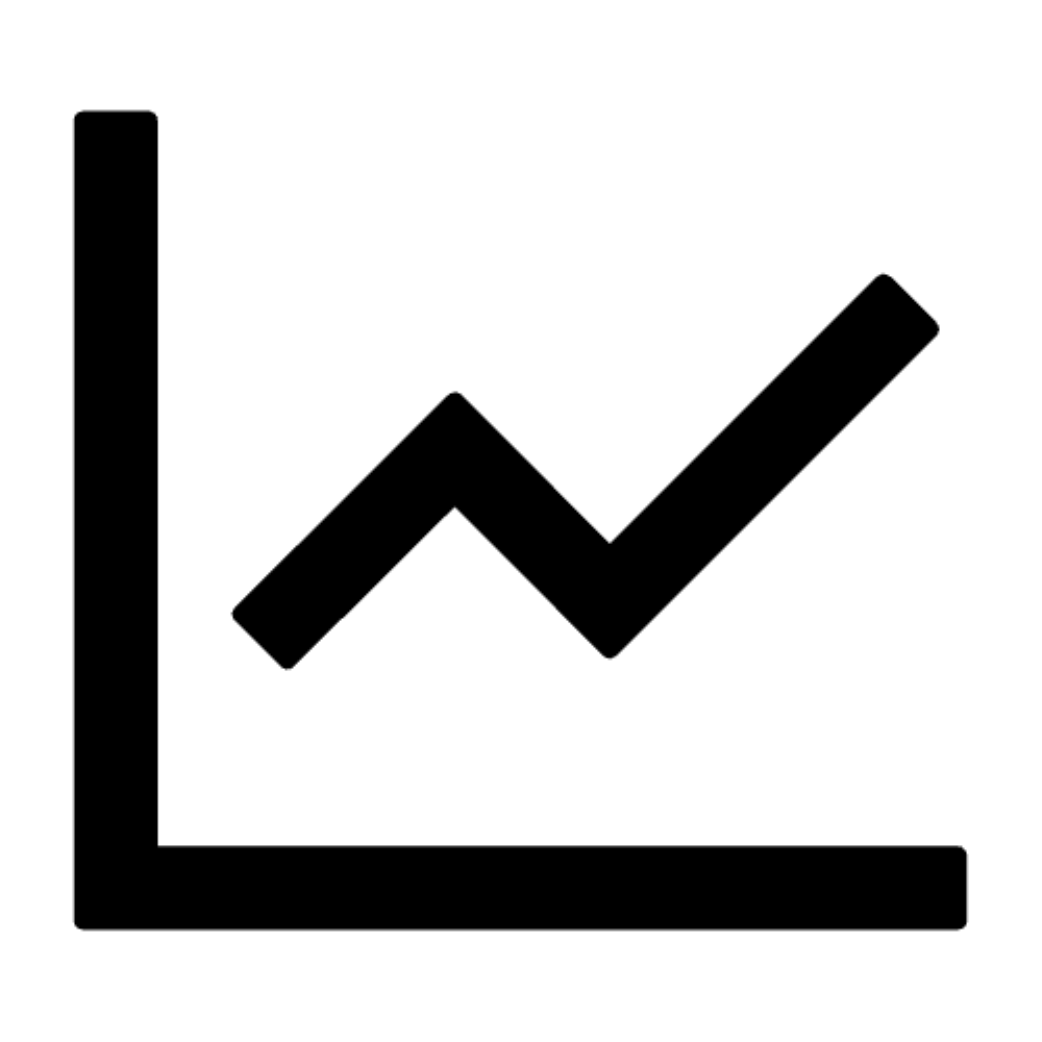}~Visualizations}\endgroup\xspace}

\newcommand{\artifactTable}{%
  \begingroup
    \setlength{\fboxsep}{2.5pt}%
    \colorbox{graybg}{%
      \includegraphics[height=0.8em]{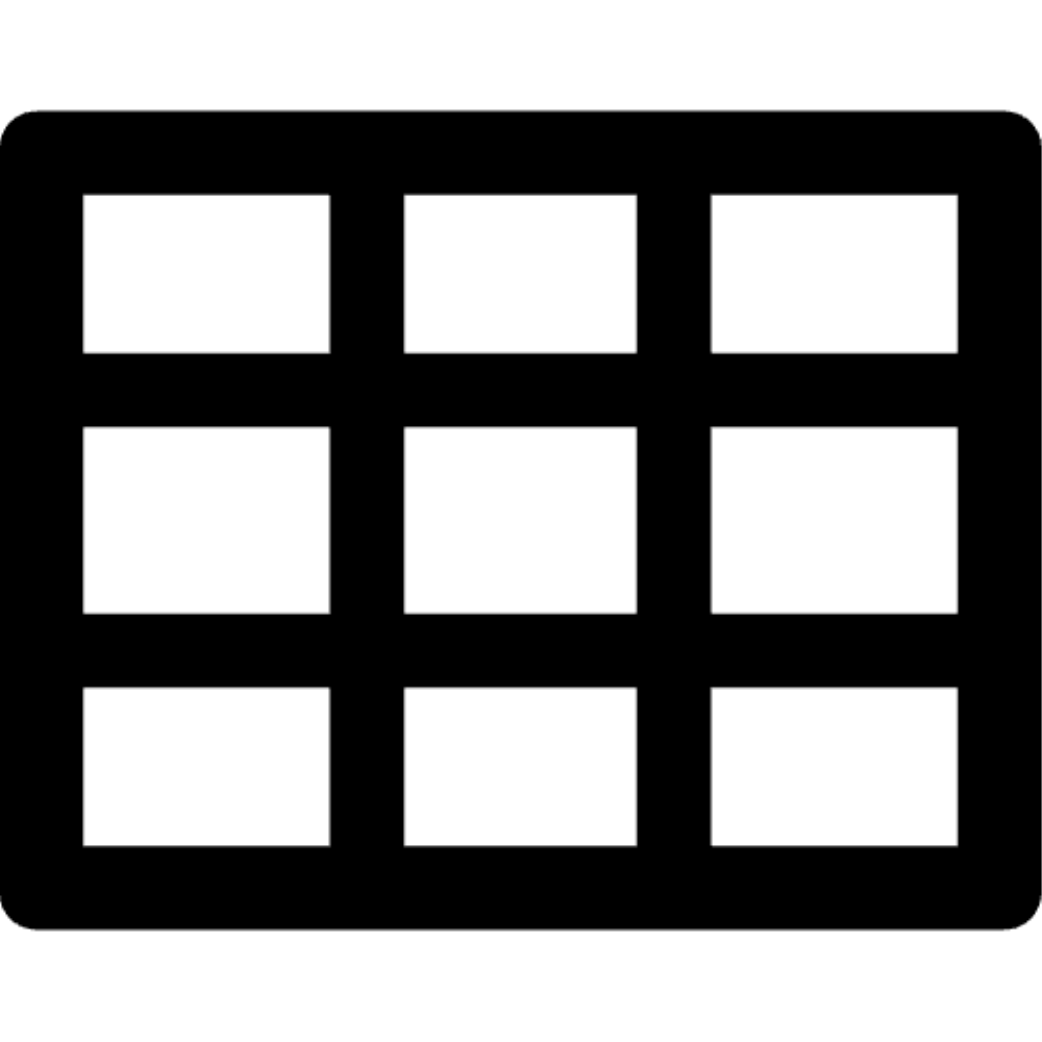}~Data Tables}\endgroup\xspace}

\newcommand{\artifactCode}{%
  \begingroup
    \setlength{\fboxsep}{2.5pt}%
    \colorbox{graybg}{%
      \includegraphics[height=0.8em]{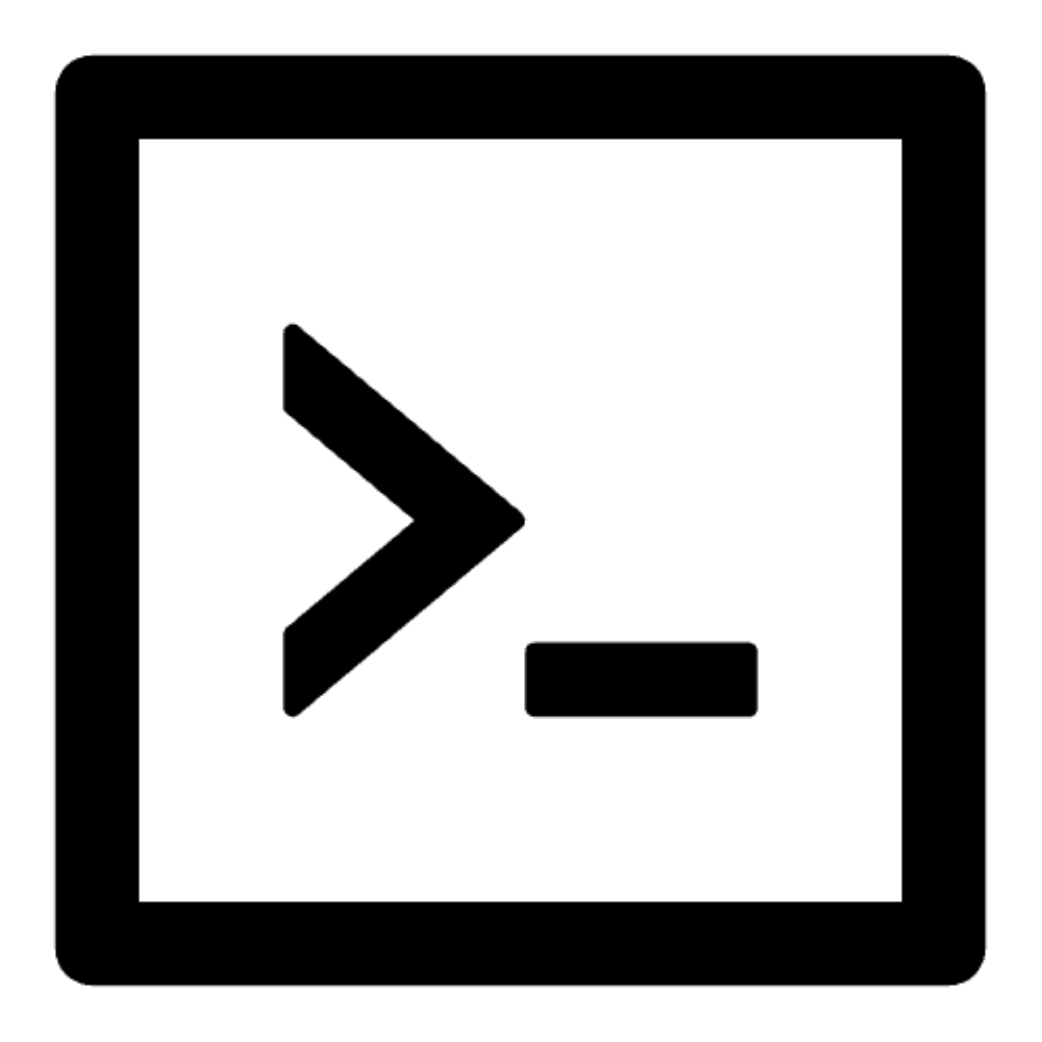}~Code}\endgroup\xspace}

\newcommand{\insightValue}{%
  \begingroup
    \setlength{\fboxsep}{2.5pt}%
    \colorbox{greenbg}{%
      \textcolor{greentxt}{%
        \includegraphics[height=0.8em]{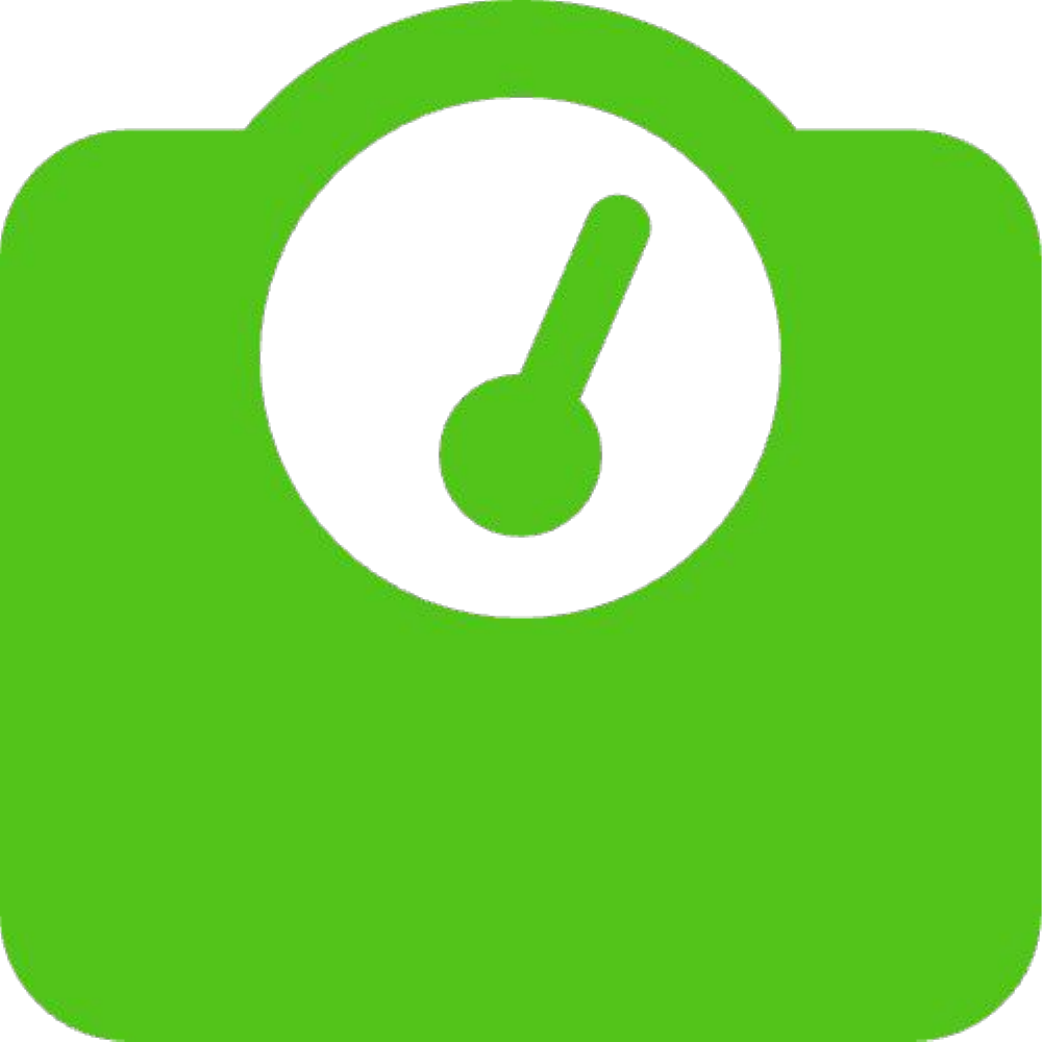}~Value}}\endgroup\xspace}

\newcommand{\speechActDeepInsights}{%
  \begingroup
    \setlength{\fboxsep}{2.5pt}%
    \colorbox{yellowbg}{%
      \includegraphics[height=0.8em]{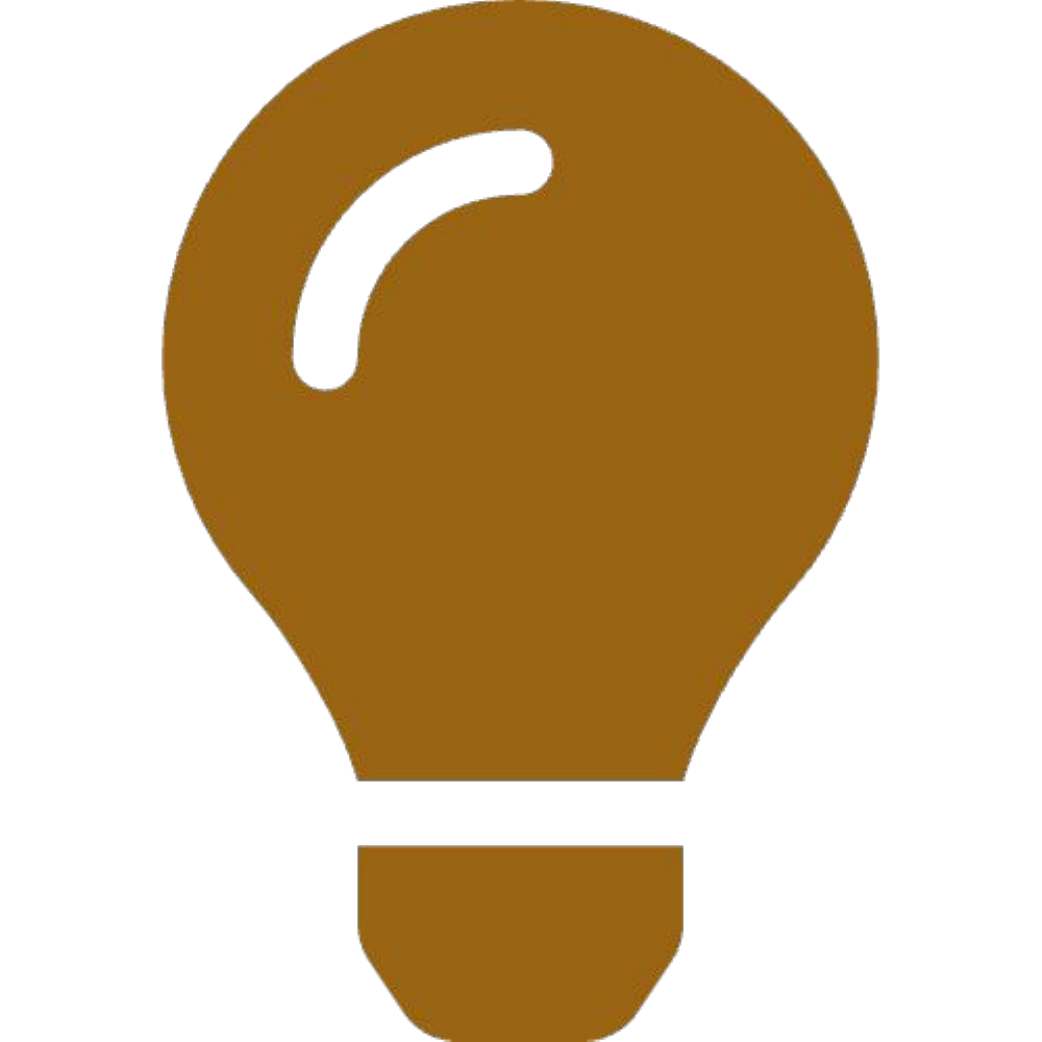}~Deep Insights}\endgroup\xspace}

\newcommand{\speechActFactFinding}{%
  \begingroup
    \setlength{\fboxsep}{2.5pt}%
    \colorbox{yellowbg}{%
      \includegraphics[height=0.8em]{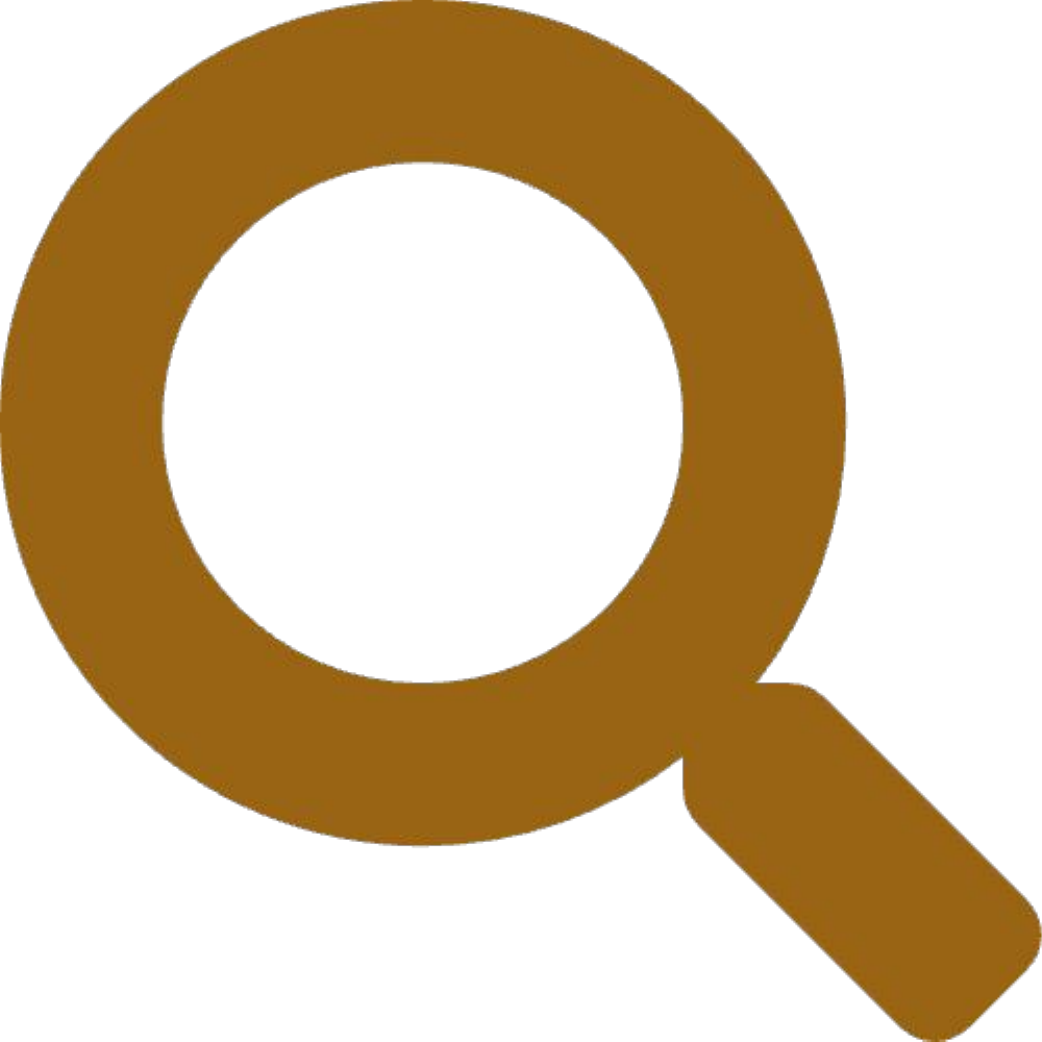}~Fact Finding}\endgroup\xspace}

\newcommand{\speechActDataTransformations}{%
  \begingroup
    \setlength{\fboxsep}{2.5pt}%
    \colorbox{yellowbg}{%
      \includegraphics[height=0.8em]{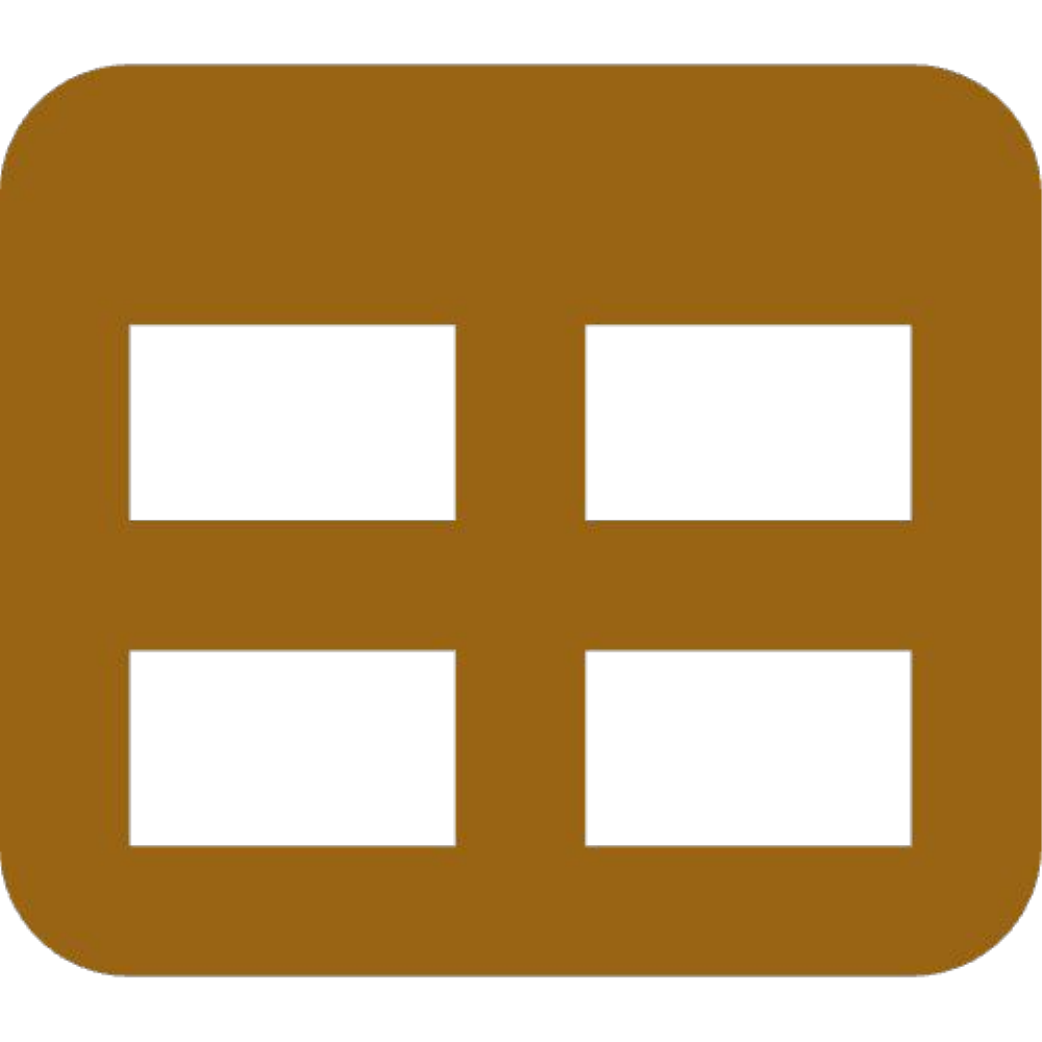}~Data Transformations}\endgroup\xspace}

\newcommand{\speechActSpecificViz}{%
  \begingroup
    \setlength{\fboxsep}{2.5pt}%
    \colorbox{yellowbg}{%
      \includegraphics[height=0.8em]{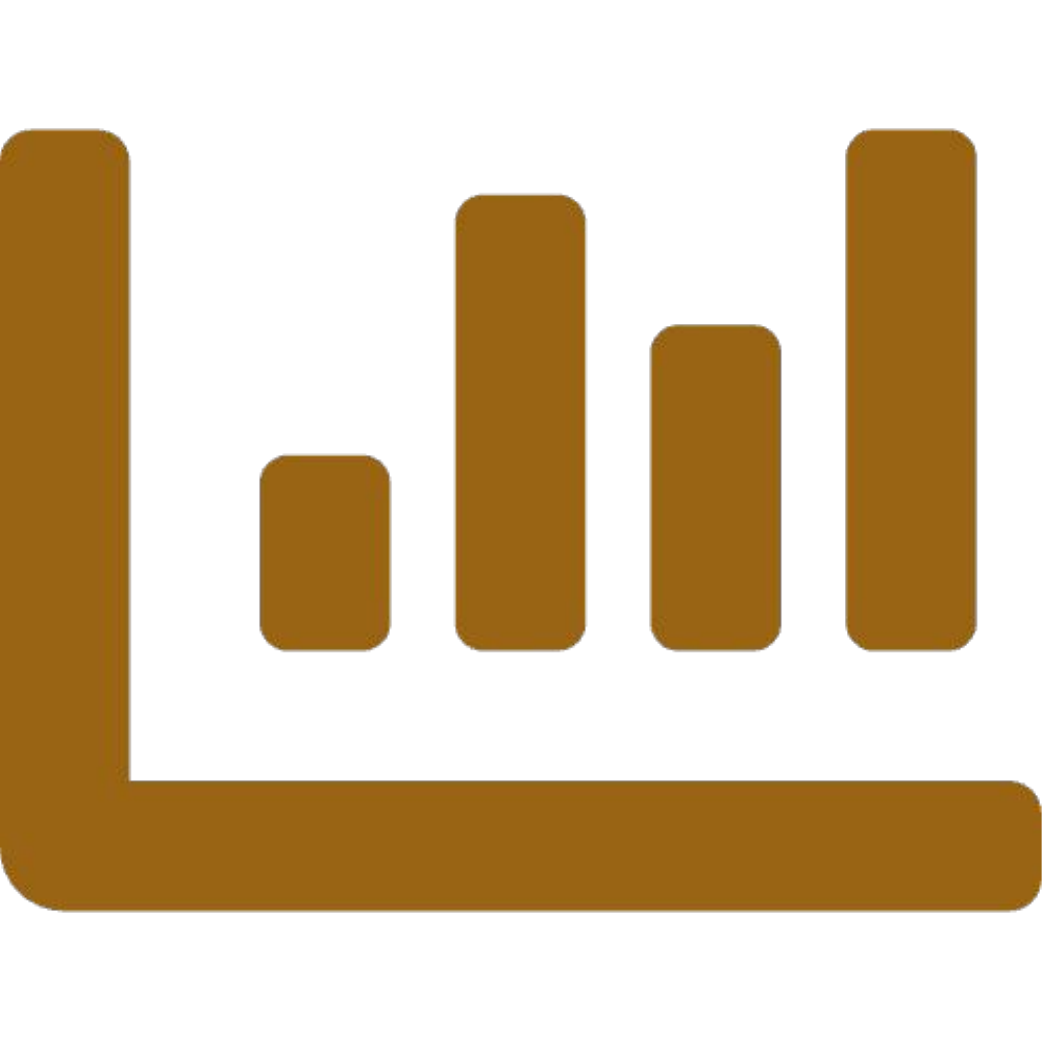}~Specific Visualization}\endgroup\xspace}

\newcommand{\speechActDebugging}{%
  \begingroup
    \setlength{\fboxsep}{2.5pt}%
    \colorbox{yellowbg}{%
      \includegraphics[height=0.8em]{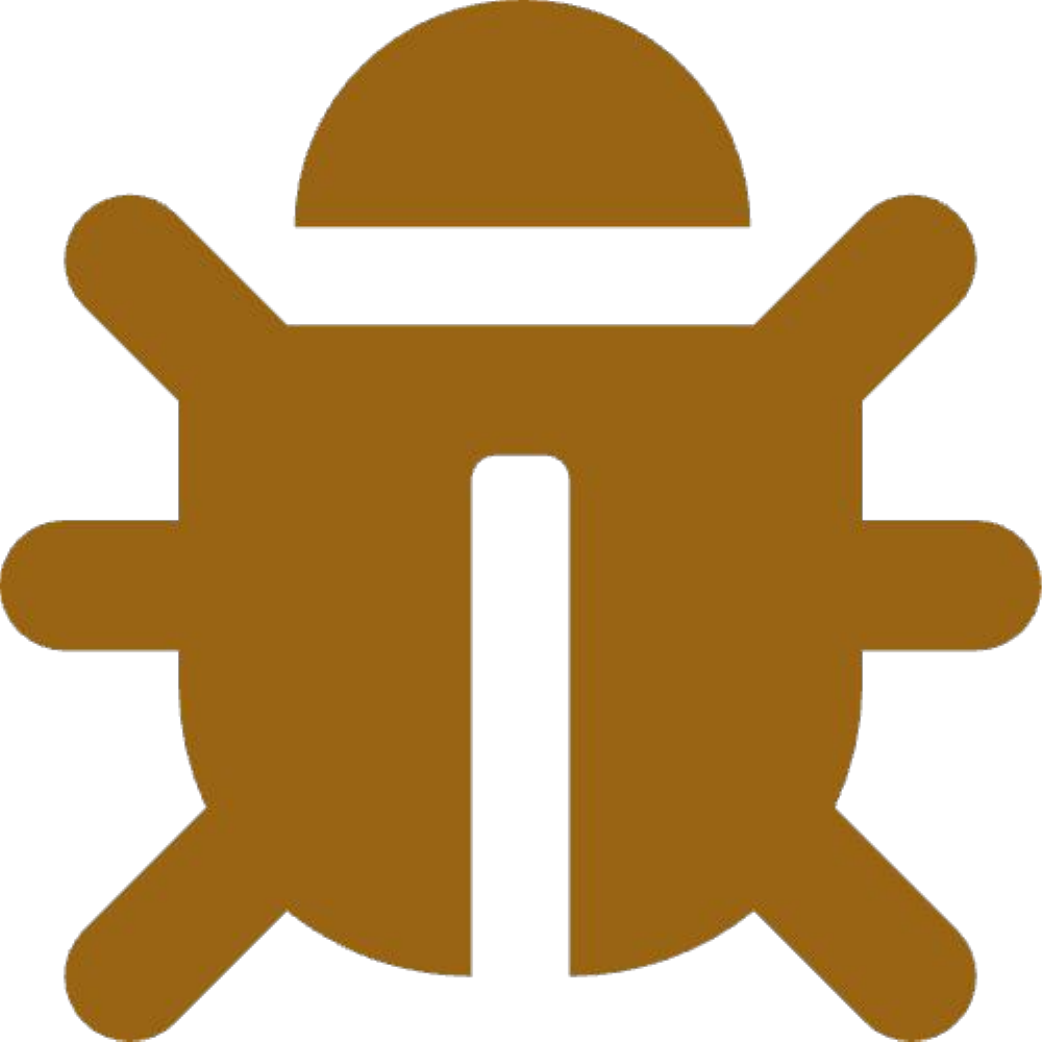}~Debugging}\endgroup\xspace}

\newcommand{\speechActRefinement}{%
  \begingroup
    \setlength{\fboxsep}{2.5pt}%
    \colorbox{yellowbg}{%
      \includegraphics[height=0.8em]{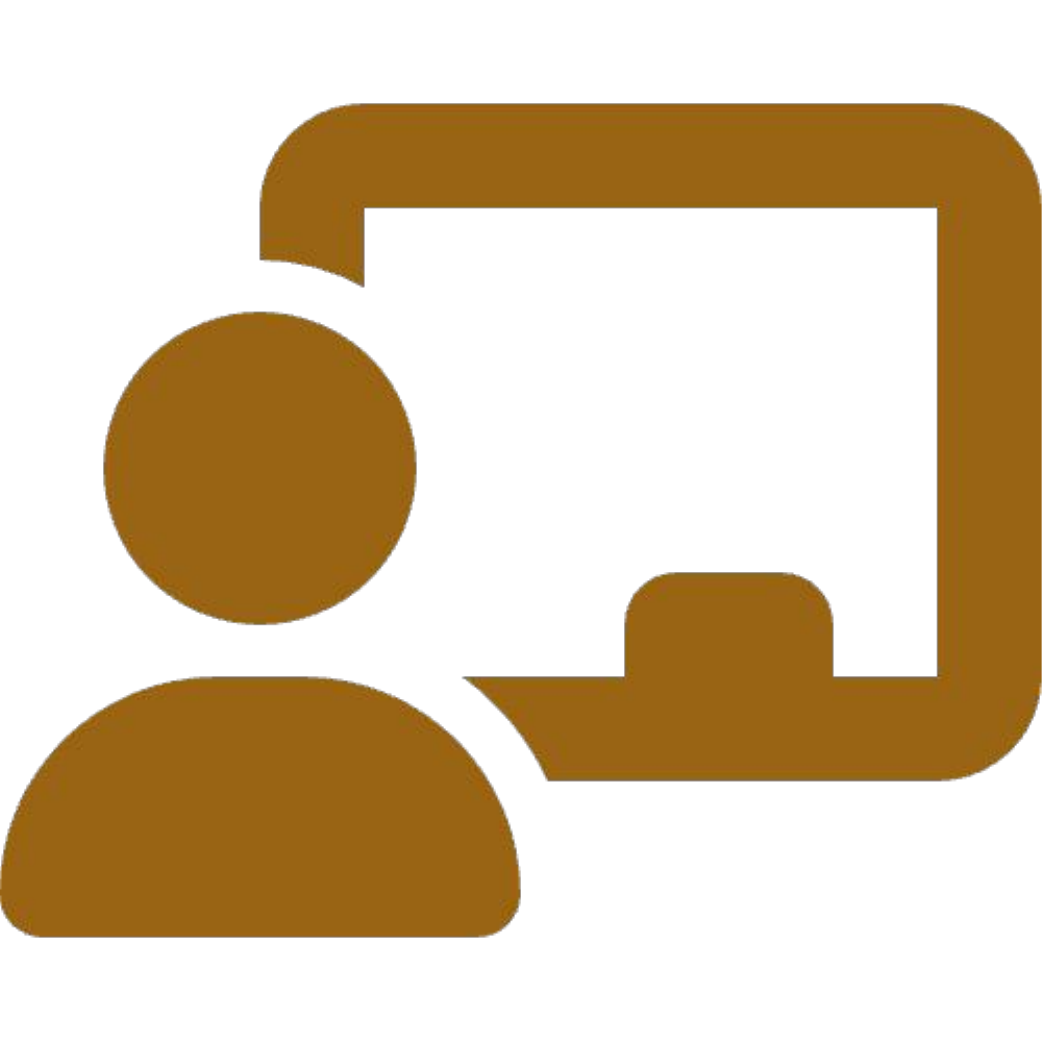}~Refinement}\endgroup\xspace}

\newcommand{\speechActRecommendation}{%
  \begingroup
    \setlength{\fboxsep}{2.5pt}%
    \colorbox{yellowbg}{%
      \includegraphics[height=0.8em]{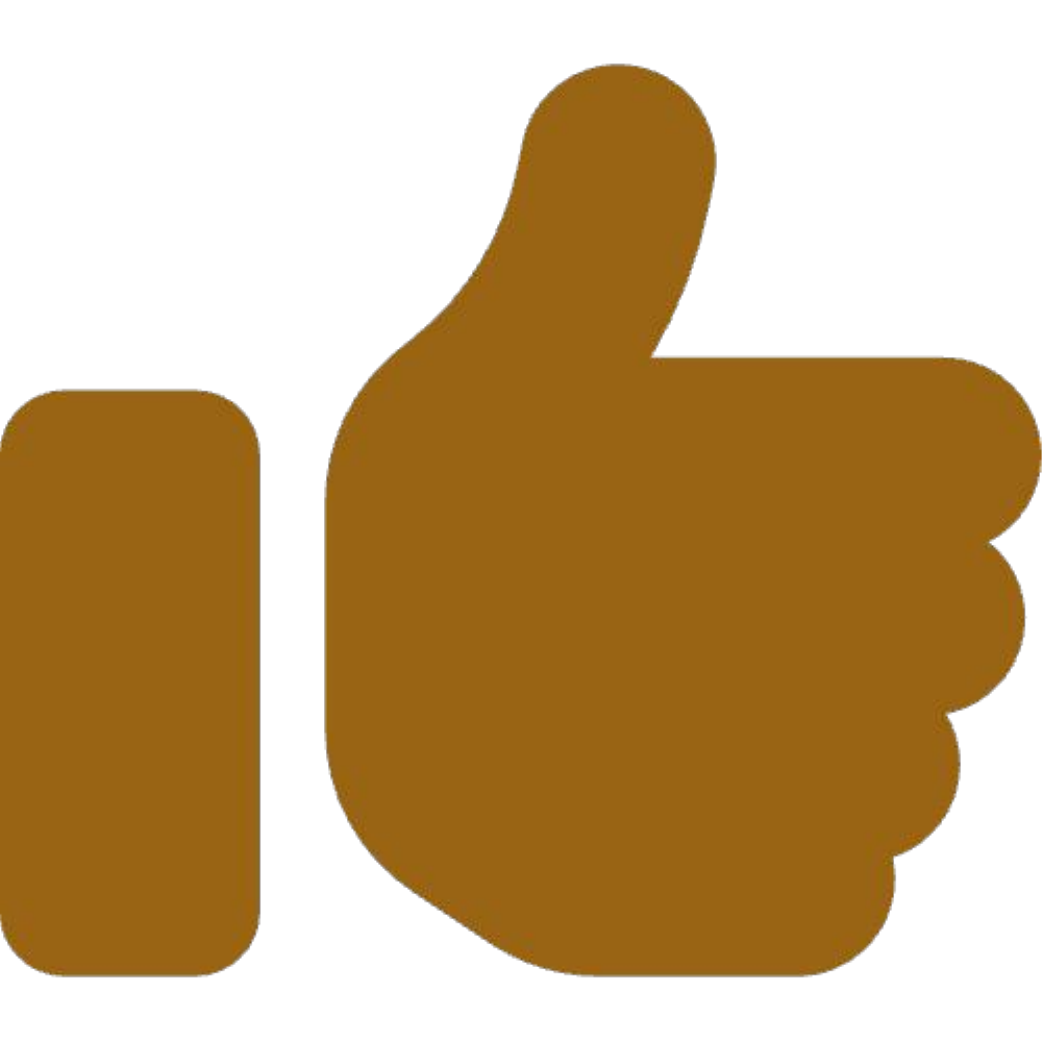}~Recommendation}\endgroup\xspace}

\newcommand{\speechActDomainKnowledge}{%
  \begingroup
    \setlength{\fboxsep}{2.5pt}%
    \colorbox{yellowbg}{%
      \includegraphics[height=0.8em]{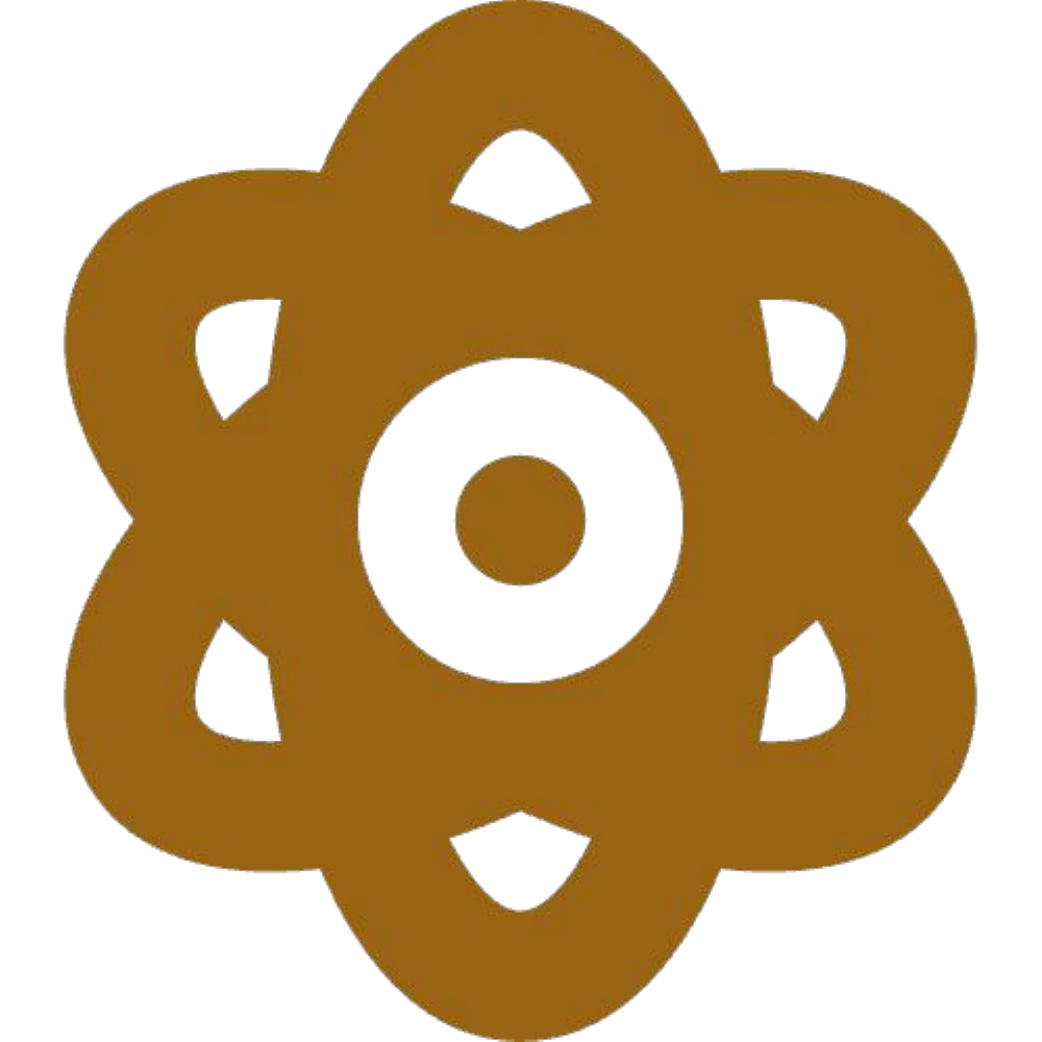}~Domain Knowledge}\endgroup\xspace}

\newcommand{\toturnbtn}{%
  \begingroup
    \setlength{\fboxrule}{0.3pt}%
    \setlength{\fboxsep}{2.5pt}%
    \fcolorbox{toturnBorder}{white}{%
      \includegraphics[height=0.7em]{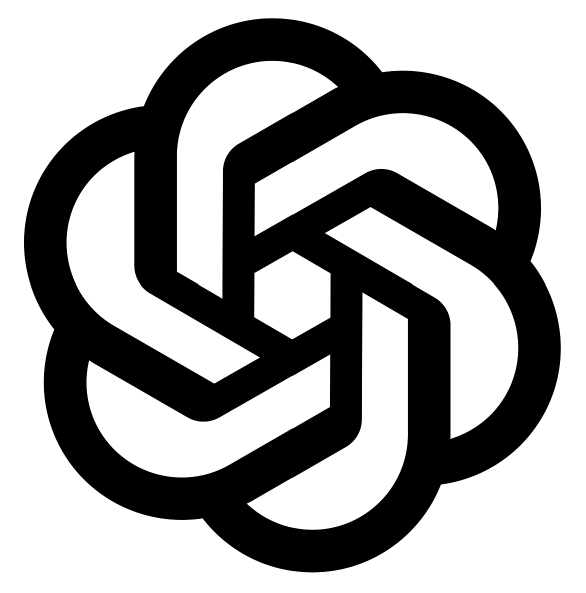}~To Turn}\endgroup\xspace}

\def\rqCharacteristics{What are the characteristics of analytical conversations?\xspace}

\def\rqInformationNeeds{How data workers’ information needs when revisiting analytical conversations shape how they re-enter and make sense of prior analyses\xspace}
\def\rqWorkflows{How data workers navigate and restructure analytical conversations to satisfy these needs, and what tensions and opportunities these workflows expose for tooling\xspace}
\def\rqAIRole{How data workers perceive the role of AI in structuring and communicating analytical conversations, and what forms of human–AI collaboration they prefer\xspace}

\definecolor{vlg}{RGB}{230,235,242}
\definecolor{impborder}{gray}{0.6} 

\newsavebox{\blockbox}

\aptLtoXcmd{\newenvironment{newimplications}{\begin{xmlelement*}{aptdispbox}\XMLaddatt{style}{background-color: \#E6EBF2; border: 0.6pt solid \#999999; padding: 6pt; display: block; }}{\end{xmlelement*}}}{
\newcommand{\implications}[1]{%
  \vspace{0.6em}
  \noindent
  {%
    \setlength{\fboxrule}{0.6pt}
    \setlength{\fboxsep}{6pt}
    \fcolorbox{impborder}{vlg}{%
      \parbox{0.95\columnwidth}{%
        \small
        #1
      }%
    }%
  }
  \vspace{0.6em}
}}

\soulregister\cite7
\soulregister\ref7
\soulregister\pageref7

\makeatletter

\newcommand{\taps@framealign}{%
  \hskip -\leftskip\relax
}

\newcommand{\taps@linespread}{%
  \setlength{\baselineskip}{1.3\baselineskip}%
}

\@ifundefined{titledbox@bg}{\newcommand{\titledbox@bg}{}}{}
\@ifundefined{titledbox@frame}{\newcommand{\titledbox@frame}{}}{}

\@ifundefined{titledbox}{%
  \newenvironment{titledbox}[3]{%
    \par\vspace{8pt}%
    \begingroup
    \def\titledbox@bg{#2}%
    \def\titledbox@frame{#3}%
    \def\FrameCommand##1{%
      \taps@framealign
      \fboxrule=0.8pt\relax
      \fboxsep=0pt\relax 
      \fcolorbox{\titledbox@frame}{\titledbox@bg}{##1}%
    }%
    \MakeFramed{\advance\hsize-\width \FrameRestore}%
    \noindent\colorbox{gray!70}{%
      \parbox{\hsize}{%
        \hspace*{8pt}\color{white}\bfseries #1\strut\hspace*{8pt}%
      }%
    }%
    \vskip 6pt%
    \begingroup
    \leftskip=8pt\relax
    \rightskip=8pt\relax
    \parindent=0pt\relax
    \taps@linespread
  }{%
    \par\endgroup
    \vskip 6pt%
    \endMakeFramed%
    \endgroup
    \par\vspace{8pt}%
  }%
}{%
  \renewenvironment{titledbox}[3]{%
    \par\vspace{8pt}%
    \begingroup
    \def\titledbox@bg{#2}%
    \def\titledbox@frame{#3}%
    \def\FrameCommand##1{%
      \taps@framealign
      \fboxrule=0.8pt\relax
      \fboxsep=0pt\relax
      \fcolorbox{\titledbox@frame}{\titledbox@bg}{##1}%
    }%
    \MakeFramed{\advance\hsize-\width \FrameRestore}%
    \noindent\colorbox{gray!70}{%
      \parbox{\hsize}{%
        \hspace*{8pt}\color{white}\bfseries #1\strut\hspace*{8pt}%
      }%
    }%
    \vskip 6pt%
    \begingroup
    \leftskip=8pt\relax
    \rightskip=8pt\relax
    \parindent=0pt\relax
    \taps@linespread
  }{%
    \par\endgroup
    \vskip 6pt%
    \endMakeFramed%
    \endgroup
    \par\vspace{8pt}%
  }%
}

\makeatother

\makeatletter
\@ifundefined{describedexamplebox}{%
  \newenvironment{describedexamplebox}{%
    \par\vspace{6pt}%
    \begingroup
    \def\FrameCommand##1{%
      \taps@framealign
      \fboxrule=0pt\relax
      \fboxsep=6pt\relax
      \fcolorbox{white}{gray!15}{##1}%
    }%
    \MakeFramed{\advance\hsize-\width \FrameRestore}%
    \taps@linespread
  }{%
    \endMakeFramed%
    \endgroup
    \par\vspace{6pt}%
  }%
}{%
  \renewenvironment{describedexamplebox}{%
    \par\vspace{6pt}%
    \begingroup
    \def\FrameCommand##1{%
      \taps@framealign
      \fboxrule=0pt\relax
      \fboxsep=6pt\relax
      \fcolorbox{white}{gray!15}{##1}%
    }%
    \MakeFramed{\advance\hsize-\width \FrameRestore}%
    \taps@linespread
  }{%
    \endMakeFramed%
    \endgroup
    \par\vspace{6pt}%
  }%
}
\makeatother

\newenvironment{promptbox}[1]{%
  \begin{titledbox}{#1}{gray!5}{gray!40}%
}{%
  \end{titledbox}%
}

\makeatletter
\@ifundefined{prompt}{%
  \newenvironment{prompt}[1][Prompt]{%
    \begingroup
    \@ifpackageloaded{lineno}{\nolinenumbers}{}%
    \promptbox{#1}%
  }{%
    \endpromptbox
    \@ifpackageloaded{lineno}{\linenumbers}{}%
    \endgroup
  }%
}{%
  \renewenvironment{prompt}[1][Prompt]{%
    \begingroup
    \@ifpackageloaded{lineno}{\nolinenumbers}{}%
    \promptbox{#1}%
  }{%
    \endpromptbox
    \@ifpackageloaded{lineno}{\linenumbers}{}%
    \endgroup
  }%
}
\makeatother

\newcommand{\messageseparator}{%
  \textcolor{lightgray}{\rule{\linewidth}{0.5pt}}\vspace{0.2cm}%
}

\newenvironment{systemmessage}{\textbf{System prompt:}\\}{}

\newenvironment{smalljinja}{%
    \small%
}{%
}

\definecolor{dimgray}{rgb}{0.41, 0.41, 0.41}

\newenvironment{describedexample}[2]{%
    \noindent\textit{#1}\\%
    \noindent%
    \begin{smalljinja}%
    \small
    \describedexamplebox
    \textcolor{dimgray}{\small\textsf{\uppercase{#2}}}\par%
    \vspace{0.4mm}%
}{%
    \enddescribedexamplebox
    \end{smalljinja}%
}

\newbool{firstelement}
\global\setbool{firstelement}{false}

\newlength{\iconwidth}
\newlength{\iconpadding}
\newenvironment{usermessage}
  {\ifbool{firstelement}{%
    \global\setbool{firstelement}{false}%
  }{\messageseparator}%
  \par\noindent
  \raisebox{-0.3\height}{\includegraphics[width=\iconwidth]{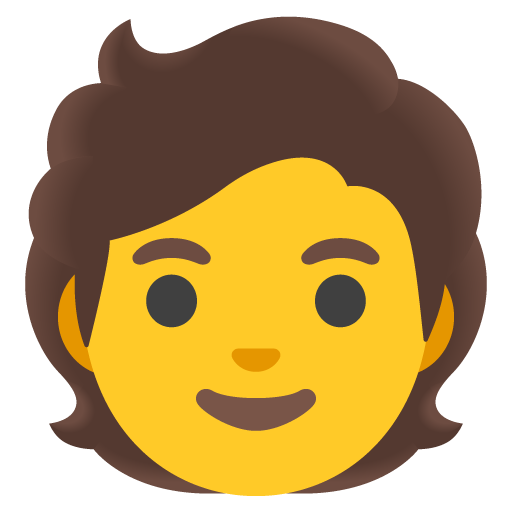}}%
  \hspace{\iconpadding}\textbf{User:}%
  \begin{adjustwidth}{\iconwidth+\iconpadding}{}%
  }
  {\end{adjustwidth}}

\newenvironment{assistantmessage}
  {\messageseparator%
  \par\noindent
  \raisebox{-0.3\height}{\includegraphics[width=\iconwidth]{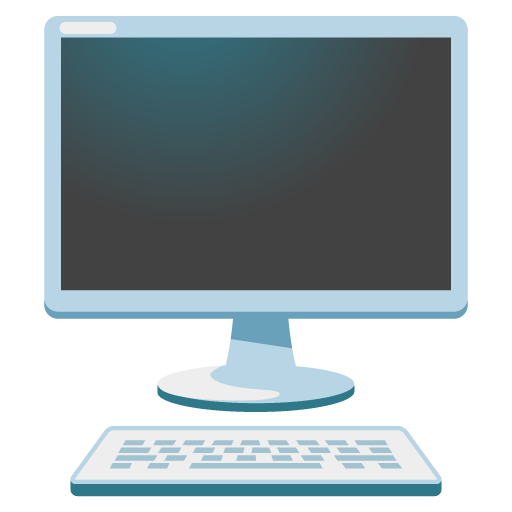}}%
  \hspace{\iconpadding}\textbf{Assistant:}%
  \begin{adjustwidth}{\iconwidth+\iconpadding}{}%
  }
  {\end{adjustwidth}}

\lstdefinestyle{jsonstyle}{
  basicstyle=\ttfamily\footnotesize,
  showstringspaces=false,
  breaklines=true,
  frame=none,
  upquote=true,
  literate=
    {:}{{{\color{purple}{:}}}}{1}
    {,}{{{\color{purple}{,}}}}{1}
    {\{}{{{\color{brown}{\{}}}}{1}
    {\}}{{{\color{brown}{\}}}}}{1}
    {[}{{{\color{brown}{[}}}}{1}
    {"}{{"}}{1} %
    {]}{{{\color{brown}{]}}}}{1},
}

\lstnewenvironment{jsonblock}[1][]%
  {\lstset{style=jsonstyle, language=Java, #1}}%
  {}

\makeatletter
\newcommand*\iftodonotes{%
  \if@todonotes@disabled
    \expandafter\@secondoftwo
  \else
    \expandafter\@firstoftwo
  \fi
}%
\makeatother

\definecolor{dandelion}{HTML}{FFD464}
\definecolor{limegreen}{HTML}{32CD32}

\theoremstyle{remark}

\newcommand*{\nameadjunct}{\relax}
\makeatletter
\renewcommand*{\NAT@nmfmt}[1]{\NAT@up #1\nameadjunct}
\makeatother

\definecolor{naivecolorname}{rgb}{0.122, 0.466, 0.70}

\definecolor{mrfcolorname}{rgb}{1.0, 0.5, 0.055}

\definecolor{mrflogitcolorname}{rgb}{0.17, 0.627, 0.17}

\definecolor{itercolorname}{rgb}{0.83, 0.153, 0.153}

\definecolor{hcbcolorname}{rgb}{0.58, 0.404, 0.747}

\pgfkeys{/csteps/fill color=black}
\pgfkeys{/csteps/inner color=white}

\ifdefined\cprotect\else\expandafter\newcommand\csname cprotect\endcsname[1][]{}\fi

\begin{document}


\title[How Data Workers Navigate, Make Sense of, and Communicate Analytical Conversations]{\textit{``I Need to Find That One Chart''}: How Data Workers Navigate, Make Sense of, and Communicate Analytical Conversations}

 \author{Ken Gu}
\orcid{0000-0002-4343-1578}
\affiliation{\institution{University of Washington}
\city{Seattle}
\state{Washington}
\country{USA}}
\email{kenqgu@cs.washington.edu}

\author{Srishti Palani}
\orcid{0000-0003-1805-7307}
\affiliation{\institution{Tableau Research}
\city{Palo Alto}
\state{California}
\country{USA}}
\email{srishti.palani@salesforce.com}

\author{Vidya Setlur}
\orcid{0000-0003-3722-406X}
\affiliation{\institution{Tableau Research}
\city{Palo Alto}
\state{California}
\country{USA}}
\email{vsetlur@tableau.com}

\renewcommand{\shortauthors}{Gu, et al.}



\begin{abstract}
Conversational interfaces are increasingly used for data analysis, enabling data workers to express complex analytical intents in natural language. Yet, these interactions unfold as long, linear transcripts that are misaligned with the iterative, nonlinear nature of real-world analyses. Revisiting and summarizing conversations for different contexts is therefore challenging. This paper investigates how data workers navigate, make sense of, and communicate prior analytical conversations. To study behaviors beyond those supported by standard interfaces (i.e., scrolling and keyword search), we develop a design probe that supplements analytical conversations with structured elements and affordances (e.g., filtering, multi-level navigation and detail-on-demand). In a user study ($n = 10$), participants used the probe to navigate and communicate past analyses, fulfilling information needs (recall, reorient, prioritize) through navigation strategies (visual recall, sequential and abstractive) and summarization practices (adding process details and context). Based on these findings, we discuss design implications to support re-visitation and communication of analytical conversations.
\end{abstract}


 

\maketitle

\section{Introduction}
\label{sec:intro}
Conversational interfaces powered by Large Language Models (LLMs) are now widely deployed as tools for data analysis~\cite{ChatGPT, microsoftCopilotExcel,TableauAgent2024, Fine2025DataScienceAgent}. These systems enable users to express diverse, often open-ended analytical goals and to iteratively explore data through natural language, without the need for programming. This significantly lowers the barriers to entry, broadening access for \textit{data workers} (i.e., individuals who access, analyze, manipulate, or communicate data to support decision-making, reporting, or operational tasks)~\cite{Crisan2020PassingTD, Muller2019HowDS, Kim2018DataSI, Kandel2012EnterpriseDA}.

While prior work has examined how data workers steer LLMs~\cite{Setlur2022HowDY, Kazemitabaar2024ImprovingSA}, execute analyses~\cite{Hong2023ConversationalAT}, and verify system outputs~\cite{Gu2023HowDA, Xie2024WaitGPTMA}, there is limited empirical understanding of how data workers interact with past analytical conversations to revisit, navigate, make sense of, and communicate their findings. These interactions are critical in real-world data workflows, such as sharing findings with business stakeholders or handing off analyses to collaborators~\cite{Pang2022HowDD, Roy2023, Piorkowski2021, Mao2019}, but are poorly supported by the linear transcripts produced by current conversational interfaces.

Beyond the known limitations of LLM-based conversational interfaces (e.g., a lack of structure, information overload~\cite{Jiang2023GraphologueEL, Suh2023SensecapeEM}), analytical conversations exacerbate common challenges of data analysis in computational notebooks. These include overlapping code variants, interleaved heterogeneous outputs (i.e., visualizations, data tables, and code execution outputs), and fragmented analysis trails
~\cite{Kery2019TowardsEF, Kery2017VarioliteSE}. These issues are further compounded by the abstraction of code in analytical conversations. As data workers no longer author code directly, they are less engaged in analytical decisions and less likely to retain a clear mental model of their analyses~\cite{Gu2023HowDA, Gu2023HowDD, Drosos2024ItsLA, Weng2024InsightLensDA}. This reduced engagement heightens the need for tools that support revisitation, navigation, and sensemaking. 

\rr{Revisitation and communication are fundamentally intertwined in analytical practice. Drawing on Pirolli and Card's sensemaking framework~\cite{Pirolli2007TheSP}, we recognize that foraging (i.e., navigating and finding information) and synthesis (i.e., constructing meaning) form an iterative loop rather than sequential stages. Communication acts as a synthesis activity that reveals gaps in understanding, prompting renewed navigation. This integration is well-established in computational notebook research. Kery et al.~\cite{Kery2018TheSI} showed that analysts develop narratives during exploration, not just afterward, the narrative in the notebook emerges through use over time. Rule et al.~\cite{Rule2018ExplorationAE} demonstrated that exploration and explanation are interleaved rather than distinct phases. Similarly, Wang et al.'s Callisto~\cite{Wang2020CallistoCT} revealed how conversational reasoning and computational narratives co-evolve.
For analytical conversations specifically, this coupling is even more critical. Unlike notebooks where code cells provide natural breaking points, conversational transcripts are continuous and linear. The same structures (threads, insights, artifacts) that support personal reorientation also enable external communication. Analysts often revisit conversations with communicative intent to prepare presentations, brief colleagues, or hand off analyses. Separating these activities would miss how analysts actually work. Thus, our study treats navigation and communication as inter-related activities sharing infrastructure (structured elements, filtering, abstraction) while serving both personal sensemaking and collaborative goals. This design choice reflects the reality that understanding one's analysis and explaining it to others are mutually reinforcing processes.}

While LLMs could be prompted to help navigate and communicate analytical conversations, their opaque generative processes offer little visibility into the provenance of their outputs, the rationale for the inclusion or exclusion of summarized content, or how that content aligns with communicative intent~\cite{Liao2023AITI, Hoque2024}. Thus, there is a critical need to understand how data workers currently navigate, make sense of, and communicate analytical conversations, and what design affordances can support these activities while preserving their agency.

In this work, we seek to address this gap through a user study that investigates the information needs and workflow strategies when \textit{navigating} and \textit{communicating} analytical conversations (\S\ref{sec:study_goals}).  We also explore the broader context of these behaviors, including how data workers engage with LLM-driven conversations and the role of AI in supporting navigation and communication.

 To observe a wider range of strategies than standard chat interfaces afford, we developed a \rr{high-fidelity} design probe, \syncsense~ \rr{for surfacing diverse behaviors and reflections.} \syncsense~augments conversational interfaces with structured elements and affordances such as filtering and multi-level navigation~(\S\ref{sec:probe}), \rr{offering participants rich scaffolding to explore how they might revisit, structure, and communicate prior analyses. Rather than evaluating a fixed solution, our goal was to utilize the tool as a lens into emerging practices}. Using this probe, we conducted a user study with 10 data workers across two sessions~(\S\ref{sec:study}). Participants revisited their analytical conversation conducted at least one week prior and summarized it for two distinct audiences (a non-technical executive and a technical analyst) across two formats: a chat message and a data report.

Our findings (\S\ref{sec:results}) reveal recurring information needs, including \textit{re-orienting} within a past conversation, \textit{prioritizing} relevant information, and \textit{recalling} specific details. Participants employed strategies such as sequential navigation, using visual cues as memory triggers, moving between levels of abstraction, and applying filters. When composing summaries, participants made manual edits such as adding audience context, reordering content, and formatting, preferring to retain control over core message curation. While participants valued AI assistance for refining tone, technicality, and length, they resisted full automation, favoring AI as a scaffolding and stylistic aid. We conclude with design implications for future conversational interfaces supporting analytical workflows~(\S\ref{sec:discussion}). 
\noindent Through this work, we make the following research contributions:
\begin{enumerate} 
  \item  Empirical observations from a two-session user study with $10$ participants. The study addresses a critical gap in understanding how data workers revisit, make sense of, and communicate findings from analytical conversations, an increasingly common interface for data analysis. To support future research, we release our dataset of participants' LLM-mediated analytical conversations.  
  \item An open-source high-fidelity design probe that supplements conversational interfaces with structured elements of analytical conversations and affordances to support observations of user needs and behaviors beyond standard chat interfaces.
  \item Design implications for future tools, emphasizing structured elements for navigation and communication, communication as a core and co-evolving analytical activity, and controllable, contextual AI assistance.
\end{enumerate}

\vspace{-3.3mm}

\section{Related Work}
Our work sits at the intersection of three key research themes: conversational interfaces for data analysis, communication challenges in collaborative data work, and sensemaking tools that support structure and dynamic abstraction.

\subsection{Conversational Interfaces for Data Analysis}
Conversational interfaces increasingly support data exploration by enabling natural language interactions that lower barriers to analysis. Early systems focused on intent parsing and visualization generation using rule-based or template-driven approaches~\cite{powerbi, ibmwatson}. These tools gradually improved to support ambiguity resolution~\cite{Gao2015DataToneMA, setlur2016eviza}, follow-up queries~\cite{srinivasan2018orko}, and more sophisticated dialogue structures~\cite{setlur2019inferencing, iris, skantze2021turn,jiang2023communitybots}. More recently, the integration of LLMs has significantly expanded conversational capabilities in analytical tools, enabling more fluid and expressive interactions across spreadsheets~\cite{microsoftCopilotExcel}, computational notebooks~\cite{Wang2025JupybaraOA, TableauAgent2024}, and analytical conversational interfaces~\cite{Hong2023ConversationalAT, Xie2024WaitGPTMA}. These systems streamline how data workers formulate analysis plans, refine workflows, and derive insights, relying on natural language as the primary interaction medium.

However, these capabilities come with new challenges. Navigation challenges emerge as conversations grow long and complex, with data workers struggling to locate specific analyses, revisit earlier findings, or understand the progression of their work~\cite{Kazemitabaar2024ImprovingSA, Xie2024WaitGPTMA}. The flexibility of natural language makes it difficult to precisely express analytical intent, hindering data workers' ability to effectively steer model behavior and leading to misaligned or incomplete analyses~\cite{Xie2024WaitGPTMA, Kazemitabaar2024ImprovingSA}.

Sensemaking challenges amplify when AI executes analysis steps directly from high-level prompts, producing outputs like code, tables, and visualizations without user intervention~\cite{Gu2023HowDD}. This adds cognitive load, as data workers must interpret both the AI's responses and the resulting artifacts. While conversational transcripts preserve entire interaction histories, their linear structure obscures the true complexity of analytical work, including branching threads, shifting topics, and evolving goals that characterize real-world data analyses~\cite{Head2019ManagingMI, Kery2018TheSI, Rule2018ExplorationAE}. Post-hoc review becomes difficult, particularly when conversations involve sequential utterances that incrementally modify queries, analyses, or visual encodings~\cite{Kazemitabaar2024ImprovingSA, Weng2024InsightLensDA, Setlur2022HowDY, Hong2023ConversationalAT}. These difficulties also create barriers when conversations must be repurposed for communication, a challenge we discuss in detail next.

\subsection{Communication and Collaboration in Data Analysis}
\looseness-1Communicating data analyses represents a central but challenging aspect of data work. Team members often differ in domain expertise, tooling familiarity, and documentation habits, making it difficult to maintain shared context~\cite{Wang2021DocumentationMH, Muller2019HowDS, Zhang2020HowDD, Crisan2020PassingTD, Bopp2017DisempoweredBD}. Breakdowns frequently arise when analytical decisions, assumptions, and uncertainties are not clearly communicated, leaving both technical and non-technical stakeholders struggling to align~\cite{Crisan2020PassingTD, Zhang2020HowDD}. Misaligned expectations around analysis depth, presentation style, and tolerance for uncertainty further complicate collaboration across roles~\cite{Wang2021WhatMA}. 

Since analyses reflect the data worker’s unique mix of statistical, domain, and coding expertise, they need to be organized and reframed for different audiences and media~\cite{Wang2021AutoDSTH, Kim2016TheER, Wang2021WhatMA, Piorkowski2021HowAD}. Traditional analytical mediums like computational notebooks and spreadsheets present well-documented challenges. Documentation is often incomplete, inconsistent, or disconnected from code and results~\cite{Wang2021DocumentationMH, Wang2021WhatMA}. Creating comprehensive documentation is time-consuming and often deprioritized~\cite{Pang2022HowDD}, which leads to difficulties when sharing work with colleagues or revisiting analyses later~\cite{Zhang2020HowDD, Smith2017}. In response, researchers have developed tools for enriching computational notebooks with better documentation~\cite{Wang2021DocumentationMH}, generating slides from notebook content~\cite{Wang2023Slide4NCP, Zheng2022TellingSF}, and supporting audience-aware communication~\cite{Wang2024OutlineSparkIA, Li:2024, iceberg:2024}. \rr{While computational notebooks remain the dominant medium for organizing and communicating analysis, our focus is on analytical conversations that increasingly precede or even replace notebooks as the primary locus of work when LLMs are involved.}

Conversational interfaces exacerbate many of these communication challenges. Unlike notebooks or spreadsheets, which modularize work into cells or sheets, conversational analyses persist as long, static, linear transcripts where code, visualizations, natural language explanations, and iterative refinements are entangled. The exploratory nature of data analysis also means that false starts and dead ends persist in the transcript, adding noise that complicates downstream communication. To this end, conversational interfaces create significant challenges for navigation, sensemaking, and communication that current tools inadequately support. To explore how these challenges might be addressed, we use a design probe to study how data workers revisit and summarize analytical conversations for different audiences and media.

\subsubsection{\rr{Analytical Conversations vs. Computational Notebooks}}

\rr{Computational notebooks and analytical conversations share various high-level similarities; both interleave code, narrative text, and visual outputs, and serve as records of exploratory analysis~\cite{Head2019ManagingMI, Rule2018ExplorationAE}. However, their structure, context, and interaction patterns differ in ways that affect how users navigate the analytical process and communicate findings.}

\rr{First, notebooks are organized around cells that the analyst can re-order, delete, or rewrite. This feature enables notebooks to be a \emph{curated} representation of work: analysts often clean up or linearize their exploratory process before sharing~\cite{Rule2018ExplorationAE, Kery2018TheSI}. In contrast, analytical conversations are organized around \emph{turns} in a dialogue, with strict temporal turn-taking between the user and the AI chat interface. False starts, dead ends, and incremental refinements persist in the transcript and cannot be easily reordered without losing conversational coherence. As a result, while analytical conversations tend to more faithfully preserve the exploratory nature of analysis, their fidelity can obscure the underlying reasoning structure, complicating efforts to summarize or communicate findings for external stakeholders.}

\rr{Second, computational notebooks typically assume a single human author who is responsible for both writing code and providing explanatory text, even when large language models (LLMs) are used to generate code snippets that are subsequently pasted into cells~\cite{Wang2025JupybaraOA}. In contrast, analytical conversations introduce a qualitatively different form of co-authorship, in which the AI not only generates code but also executes it, suggests lines of analysis, and provides narrative interpretations inline. This blurs the boundary between user intent, system action, and explanation, and means that key analytical decisions are distributed across many turns, rather than encapsulated in a small number of curated cells. Existing notebook tools for documentation, slide generation, and audience-aware communication~\cite{Wang2021DocumentationMH, Wang2023Slide4NCP, Wang2024OutlineSparkIA} operate on the assumption that the analytic record is already structured into cells; they do not help users recover structure from raw conversational transcripts where that assumption does not hold.}

\rr{Third, interaction patterns differ. Notebooks support \emph{direct manipulation} of code and artifacts: users execute cells, inspect outputs, and annotate results in place. By contrast, analytical conversations rely on \emph{natural language} turn-taking to steer analysis; users describe what they want, and the AI responds with code, tables, and visualizations~\cite{Gu2023HowDD, Hong2023ConversationalAT}. This indirection makes it harder to specify precise analytical operations~\cite{Xie2024WaitGPTMA, Kazemitabaar2024ImprovingSA} and, when revisiting a conversation, harder to see how particular artifacts relate to evolving goals. Our findings around orienting, recalling, and prioritizing (\S\ref{sec:results_info_needs}) highlight that users face distinct information needs when revisiting conversational logs that are not addressed by notebook execution histories or provenance tools alone~\cite{Wu2020B2BC, Wang2022DiffIT}.}

\rr{While prior work improves how analysts document, refactor, and communicate notebooks once the analysis has been curated into cells~\cite{Wang2021DocumentationMH, Wang2021WhatMA, Wang2023Slide4NCP}, our work studies how data workers revisit and summarize \emph{analytical conversations} \textit{before} any such curation may exist: when the primary record of work is a long, entangled chat with an LLM. Our design probe and empirical study focus on this conversational substrate, examining how structure and dynamic abstraction can help users reconstruct, navigate, and communicate analyses directly from conversational histories, rather than assuming a notebook-style representation is already in place.}

\subsection{Structure and Dynamic Abstraction Support Navigation and Sensemaking}
Sensemaking involves not only accessing information, but also the ability to organize and abstract it to reveal meaning and support decision-making~\cite{paulmorris:2011}. Research across domains in HCI has consistently shown that introducing structure, whether system-defined or user-defined, significantly enhances users' ability to make sense of complex content~\cite{Sultanum2021TextVA, Pavel2015SceneSkimSA, Fok2023QlarifyRE, August2022PaperPM}. In analytical workflows, this need is well-documented in computational notebooks, where non-linear exploration and fragmented reasoning can obscure analytical intent~\cite{Head2019ManagingMI, Kery2019TowardsEF, Kery2017VarioliteSE, Rule2018ExplorationAE, Wu2020B2BC, Wang2022DiffIT}.

Critical to sensemaking is the ability to dynamically abstract information at multiple levels of granularity~\cite{Suh2023SensecapeEM, Zhang2017WikumBD, Zhang2018MakingSO}. Rule et al.~\cite{Rule2018ExplorationAE} demonstrated how visual hierarchy and structural cues support both comprehension and collaborative review. Recent work has begun exploring structure in analytical conversations specifically. InsightLens~\cite{Weng2024InsightLensDA} introduced structured representations around insights extracted from analytical conversations, though it remained focused on isolated insight units. DS-Agent~\cite{ds-agent:guo2024} demonstrated how structuring analytical workflows can scaffold LLM performance, suggesting benefits for both human comprehension and automated support. Our work builds on this research by examining how users engage with the broader conversational structure, including speech acts, conversation turns, and analytical threads, and how such structure can support user-driven sensemaking and communication goals.

\section{Structured Elements of Analytical Conversations}
\label{sec:components}
To ground our investigation, we examine the structure of analytical conversations to identify where users may benefit from support. Drawing from discourse analysis, conversational systems, and data analysis workflows, we \rr{consolidate elements that reveal the layered, heterogeneous nature of these conversations. Specifically, we adapted components such as speech acts (e.g., inform, request, justify) from discourse and pragmatic analysis to capture the functional intent of utterances; topic threads and turn-taking structures from conversation analysis to represent shifts in focus or sub-goals; and insights and artifacts (e.g., charts, tables, filters) from prior work on analytic reasoning and notebook use to highlight the products and intermediate results of analysis~\cite{Setlur2022HowDY,Wang2019HowDS}. We then reviewed these taxonomies and merged overlapping constructs into a coherent schema focused on revisitation and communication. Rather than aiming for exhaustiveness, our goal was to provide a practical scaffold that helps participants interpret, organize, and act on prior conversational content.}

\xhdr{Turns} A \textit{turn} is the basic unit of an analytical conversation, i.e., a user query and a conversational interface response~\cite{2023ChatbotsD}. An analytical conversation is a sequence of \textit{turn-taking} between the user and the conversational interface, often stretching across dozens of back-and-forth exchanges. 

\xhdr{Threads} Across these turns, \textit{analysis threads} group related turns around a shared analytical goal or topic. These can nest within one another and often reflect the iterative, non-linear nature of data exploration~\cite{Kery2018TheSI, Grolemund2014ACI, Klein2007ADT, Koesten2019TalkingDU, Pirolli2007TheSP}. 

\xhdr{Speech Acts} User utterances take different functional roles, or \textit{speech acts}~(i.e., utterances that perform an action by the user~\cite{Austin1962HowTD, Searle1969SpeechAA}). In analytical conversations, there can be a wide range of analysis-specific \textit{speech acts} from ``fact-finding'' to seeking ``specific visualizations'' to ``debugging''~\cite{Setlur2022HowDY}.

\xhdr{Analysis Artifacts}
Many responses include structured outputs (i.e., visualizations, data tables, and code) separate from natural language text that the analytical conversational interface generates. These~\textit{artifacts} embody critical steps in the analysis, but are intermixed with long passages of text and can be difficult to locate later. 

\xhdr{Analysis Insights} Finally, conversations surface~\textit{data insights} that represent critical outcomes of an analysis~\cite{Weng2024InsightLensDA, Ding2019QuickInsightsQA, Keim2002InformationVA, Chen2009TowardEI}. In analytical conversations, prior work underscored the challenge of uncovering insights, as they are often buried within back-and-forth exchanges, lengthy text, and embedded artifacts~\cite{Weng2024InsightLensDA}.

\section{Study Goals and Design Rationale}
\label{sec:study_goals}
Our study explores four research questions:
\begin{itemize}
    \item \textbf{RQ1} \rqCharacteristics
    \item \textbf{RQ2} \rqInformationNeeds
    \item \textbf{RQ3} \rqWorkflows
    \item \textbf{RQ4} \rqAIRole
\end{itemize}

\rr{These research questions intentionally couple navigation (i.e., revisiting past analyses) with communication (i.e., summarizing for audiences) because of their overlapping demands, situated relevance, and theoretical interdependence. Both tasks require similar affordances (e.g., filtering, abstraction, content selection), suggesting shared underlying infrastructure. From a practical perspective, revisitation and communication frequently co-occur in real-world settings; analysts often return to prior conversations in order to prepare presentations or brief stakeholders, making it artificial to study navigation in isolation~\cite{Mao2019, Pang2022HowDD, Piorkowski2021}. Moreover, theoretical frameworks such as sensemaking models~\cite{Pirolli2007TheSP} and prior research on computational notebooks~\cite{Rule2018ExplorationAE, Kery2018TheSI, Wang2020CallistoCT} emphasize that understanding and explanation are not discrete steps, but are rather iterative and mutually reinforcing processes. By examining these activities together, we can better understand how the structured elements support both personal sensemaking and external communication, while also revealing where user needs diverge.}

Standard conversational interfaces, limited to linear scrolling and keyword search, constrain users' ability to revisit, structure, or communicate analytical conversations. To investigate how data workers engage with these conversations, we developed a high-fidelity design probe that augments a conventional interface with structured navigation and summary authoring features. Rather than serving as a solution, this probe functions as a research instrument that structures elements of an analytical conversation (\textbf{RQ1}) and elicits diverse usage behaviors. Specifically, the probe enables us to investigate data workers' information needs (\textbf{RQ2}), workflow strategies (\textbf{RQ3}), and preferred roles of AI (\textbf{RQ4}) in analytical conversations.

\subsection{Design Considerations for the Probe}

\rr{Our research probe serves as shared infrastructure for both navigation and communication tasks. When designing \syncsense, we incorporate affordances drawn from established analytical tools like Jupyter notebooks, spreadsheets, and code repositories, but these features remain unexplored in the context of analytical conversations. The following design considerations prioritize flexible interaction methods that expand beyond scrolling and keyword search without prescribing specific usage patterns, enabling data workers to naturally adopt their preferred approaches for both personal reorientation and external communication. This approach allows us to observe where navigation and communication share needs (RQ2, RQ3) and where they diverge.}

\xhdr{DC1: Layer structured elements over raw conversations}
Prior work demonstrates that structured representations of unstructured discussions (e.g., forums, clinical notes) and LLM outputs (e.g., hierarchical views, structured fields) reveal different user behaviors and navigation patterns compared to raw logs or flat summaries~\cite{Zhang2017WikumBD, Sultanum2019DoccurateAC, Suh2023SensecapeEM, Jiang2023GraphologueEL}. Our probe should layer structured representations of analytical conversation elements (\S\ref{sec:components}) over the original conversation transcript, while preserving full access to the raw conversation. This design allows data workers to toggle between structured and unstructured views, enabling us to observe how they engage with different forms of representation for navigation, sensemaking, and communication.

\xhdr{DC2: Support overview and detail on-demand}
To manage long, heterogeneous conversations, data workers need flexible control over information granularity. Prior work has demonstrated that overview structures help users re-orient to the conversation's scope~\cite{Shneiderman1996TheEH, Suh2023SensecapeEM}, filters surface relevant content and reduce cognitive load~\cite{Hearst2006Faceted, Bates1989Berrypicking}, and progressive disclosure reduces overload by revealing details incrementally~\cite{Furnas1986Fisheye}. Our probe should integrate these affordances to let data workers fluidly shift between high-level overviews and deeper exploration, allowing us to observe diverse strategies for managing conversational complexity.

\xhdr{DC3: Support modular composition from conversation elements} 
Modularity and reuse support flexible re-composition for varied goals~\cite{Shipman1999Formality, Marshall2005Encountered, Zheng2022NB2Slides, Lin2024RamblerSW}. The probe should enable participants to select and recombine structured conversation elements into new compositions. This interaction will enable us to observe what users prioritize, how they filter and organize content, and which strategies they use to communicate insights to varying audiences.




\begin{figure*}[t!]
    \centering
 \includegraphics[width=0.98\textwidth]{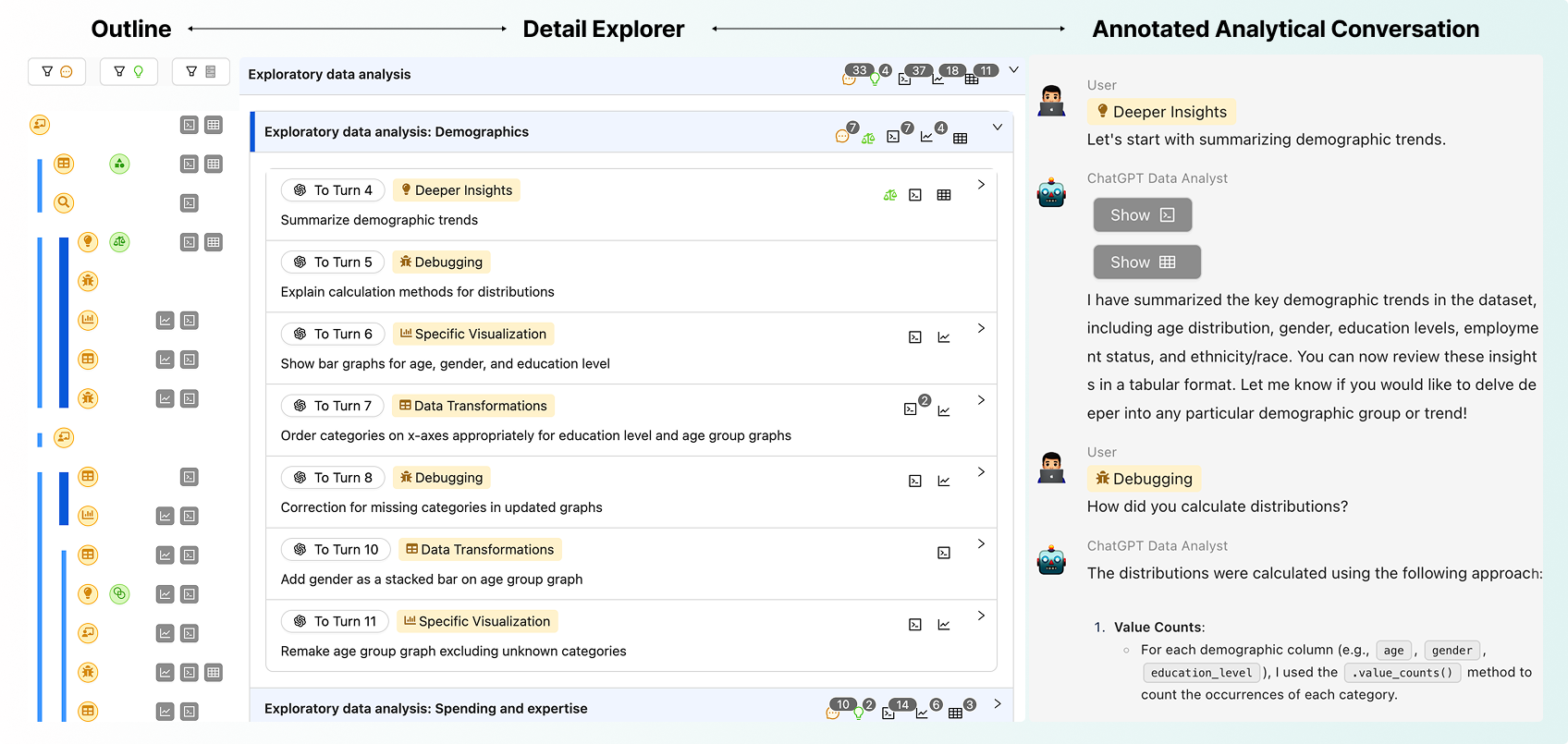}
  \caption{\syncsense~extracts and presents elements of an analytical conversation at varying levels of abstraction. The \chatTimeline panel provides a high-level overview of the conversation elements, the \chatContents panel displays additional details about each element, and the \origChat panel shows the raw conversation annotated with each component. These three synchronized panels are connected visually and interactively through the structured elements of an analytical conversation (\S\ref{sec:adding_structure}).}
  \label{fig:teaser}
\end{figure*}

\section{Structured Interface for Revisiting Conversations}
\label{sec:probe}
We introduce \syncsense, a design probe that augments analytical conversations with structured elements to support revisitation and summarization. Rather than offering a fully featured or optimized product, \syncsense~intentionally exposes multiple alternative representations of an analytical conversation, i.e., threads, artifacts, insights, and authored summaries within a four‑panel interface. By surfacing content at different levels of abstraction and placing revisitation and communication side‑by‑side, the probe is designed to provoke reflection and elicit diverse strategies for orienting, navigating, and shaping analytical narratives. Its purpose is not to prescribe an ideal workflow, but to create conditions in which participants reveal their information needs, breakdowns, and preferred patterns of interaction that would otherwise remain hidden in conventional chat interfaces or more polished prototypes. 

\subsection{Adding Structure to an Analytical Conversation}
\label{sec:adding_structure}
To surface analytical conversation elements across the interface (\textbf{DC1}), \syncsense~extracts components identified in \S\ref{sec:components} and displays them using icons and thematically colored tags. These visual cues help foreground latent structures in the conversation to elicit a range of user behaviors and reflections.

\xhdr{\componentThreads} Threads and nested sub-threads (where a thread can be nested within a larger thread) offer natural levels of abstraction to manage growing conversation complexity~\cite{Hong2023ConversationalAT}. As analytical conversations unfold, they often involve topic shifts, branching inquiries, exploratory detours, and iterative refinements; properties that often challenge linear models of interaction. To surface this non-linearity, \syncsense~automatically groups related (potentially non-consecutive) turns into threads and organizes them into higher-level parent threads. This feature was introduced to probe how participants perceive, interpret, and traverse complex conversation structures.

\xhdr{\componentSpeechAct} To contextualize and organize conversations, \syncsense~categorizes user prompts into analysis-specific speech acts. This tagging helps surface functional distinctions between turns, such as data requests, follow-ups, visualizations, or refinement prompts, making it easier for users to locate relevant parts of the conversation. Building on Setlur et al.'s taxonomy~\cite{Setlur2022HowDY}, we extend the categorization to better capture the broader range of low-level and iterative interactions that commonly arise in LLM-assisted workflows~\cite{Gu2023HowDA, Kazemitabaar2024ImprovingSA, Rogers2023TracingAV}. Table~\ref{tab:speechacts} shows the speech act categories classified in~\syncsense. By revealing the functional intent of utterances through speech act tags, the probe invites participants to interpret not just the content but the purpose behind each turn. This feature helps elicit insights into how users mentally organize their analytical process, whether they favor goal-driven exploration, iterative refinement, or opportunistic branching.


\begin{table}[h]
\centering
\small
\setlength{\tabcolsep}{4pt}
\begin{tabular}{@{}p{0.35\columnwidth} p{0.64\columnwidth}@{}}
\toprule
\textbf{Speech Act Type} & \textbf{Definition and Example} \\
\midrule
\speechActFactFinding & Expecting a single response like a number, yes/no etc. (e.g., \textit{``Did any passengers survive?''}) \\

\speechActDeepInsights & Seeking deeper insights (e.g., \textit{``How much more likely was survival in first class?''}) \\

\speechActSpecificViz & Requesting a specific type of visualization (e.g., \textit{``Plot a bar chart of passengers by class.''}) \\

\speechActDataTransformations & Asking for data manipulations such as filtering, sorting, or aggregating (e.g., \textit{``Can you filter to only show first class passengers?''}) \\

\speechActRefinement & Refining a previous query or asking for more detail (e.g., \textit{``Show survival rates by age group.''}) \\

\speechActRecommendation & Requesting recommendations or predictions based on the data (e.g., \textit{``What is the best course of action for the airline?''}) \\

\speechActDomainKnowledge & Asking about domain knowledge not directly present in the dataset (e.g., \textit{``What does first class mean in this context?''}) \\

\speechActDebugging & Debugging code or visuals (e.g., \textit{``Why are there very few counts for first class passengers.''}) \\
\bottomrule
\end{tabular}
\caption{Speech act types, definitions, and examples.}
\label{tab:speechacts}
\end{table}

\xhdr{\componentArtifact} \syncsense~distinguishes between analysis artifacts (i.e., \artifactViz, \artifactTable, and \artifactCode). \textit{Visualizations} and \textit{data tables} serve as essential elements in the presentation of data analyses~\cite{Wang2023Slide4NCP} while \textit{code and execution outputs} support provenance and verification of how visualizations and tables were derived~\cite{Gu2023HowDA,Xie2024WaitGPTMA}. \rr{Within the design probe, making these artifacts explicitly visible and filterable allows us to investigate how users revisit and prioritize different outputs depending on their goals (e.g., reporting vs. verification). This design choice lets us observe whether users gravitate toward concrete outputs when orienting themselves in long conversations, and how these anchors shape summarization workflows.}

\xhdr{\componentInsight} \syncsense~captures analytical insights as structured elements, linking them with one or more conversation turns. Insights are categorized using the taxonomy proposed by Wang et al.~\cite{Wang2020DataShotAG} (Table~\ref{tab:insighttypes}), which was derived from analyzing $793$ insights across $245$ fact sheets focused on tabular data. \rr{Structuring insights in this way invites participants to reflect on the nature and diversity of their findings and helps supports abstraction at the right level of granularity.}


\begin{table}[h]
\centering
\small
\setlength{\tabcolsep}{4pt}
\begin{tabular}{@{}p{0.25\columnwidth} p{0.7\columnwidth}@{}}

\toprule
\textbf{Insight Type} & \textbf{Definition and Example} \\
\midrule
\insightValue & Retrieve the exact value of data attribute(s) under specific criteria (e.g., \textit{``46 horses have won two out of three Triple Crown races.''}) \\

\insightProportion & Measure the proportion of selected data attribute(s) within a set (e.g., \textit{``Protein takes 66\% in the diet.''}) \\

\insightDifference & Compare two or more data attributes (e.g., \textit{``There are more blocked beds in the Royal London Hospital compared with the UK average.''}) \\

\insightDistribution & Show how values are shared across or broken down by data attributes (e.g., \textit{``The distribution of the unicorn companies is approximately normal over their age.''}) \\


\insightTrend & Present a general tendency over a time segment (e.g., \textit{``The Border Patrol budget rose from 1990 to 2013.''}) \\

\insightRank & Sort data attributes by their values and show rankings (e.g., \textit{``The top reason for consumers to engage in showrooming is 'the price is better online'.''}) \\

\insightAggregation & Calculate descriptive statistics (e.g., average, sum, count) (e.g., \textit{``The average price for gas is \$4.06.''}) \\

\insightAssociation & Identify relationships between data attributes (e.g., \textit{``There is a negative correlation between the number of food vendors and their distance from the market.''}) \\

\insightExtreme & Identify extreme data cases within a certain range or along attributes (e.g., \textit{``The character with the most epigrams is Oscar Wilde himself, with 12.''}) \\

\insightCategorization & Select data attribute(s) that satisfy specific conditions (e.g., \textit{``Josh and Sam are two popular names in 2004.''}) \\

\insightOutlier & Explore unexpected or statistically deviant data points (e.g., \textit{``‘Lucy' has the highest word count.''}) \\
\bottomrule
\end{tabular}
\caption{Insight types, their definitions, and examples.}
\label{tab:insighttypes}
\end{table}

\begin{figure}[t!]
    \centering
    \includegraphics[width=0.93\linewidth]{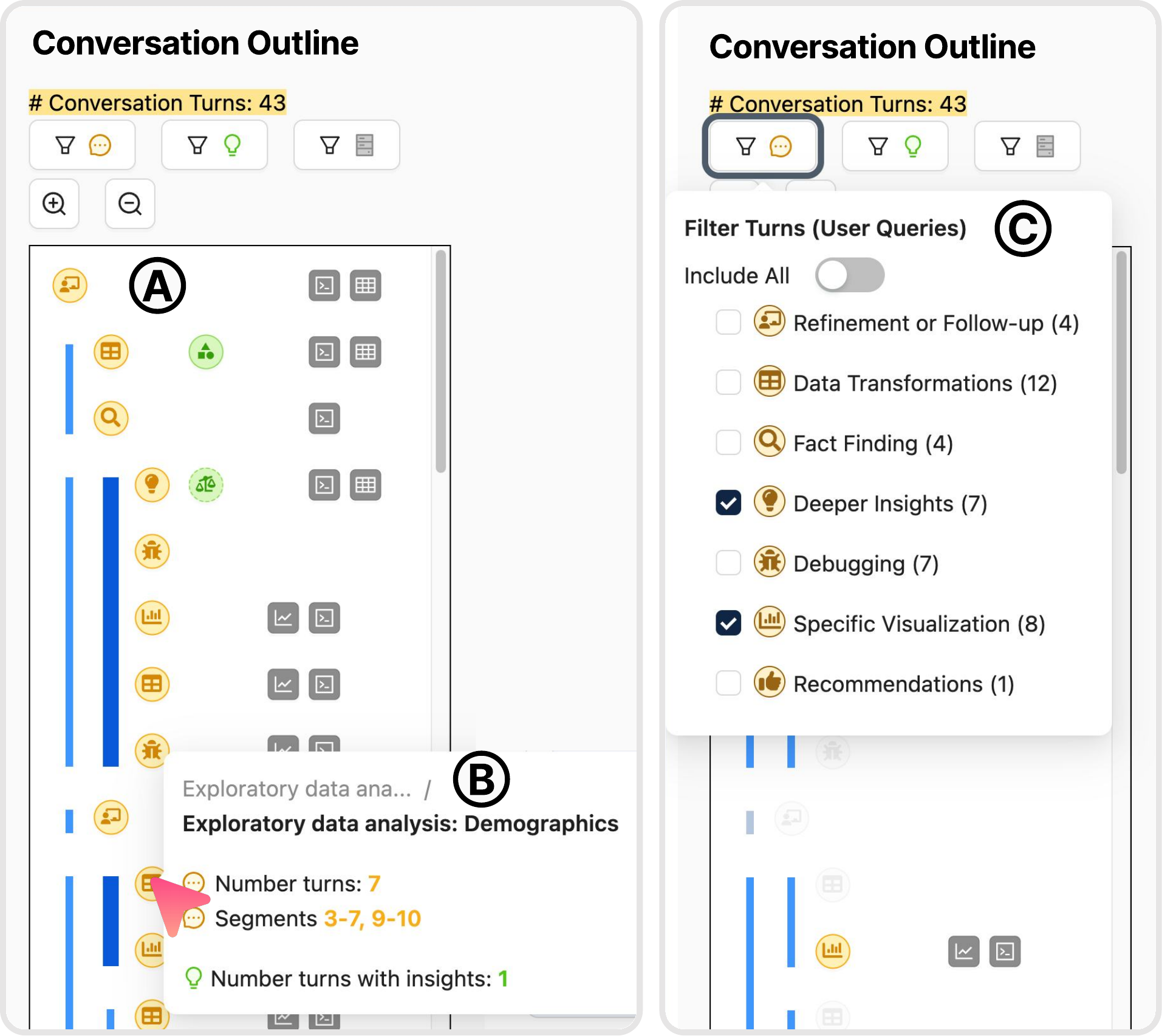}
    \caption{The \chatTimeline panel offers a visual overview of the conversation by displaying components (i.e., \componentThreadsShort,~ \componentSpeechActShort,~ \componentInsightShort,~ and~ \componentArtifactShort) as vertically arranged icons following the dialogue \includegraphics[height=0.8em]{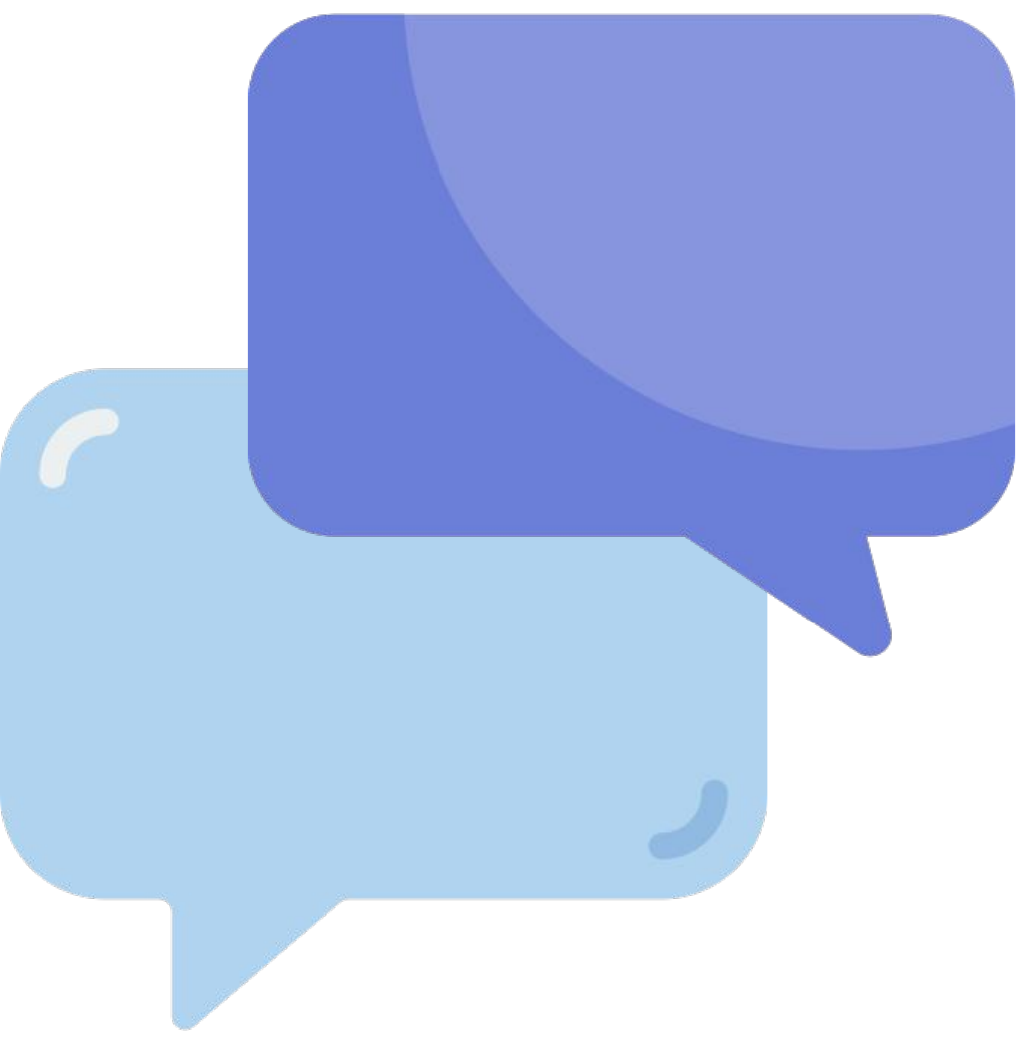}~ turns (A). Hovering over icons reveals additional details (B). Users can filter by speech acts, insights, and artifacts which update the outline view and the contents in the \chatContents(C).}
    \label{fig:timeline_panel}
\end{figure}

\subsection{Structured Conversation Content Panels}
\label{sec:structured_content_panels}
To support overviews and detail-on-demand (\textbf{DC2}),~\syncsense~organizes conversation content across three synchronized panels, arranged from left to right (Fig.~\ref{fig:teaser}). These panels are linked through structured analytical conversation elements (\S\ref{sec:adding_structure}), using consistent iconography and color coding for visual coherence. The \chatTimeline panel offers a high-level overview of the conversation elements (\S\ref{sec:timeline_panel}), the \chatContents~panel displays mid-level detail for selected elements (\S\ref{sec:contents_panel}), and the \origChat~panel shows the raw conversation, annotated with speech acts and artifacts for each turn (\S\ref{sec:origchat_panel}).

\subsubsection{\chatTimeline Panel}
\label{sec:timeline_panel}
The \chatTimeline~panel offers a visual scaffolding of the entire conversation by aligning structured elements along a vertical sequence that mirrors the original flow (Fig.~\ref{fig:timeline_panel}A). Each conversation turn appears as a row, with associated elements (i.e., \componentSpeechActShort, \componentArtifactShort, and \componentInsightShort) presented horizontally in the row corresponding to that turn. \componentThreads are indicated in blue as vertical bars spanning relevant turns, with nested sub-threads indented to the right of their parent thread. Hovering over the elements reveals contextual details (Fig.~\ref{fig:timeline_panel}B) such as turn information or thumbnails of data tables and visualizations. Clicking an icon displays the corresponding element in the \chatContents panel, allowing users to navigate directly to the relevant content for more details and trace back to the original conversation. Artifact icons also open a popover showing artifacts of that type for the selected turn. The popover includes buttons for quick navigation to the related content in both the \chatContents and \origChat panels. Users can filter to specific elements, i.e., the speech acts, insights, and artifacts, using the controls at the top of the panel (Fig.~\ref{fig:timeline_panel}C). Filters apply across both the \chatTimeline and \chatContents panels (\S\ref{sec:contents_panel}), updating the visible icons accordingly. \rr{
Observing users' interactions with this structured overview helps elicit design-relevant behaviors, such as the value of summarizing via structure, the role of threads in organizing workflows, or how compact representations affect information prioritization.}

\begin{figure*}[t!]
    \centering
  \includegraphics[width=0.93\linewidth]{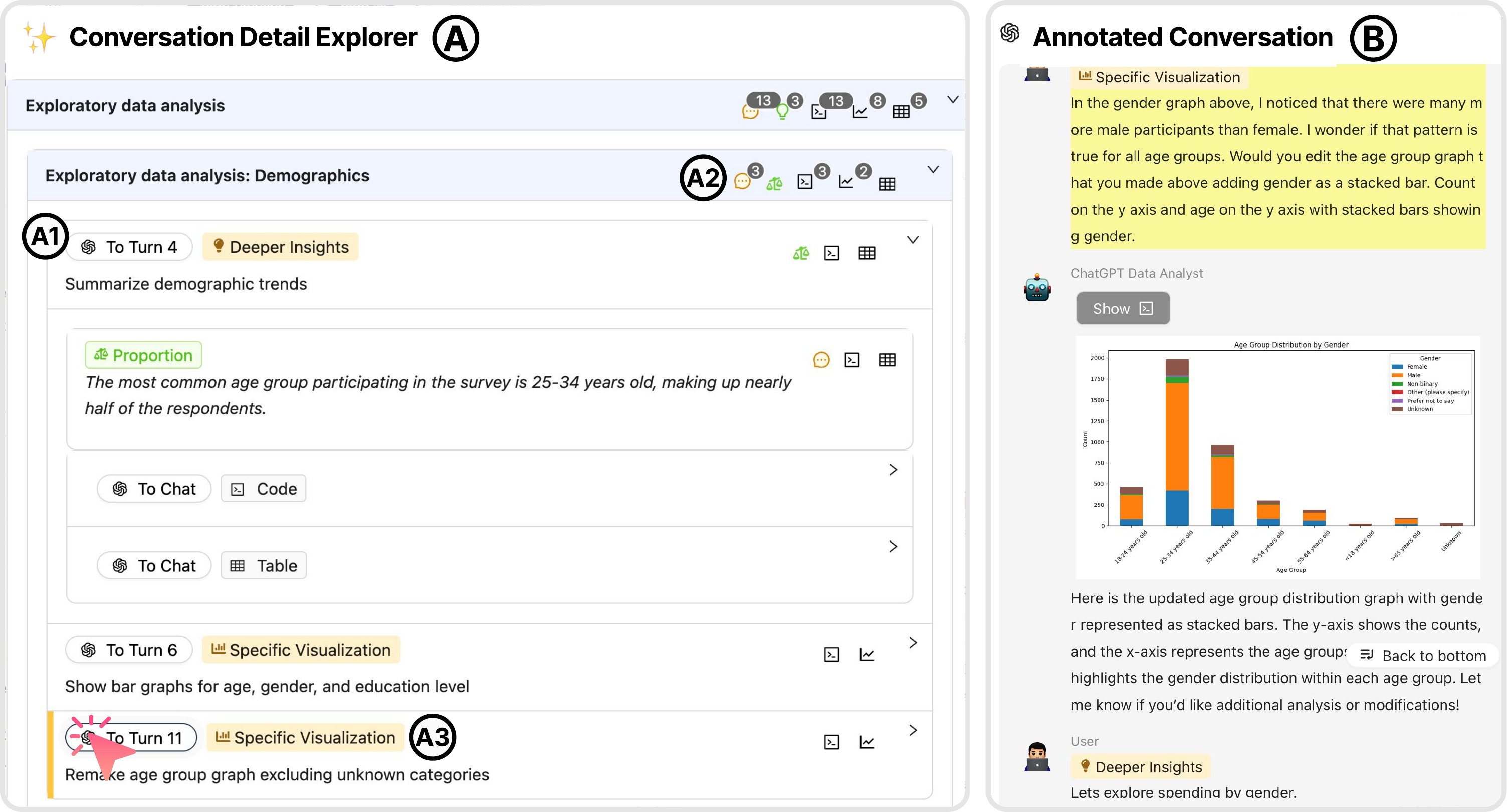}
  \caption{\chatContents and \origChat panels. The \chatContents panel presents a structured, mid-level view of the conversation (A). Threads can be expanded to reveal individual turns and elements (A1). Icons with content counts summarize nested elements (A2). Clicking the \toturnbtn button navigates to the corresponding turn in the \origChat panel (A3). The \origChat panel displays the full raw conversation, annotated with speech act tags and artifact icons to support provenance and traceability (B). }
  \label{fig:content_panels}
\end{figure*}

\subsubsection{\chatContents Panel}
\label{sec:contents_panel}
The \chatContents panel provides a mid-level abstraction of the conversation; more detailed than the \chatTimeline panel but more concise than the raw chat conversation (Fig.~\ref{fig:content_panels}A). Turns and their corresponding components are grouped under threads (Fig.~\ref{fig:content_panels}A1), with nested elements expandable via the dropdown (``>'') button. To help users get a sense of the contents nested within a content block, icons representing the content (i.e., turns, artifacts, or insights) with the counts of that content are presented (e.g., in Fig.~\ref{fig:content_panels}A2). To support orientation, hovering over the contents in the \chatContents panel triggers an increase in the size of the corresponding icon in the \chatTimeline. For detail on demand, users can click the \toturnbtn~button, which opens the~\origChat and scrolls to the corresponding turn (Fig.~\ref{fig:content_panels}A3). \rr{This panel was designed to elicit user strategies for managing analytical complexity, e.g., how they interpret thread groupings, which content types they prioritize, and how they trace reasoning across nested structures.}

\subsubsection{\origChat Panel}
\label{sec:origchat_panel}
To preserve provenance and provide access to full conversational detail, \syncsense~includes the original raw analytical conversation in the \origChat panel. Each turn is annotated with speech act tags and artifact icons to aid browsing (Fig.~\ref{fig:content_panels}B) \textbf{(DC1)}.


\begin{figure*}[t!]
    \centering
  \includegraphics[width=0.93\linewidth]{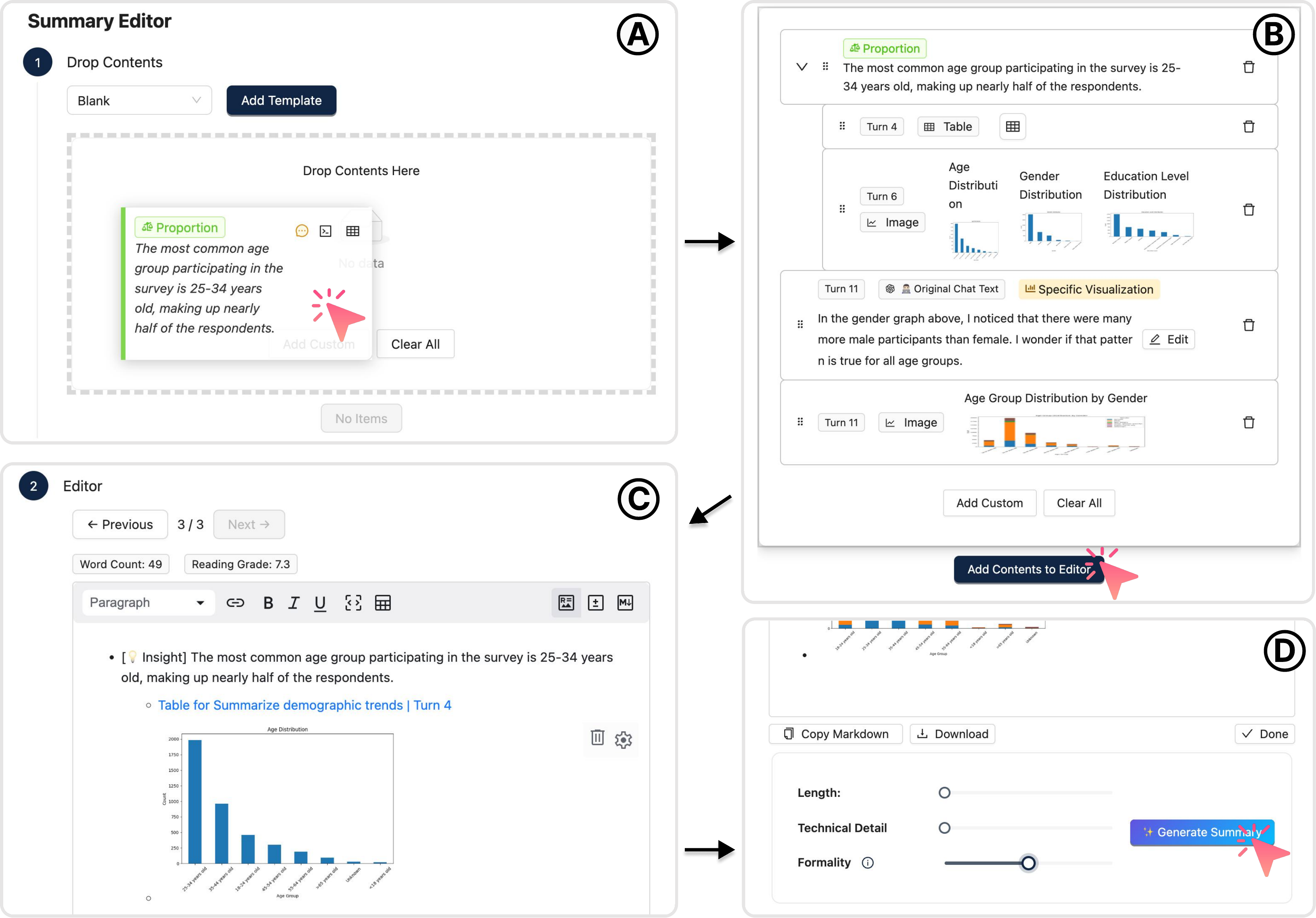}
  \caption{Summary Authoring Process in the Authoring Panel. Users can drag items from the \chatContents or \origChat panels into the drop container within the \authoring panel (A). Within the drop container, the items can be rearranged by dragging to adjust the order or nesting structure (B). The contents are serialized in the post-drop editor, where users can utilize an LLM to reformat the summary based on specified parameters such as \textit{length}, \textit{technical detail}, and \textit{formality} (C). Finally, the LLM-generated output is displayed in the post-LLM editor so users can make any final edits.}
  \label{fig:authoring_process}
\end{figure*}

\subsection{Summary Authoring Panels}
\label{sec:authoring_panel}
To support flexible composition from conversation elements (\textbf{DC3}), \syncsense~includes an Authoring Panel (Fig.~\ref{fig:authoring_process}). The panel allows users to drag and drop elements into a workspace, organize them into summaries, and optionally, refine the text with LLM assistance (\textbf{DC4}). \rr{This interaction supports reflection on what content users value, how they organize it, and how much control they desire.}

\subsubsection{Composing Summary Content}
Users can drag any chat element (e.g., thread, turn, insight, or artifact) from the \chatContents panel or \origChat into the composition container (Fig.~\ref{fig:authoring_process}A). Dropped elements retain their nested structure, with each appearing as a block and sub-elements indented accordingly. Users can reorder blocks, edit their text directly, or insert custom notes using freeform blocks. Once users are satisfied with the content, they can click the \textit{``Add Contents to Editor''} button (Fig.~\ref{fig:authoring_process}B). This action serializes the chat contents into markdown format. Each block becomes a bullet point, with nested elements rendered as indented bullets. For some chat elements, metadata tags indicate the type of content (e.g., an insight will be prefixed with ``[Insight]"). 

\subsubsection{Editing and Refining Summaries} \syncsense~provides multiple levels of AI assistance for summary composition (\textbf{DC4}).
Users can directly edit markdown text in the editor (Fig.~\ref{fig:authoring_process}C), with code and tables linked as raw files and visualizations embedded inline. To refine the summary, users adjust sliders for length, technical detail, and formality. These settings reflect both structural (length) and contextual (audience and purpose) needs~\cite{Hoeve2020WhatMA, SprckJones1998AutomaticSF}. When the \textit{``Generate Summary''} button is clicked, a refined summary (generated by an LLM based on these settings) is written into the editor. \rr{This probe feature elicits preferences around AI-assisted communication, helping us understand where users welcome automation and where they seek fine-grained control.}

\subsection{Implementation Details}
The \syncsense~design probe is implemented in the React~\cite{MetaOpenSource2024} framework based on Typescript using the Ant Design library~\cite{AntDesign} for UI components. 
The backend is implemented in Python using FastAPI~\cite{fastapi}, which is designed to extract the chat contents from ChatGPT's conversation JSON object, execute the code within the conversation, and generate summaries. To ensure consistent artifact rendering, code execution occurs in a sandboxed Python environment that mirrors OpenAI's code interpreter setup, using the same libraries inferred via prompting ChatGPT\footnote{OpenAI's shared chat currently does not include generated visualizations.}. For all LLM-based functionalities, we use OpenAI's GPT-4o model~\cite{openai-gpt4o}. Specifically, these tasks include (1) extracting threads from the conversation, (2) extracting the speech act from each user query, (3) extracting and categorizing insights from the conversation, and (4) generating the updated summary given the markdown input. Prompts were iteratively refined and tested using two hand-labeled conversation transcripts and grounded in definitions from prior work (i.e., speech acts, insight types). We include all prompts in our supplementary materials.

\section{User Study}
\label{sec:study}
We conducted a user study that combined the use of ~\syncsense~as a design probe~\cite{Boehner2007HowHI} with task-based, semi-structured interviews. This approach enabled us to observe realistic user behavior while eliciting participant reflections on our research questions (\S\ref{sec:study_goals}).

\subsection{Participants}
We recruited 10 participants (six male, four female) via mailing lists at a data analytics software company. This sample size aligns with HCI norms for design probe studies~\cite{Caine2016LocalSF,GAMUT, ZamfirescuPereira2023WhyJC, rachatasumrit2021forsense}
. Participants reported an average of 11 years ($\sigma = 6.4$) of data analysis experience. All participants engaged in data analysis at least monthly, with six doing so daily, typically using data visualization tools (e.g., Tableau, Power BI) in their workflows. With the exception of one participant, all reported using conversational AI tools (e.g., ChatGPT, Claude, Perplexity) at least a few times per month.

\subsection{Procedure \& Task}
To balance realistic usage scenarios where data workers may need to revisit and summarize past analyses, the study was conducted in two sessions: (1) an open-ended analysis session with ChatGPT Data Analyst, and (2) a summary authoring session held more than a week later, where participants revisited their prior conversation and used~\syncsense~to create summaries for different audiences. This delayed two-part design enabled us to observe behaviors relevant to both long-term revisitation and communicative reframing, such as delayed summarization and tailoring outputs for different stakeholders. 

\subsubsection{Part 1: Data Analysis Session}
\label{sec:dataAnalysisSession}
Participants analyzed the \textit{Great American Coffee Taste Test} dataset from TidyTuesday~\cite{great-american-coffee-taste-test-2024}, a survey of 4,042 respondents' coffee preferences across 57 columns. The dataset combines categorical, numerical, and free-text variables (e.g., brewing methods, taste ratings, spending habits, demographics). The dataset presents realistic challenges, such as over 10 columns with >80\% missing values, primarily from optional free responses. We selected this dataset because it is broadly relatable while still complex enough to support meaningful analysis. Participants were provided the following scenario: 
 \textit{``You and your business partner, both trained data analysts, are planning to open a coffee shop in your city. To make informed decisions, you’ll analyze data to guide key strategies, such as which coffee to source, pricing, and target customer demographics. Your goal is to uncover actionable insights to shape your business plan. To assist with your analysis, you'll use ChatGPT Data Analyst as your analytical tool.''}

To balance ecological validity with experimental control, all participants worked with the same dataset. They were instructed to conduct a well-rounded analysis by uncovering multiple insights, engaging with varied data types (ordinal, categorical, continuous, and unstructured text), and producing different artifacts (e.g., visualizations and tables). The task was designed to take approximately 45 minutes. Rather than priming participants with a detailed prompt, we framed the activity as a business decision-making task, mirroring the goal-oriented style of Jeopardy-style evaluations~\cite{Gao2015DataToneMA}.

\subsubsection{Part 2: Revisiting the Analytical Conversation}

The second session was conducted remotely, with participants sharing their screen over video conferencing (Google Meet) on their own computers. This 90-minute session included: (i) a tutorial walkthrough of~\syncsense~(\textasciitilde 20 minutes), (ii) two recall and summarization tasks (\textasciitilde 50 minutes), and (iii) a semi-structured interview (\textasciitilde 20 minutes). Tutorial materials are included in the supplementary material.

Participants completed two summarization tasks, based on their prior analytical conversation. For each task, we asked participants to craft a summary tailored to a specific audience and output medium. Audience options were either an \textit{investor}, who is non-technical, seeking a persuasive business case, or an \textit{analyst business partner}, who is technical and potentially continuing the project. The output format was either a detailed \textit{data report} or a \textit{direct message}, with both audience and medium counterbalanced across participants. During the tasks, participants were encouraged to think aloud and explain their interactions with the probe. After each task, they reflected on their content selection and emphasis decisions. A concluding semi-structured interview explored their information needs, summarization strategies, and perspectives on the role of AI in supporting these activities. 

\subsection{Analysis}
To study the characteristics of an analytical conversation (\textbf{RQ1}), we examined the raw conversation data exported from ChatGPT~\cite{openai-chatgpt-export-2025}
for all participants and recorded the speech acts and insights extracted by our backend LLM modules. For summary-related patterns, we examined elements added to the \authoring panel and analyzed the final authored summaries. To identify recurring themes, we reviewed the transcripts and video of the participants' screens while conducting the tasks. We performed iterative coding on participants' actions, the intent behind their actions, think-aloud comments, and interview responses guided by (\textbf{RQ2-4}). Three researchers independently open-coded one participant session, discussed findings, and created an initial code set. They repeated this for a second session to refine consistency and finalize the codebook. Two researchers then independently applied the final codebook to all sessions. The codebook is available in the supplementary material. Findings in \S\ref{sec:results} are organized by themes derived from this coding process. 

\section{Findings}
\label{sec:results}

We present our study findings organized around the four research questions: characteristics of analytical conversations (\textbf{RQ1}), participants' information needs (\textbf{RQ2}), their navigation and summarization workflows (\textbf{RQ3}), and their preferences on AI's role in this process (\textbf{RQ4}). For RQ2 and RQ3, we frame our findings using a codebook, with direct references to thematic codes shown \textbf{in bold}.


\begin{figure}[h]
    \centering
  \includegraphics[width=\linewidth]{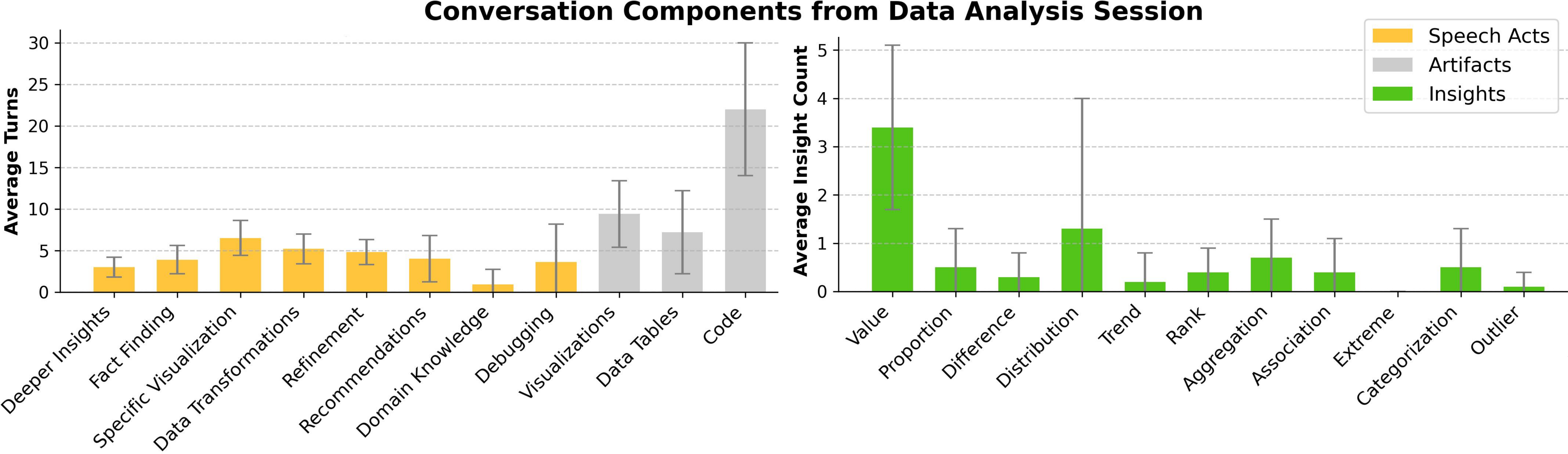}
  
  \caption{Analytical Conversation Elements from Session 1 (Sec.~\ref{sec:dataAnalysisSession}). Across participant conversations, all \componentSpeechActShort and \componentArtifactShort types were observed. Except for \insightExtreme, all types of \componentInsightShort appeared. The top chart shows the average number of turns each element was involved in, while the bottom chart shows the average count of unique insights per conversation.
  }
\label{fig:results_convo_characteristics}
\end{figure}

\subsection{RQ1: \rqCharacteristics}
\label{sec:results_convo}

While participants pursued the same analytical goal using the same dataset, their interactions with the analytical conversational interfaces were diverse, lengthy, and artifact-rich. On average, participants spent $69$ minutes engaging in these analytical sessions ($\sigma = 40$ minutes). The conversations were notably long, averaging $29.5$ turns per session ($\sigma = 10.4$), and heavily skewed toward assistant responses. Specifically, user queries averaged $24.4$ tokens ($\sigma = 14.2$), while assistant replies averaged $319.7$ tokens ($\sigma = 177.7$), excluding the artifacts. These conversations were substantially longer and more detailed than typical LLM conversations in the wild. For comparison, conversations in two large-scale real-world LLM datasets, LMSYS-Chat-1M~\cite{Zheng2023LMSYSChat1MAL} and WildChat~\cite{Zhao2024WildChat1C}, average only $2.02$ and $2.54$ turns, respectively.

Conversations featured a variety of artifacts: on average, $22$ turns involved code ($\sigma = 8.0$), with $29.4$ unique code ($\sigma = 7.5$). $7.2$ turns included one or more data tables ($\sigma = 5.0$), resulting in $7.7$ unique tables per conversation ($\sigma = 7.5$). Visualizations appeared in an average of $9.4$ turns ($\sigma = 4.0$), with $16.3$ unique visualizations generated ($\sigma = 9.1$). The most common speech acts classified by our probe involved requests for \speechActSpecificViz, \speechActDataTransformations, \speechActRefinement, \speechActFactFinding, and \speechActDeepInsights. Except for \insightExtreme, all insight types also appeared in the conversations. Figure~\ref{fig:results_convo_characteristics} summarizes the occurrences of each artifact, speech act, and insight types.

\begin{figure}[h]
    \centering
  \includegraphics[width=0.9\linewidth]{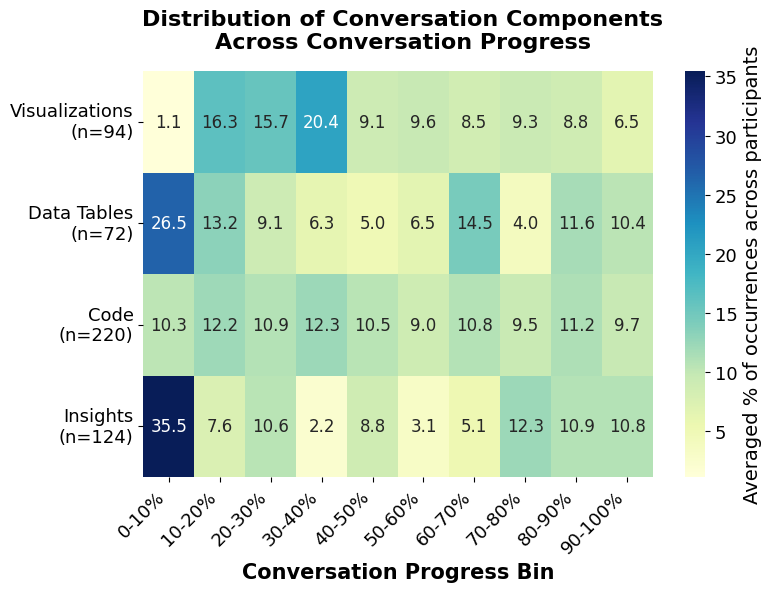}
    \caption{Data tables and insights typically appeared within the first 10\% of conversation turns, while visualizations were more frequent in the following 30\%. Code occurrences were distributed evenly throughout. Frequencies are normalized by turn decile within each element type (i.e., rows sum to 100\%).
    }
  \label{fig:results_component_heatmap}
\end{figure}

Code was distributed evenly throughout the conversation, suggesting participants consistently engaged in code execution across all stages. In contrast, data tables and insights appeared more often within the first 10\% of turns. Visualizations were more frequent in the following 30\% (Fig.~\ref{fig:results_component_heatmap}).


\xhdr{Case Study of P7's Analytical Conversation} To help contextualize these statistics, let us consider P7's conversation\footnote{P7's Analytical Conversation: \url{https://chatgpt.com/share/675c77f9-570c-800a-8d2e-a24bb7eb2df8}} which contained $30$ turns, $29$ of which involved code, $13$ with tables, and $10$ with visualizations. The conversation begins with a dataset exploration (generating a data table), including a demographic overview that generated two visualizations. The participant then shifts focus to identifying which \texttt{coffee type} to buy next year, leading to a deeper investigation of \texttt{preferences} across \texttt{age} groups. This generated a visualization of \texttt{coffee preferences} by \texttt{age}. The participant spent several turns refining this visualization, before exploring additional dimensions such as \texttt{gender}, \texttt{urban} vs. \texttt{rural residence}, and \texttt{political affiliation}. Throughout, they go back to seeking high-level summaries of the dataset, such as column names and counts. This example reflects a broader pattern we observed across participants: analytical conversations are iterative, shift in focus over time, and involve a mix of exploratory and deeper engagement behaviors. We include the URLs and statistics of all conversations in our supplementary material.


\subsection{RQ2: \rqInformationNeeds}
\label{sec:results_info_needs}
When reviewing and summarizing their conversations, participants' information needs centered around three key objectives: \textit{orienting} themselves to understand the structure and content of the conversation, \textit{recalling} specific conversation elements, and \textit{prioritizing} the information most relevant for review and summarization. 

\xhdr{Orienting} \textit{Orienting} refers to when a participant needs to situate themselves within the broader context of the conversation. All participants (10/10) demonstrated this need, especially when revisiting the conversation after at least a week. The purposes of orienting involved familiarizing themselves with the conversation, jogging their memory of certain elements, and identifying quickly the important main pieces of the conversation.  For example, while using the \chatContents panel, P2 expressed needing further detail to facilitate their understanding of the conversation: \textit{``All right, I did data cleaning. Let's see here. Coffee consumption. I think I need a little more detail to remember what these things were.''} Similarly, P10, while noting that they \textit{``remember a few of the key takeaways or insights,''} described wanting an overview that could \textit{``help recap how things are going on.''} These cases illustrate both the variability in participants' memory of prior conversations and the differing levels of detail or overview they sought to fulfill their orienting needs.

\xhdr{Recalling} In contrast to orienting, where participants sought a general understanding of the conversation, \textit{recalling} refers to retrieving specific information from memory.  Several participants (4/10) expressed this need [P1, P2, P3, P10], often aiming to find specific visualizations, insights, or sub-threads of analysis. For example, while searching for a specific chart they had generated, P1 stated, \textit{``There was one with the coffee spend. That's the one I'm trying to find where I zoomed in...I need to find that one chart.''} P2 was looking for a very specific conversation turn analyzing neighborhood recommendations for opening a coffee shop: \textit{``Oh, there was my neighborhoods. There we are, the very last one.''}

\xhdr{Prioritization} \textit{Prioritization} refers to the need to determine which parts of the analytical conversation to focus on. This need was especially salient when summarizing conversations and working with long, complex transcripts. All participants (10/10) emphasized the importance of selectively identifying key content; both to decide what to include or exclude in their summary communication, and to support their own re-engagement with the analysis. For example, P1 chose to prioritize analysis findings over initial data-cleaning turns: \textit{``I need to scroll down a little bit and go to where I remember actually getting something that would be appropriate for this analysis.''} Similarly, P6 underscored the need to gather key insights and artifacts first, explaining: \textit{``First important part is to [find] whatever the insights were from the conversation and the data itself, what are important key artifacts I need to look over.''} Notably, prioritization often worked in tandem with orienting, helping participants better situate themselves within the conversation by narrowing their focus to what mattered most to them.

\rr{Overall, the findings of RQ2 show that revisiting analytical conversations is not a single ``summarization" task, but a progression from re-establishing context, to locating specific artifacts, to curating what should be surfaced or communicated. Each stage foregrounds different cognitive work and suggests different forms of support.}


\subsection{RQ3: \rqWorkflows}
\subsubsection{Navigation Strategies}
\label{sec:results_strategies_info_needs}

To satisfy these information needs, participants employed a range of navigation strategies which served the foraging phase of the sensemaking process~\cite{Pirolli2007TheSP}.

\xhdr{Sequential navigation}  
\textit{Sequential navigation} refers to participants scanning through the analytical conversation chronologically to re-familiarize themselves with the analysis. All participants (10/10) employed this strategy, but at varying levels of abstraction. Some (6/10) participants relied on the high-level overview and quick entry points from the \chatTimeline panel [P1, P2, P3, P4, P7, P10]. Others (8/10) navigated using the intermediate \chatContents panel [P1, P2, P3, P4, P5, P6, P10], and several (7/10) scrolled directly through the detailed \origChat [P2, P3, P4, P6, P8, P9, P10]. For example, P6 adopted a fine-grained approach, reading through the entire raw chat: \textit{``I don't remember the conversation that I had so I actually went to the whole conversation and then I was looking at, okay, what was the points or what was the queries I made and then what was the result.''}  
In contrast, P4 opted for a more abstract strategy, scanning the \chatTimeline by \textit{``looking at the left-most set of blue lines''} representing the threads. 

\xhdr{Abstractive navigation}
\textit{Abstractive navigation} refers to participants' strategy of moving between different levels of abstraction (e.g.,  moving from the \chatTimeline to specific conversation turns' details in the \chatContents, to then zoom in or out for detail or context). This horizontal form of navigation often occurred within the broader flow of sequential navigation,  enabling participants to zoom in or out as needed for additional context or detail. Participants commonly employed abstractive navigation to drill-down into specific segments or retrieve supporting context [P1, P2, P3, P4, P7, P10]. For example, P1 found a relevant table in the \chatContents~panel and wanted greater detail in the \origChat: \textit{``So let me look at this table and code. When I scroll to turn 10, I want to find it on the side so I can go right to the table...I can just click on this [to turn] box.''} Similarly, P7 described using the \chatTimeline as an entry point to access additional detail to support their orienting: \textit{``I really like that [Outline] view of the different steps that I took and being able to drill-in and see, like, exactly what I did.''} Abstractive navigation was used to support all three types of information needs depending on the participant’s intent.

\xhdr{Visual recall}
\textit{Visual recall} refers to the use of visual elements, such as the element tags, icons, and visualizations as cognitive cues to retrieve specific information. For orienting needs, visual recall was often combined with sequential navigation to make sense of the conversation [P1, P2, P3, P4, P5, P8]. This strategy was supported both by finding existing visualizations in the conversation, and \syncsense's structured tagging and iconography. For example, P3 emphasized the value of these features in helping them make sense of the conversation: \textit{``I'm having to use it to navigate to remind me what we're doing now [and] if that's the case, I really like these icons.''} Similarly, P8 expressed a preference for visual recall over browsing text: \textit{``I have a very visual memory, and so, like, pictures and things can trigger more than just, like, reading text.''}

\xhdr{Filtering}
\textit{Filtering} refers to the strategy of including or excluding content based on specific criteria. Several participants (5/10) used filters primarily to support prioritization [P2, P3, P4, P6, P7]. For example, P3 methodically filtered the conversation to only a few relevant speech acts: \textit{``I'm not interested in recommendations. I'm not interested in fact finding...I don't need comparisons. I don't want my visualizations... Okay, I believe I've now filtered down to just what I need.''} Likewise, prioritization to support communicative goals motivated P4's decision to filter: \textit{``I'd want to show him like what stands out in terms of coffee preferences and how I want to market to those people... based on that, I can just filter the conversation to insights or recommendations.''} Meanwhile, P2 filtered the conversation to just \texttt{insights} before proceeding with sequential navigation.

\rr{Overall, RQ3 reveals that users rely on a combination of structural overviews, semantic cues, and visual landmarks to forage within long analytical conversations. A key tension is between maintaining the narrative continuity of the original chat and efficiently jumping to semantically relevant segments.}

\subsubsection{Strategies for Communicating Analytical Conversations}
\label{sec:results_strategies_summary_communication}
To tailor their summaries for different audiences and media, participants made or expressed a desire to make a range of edits using both manual changes and LLM assistance. These edits reflected distinctive communicative goals and preferences, often involving multiple iterations of refinement. Here, we include the most prominent strategies participants employed.


\xhdr{Add process details}
\textit{Adding process details} involves supplementing the summary with background context or outlining specific steps taken during the analysis. Participants achieved this either by manually editing the summary (Fig.~\ref{fig:authoring_process}B and C) or by using the ``\textit{Generate Summary}'' button and summary sliders (Fig.\ref{fig:authoring_process}D). This strategy was employed by 7 out of 10 participants [P1, P2, P3, P4, P5, P9, P10]. For instance, P1 clarified how missing data was handled during cleaning: \textit{``Note on steps: The original cleaning added mean values to replace the missing values. I asked it to remove those average values. The new CSV should not have those values. Suggest double-checking the cleaned data file.''} 

\xhdr{Add action-oriented context}
Beyond procedural or analytical details, participants emphasized the importance of framing summaries to support interpretation and downstream use. \textit{Adding action-oriented context} positions the analysis in terms of its goals, relevance, and audience, helping others understand not just what the analysis shows, but why it matters and what to do with it. When aspects of this context existed in their original conversation, participants (4/10) directly selected it during composition [P2, P3, P4, P6]. For example, P4 explained a decision about including non-binary gender responses: \textit{``A small sample size for responses by gender. So maybe that's something that should be considered as to whether we group the people who didn't put male or female.''} However, such cues were rarely sufficient. Because context was typically absent from element serializations, participants usually wrote additional notes themselves [P2, P3, P4, P5, P6, P7, P9]. P4 explained that beyond the gender reasoning, they needed to \textit{``put some of my own thoughts in there''}. Similarly, P7 revised their summary's introduction to foreground stakeholder goals:
\textit{``[For] handing this off ... here's the questions my stakeholders were asking, here's why I was working on it.''} They added that such framing rarely emerges from AI interactions alone:
\textit{``That's something that I don't think would necessarily come from my chats with the AI. That would probably just have to be some human color that I put into it.''}

\xhdr{Adjust overall length}  
Participants (6/10) also \textit{adjusted the length of their summaries} either manually or by using the slider to suit their audience [P4, P5, P6, P7, P8, P10]. Participant-authored summaries averaged 245 words ($\sigma$ = 172) and 134 unique terms ($\sigma$ = 76), but varied widely across the 20 summaries. P4, for example, noted the need to provide a fuller picture when writing for an investor: \textit{``You gotta hook them up front, right? But you gotta give them some level of detail to make them feel comfortable. When I regenerated the summary, [it was] very short and concise; just feels like I'm not giving them enough information.''} These adjustments reflect a desire for audience-sensitive communication, where participants weighed informativeness against conciseness to align with reader expectations.

\xhdr{Reorder content and modify formatting}
Participants sometimes \textit{reordered summary content or adjusted its formatting} to improve logical flow, readability, or visual clarity [P2, P3, P6, P7, P10]. These structural adjustments helped draw attention to key insights or made the structure more digestible for their intended audience. For example, P6 noted a desire to emphasize important ideas through formatting:
\textit{``These are important points to me, and I would like to focus and just make it bold, make it in bullets. That type of thing.''} This behavior was prevalent across participants, with 15 of the 20 authored summaries including either section headings or lists, and 11 including both, demonstrating a strong preference for well-structured and visually scannable summaries.

\rr{Together, these strategies show that summaries of analytical conversations are not merely compressions of chat logs. Instead, participants treat them as authored narratives that must balance process transparency, interpretive framing, and audience expectations.}

\subsection{\rr{Characteristics of Authored Summaries}}
\rr{The $20$ authored summaries varied in length and structure but shared several features that speak to their perceived quality and usefulness (see Appendix \ref{appendix:result_codes} Figure \ref{fig:results_summary_composition}). On average, summaries were $245$ words long ($\sigma$= 172) and contained $134$ unique terms ($\sigma$= 76), indicating that participants did more than copy the raw transcript; they condensed and rephrased the original conversation into a more compact representation. Most summaries were structurally organized, where $15$ of $20$ included either section headings or bullet lists, and $11$ included both, reflecting a strong preference for scannable, presentation-ready formats. In terms of content, the majority of summaries combined insights and artifacts: visualizations and insights appeared in over 75\% of the compositions in the Authoring Panel just prior to serialization, and many participants also included conversational text to carry interpretive context. Rather than simply enumerating all artifacts, participants prioritized a small set of anchor visualizations or tables and surrounded them with interpretive text and audience-specific framing. These patterns suggest participants produced summaries that were concise, structured, and insight-centric, aligning with prior notions of useful analytic documentation and handoff materials.}
\vspace{-2mm}

\subsubsection{\rr{Relationship between Navigation Strategies and Summary Outputs}}
\rr{While our study was not designed to isolate causal effects, we observed descriptive patterns suggesting that how participants navigated the conversation shaped the kinds of summaries they produced. Participants who engaged in abstractive navigation (i.e., moving between overview and detailed views across panels) were also those who most frequently enriched their summaries with process details and action-oriented context. Five of the six participants coded with abstractive navigation also added process notes (e.g., data cleaning decisions, caveats about missing data) to their summaries, and four added audience-oriented framing about stakeholder goals or next steps. In contrast, participants who remained mostly in the raw conversation view tended to produce more chronological overview summaries that closely followed the turn order of the chat, with less re-framing of the analysis as a coherent story for a specific audience. Similarly, summary reordering and formatting (e.g., restructuring content into sections and bullet lists) was strongly associated with more selective navigation: four of the five participants who reordered or reformatted their summaries also relied on filtering to narrow the conversation to relevant speech acts or artifacts. This pattern suggests that participants who treated navigation as an opportunity to curate the conversation, by filtering, zooming between abstraction levels, and using visual cues, also treated summarization as an opportunity to author a new, audience-appropriate narrative rather than simply condensing the transcript. These observations empirically ground our framing of navigation and communication as mutually reinforcing activities rather than separate stages.}

\subsection{RQ4: \rqAIRole}
\label{sec:results_role_ai}
Given participants' experience with~\syncsense's LLM-enabled features for supporting navigation and summarization, we examine their preferences regarding the role of AI in structuring and summarizing analytical conversations. With the exception of P8, who preferred to communicate by directly sharing the original analytical conversation rather than composing a separate summary, all participants expressed a desire to remain involved in the summarization process. Here, we unpack the nuances of these preferences, focusing on how participants balanced automation with control, and where they saw value in human versus AI contributions.

\xhdr{Retaining editorial control in high-stakes communication}
Participants (4/10) expressed the importance of maintaining control over summary content, particularly in high-stakes contexts such as presentations or external reports [P1, P2, P5, P7]. Authorship and narrative framing were closely tied to their sense of professionalism and responsibility. For example, P7 emphasized the importance of having \textit{``tight control''} over the final wording, explaining that they would not feel comfortable sharing output that did not reflect their personal communication style. Likewise, P2 noted that the level of AI involvement should align with the stakes of the task: \textit{``If I’m going to do this big, important presentation, it’s more important that I can stand behind what I’ve done.''}

\xhdr{Supporting selective and structured authoring}
Participants expressed a desire to retain control over which elements were included in summaries, while still benefiting from AI-generated structure [P2, P3, P4, P6, P10]. For example, P6 emphasized the need to selectively include details that might otherwise be overlooked: \textit{``I want to be very selective and concise… having this control will be helpful.''} P10 described a workflow where AI provides a structured starting point, paired with tools to support organization and memory:
\textit{``I would want to give it everything, let it do a first pass, and then work through it... I would almost want a bookmark feature within the interface, because looking through the entire thing is a lot of work. I kind of lose track.''} These reflections highlight participants' preference for a collaborative workflow, where AI accelerates organization and drafting, but users remain the ultimate curators of what gets communicated.

\xhdr{Using AI for acceleration, not automation}
Some participants (4/10) viewed AI as useful for drafting or surfacing insights quickly, especially under time constraints [P2, P5, P6, P10]. For instance, P5 appreciated the time-saving benefits when responding to stakeholders: \textit{``It would save me so much time… it's a really valuable tool.''} Despite the need for control, participants welcomed AI as a productivity partner, augmenting rather than replacing their work per se, supporting efficiency without taking over key decision-making tasks during data analysis workflows~\cite{Gu2023HowDD, Mcnutt2023OnTD}.

\xhdr{Scaffolding iterative and ongoing analyses}
While participants primarily focused on summarization, several also highlighted the need for better support during the analysis process itself. These reflections reveal a broader desire for AI to facilitate not just output generation, but also real-time structuring, prioritization, and iterative communication throughout the analytical workflow [P2, P3, P10]. For instance, P2 described wanting lightweight structuring features embedded directly in the ChatGPT interface to help manage and compartmentalize their workflow:
\textit{``It would have been nice if I could say, okay, just forget this, we're not getting anywhere. Or, pin this for later but I don’t want to do it right now. I want to compartmentalize my experience as I’m doing it, rather than retrofitting the organization after the fact.''}
P10 similarly emphasized the need for tooling that supports the iterative \textit{``back and forth between analysis and presentation''}, allowing the user to switch between communication and analysis \textit{``modes.''}

\rr{A consistent theme is that participants want AI that is opinionated enough to structure and accelerate their work, but deferential enough to preserve human authorship, especially when communicating outward.} 



\section{Discussion}
\label{sec:discussion}
We situate our findings within the broader literature, highlighting design implications for future tools  (\S\ref{sec:implications}), and reflecting on limitations of our study (\S\ref{sec:limitations}).

\paragraph{RQ1: Structure of analytical conversations.}
\rr{RQ1 examines how analytical conversations with LLMs unfold and what their structure reveals in comparison to typical LLM chat interactions. As detailed in \S\ref{sec:results_convo}, these conversations are substantially longer than those observed in large-scale chat corpora~\cite{Zheng2023LMSYSChat1MAL, Zhao2024WildChat1C}, and are densely interwoven with code, tables, visualizations, and insight statements. We observe distinct temporal patterns: early stages emphasize data tables and high-level insights, while later phases shift toward visualization-driven exploration, with code consistently present throughout. These observations address RQ1 by positioning analytical conversations as artifact-rich analytic sessions rather than brief Q\&A exchanges. This complexity aligns with prior studies on the inherently iterative and fragmented nature of computational notebooks~\cite{Head2019ManagingMI, Kery2017VarioliteSE, Rule2018ExplorationAE}, and reifies the need for structured representations of LLM-driven analysis~\cite{Suh2023SensecapeEM, Weng2024InsightLensDA}. These insights motivate a shift in perspective; chat histories should be treated as evolving analytic documents requiring dedicated navigation and structuring support, rather than as flat, linear transcripts.}\vskip10pt

\aptLtoX{\begin{newimplications}
\rr{\textbf{Implication: Reframing analytical conversations as evolving documents.} Analytical conversations with LLMs are substantially longer, more artifact-rich, and more structurally varied than typical chat interactions. Tools should therefore treat these sessions less as ephemeral chats and more as complex analytical documents that unfold over time. In particular, interfaces need to (i) surface and organize heterogeneous artifacts (code, tables, visualizations, insights) as first-class objects, (ii) recognize that different phases of the conversation emphasize different artifacts, and (iii) provide navigation and summarization support that respects this temporal structure rather than assuming a flat, turn-by-turn log.}
\end{newimplications}}{
\implications{\rr{\textbf{Implication: Reframing analytical conversations as evolving documents.} Analytical conversations with LLMs are substantially longer, more artifact-rich, and more structurally varied than typical chat interactions. Tools should therefore treat these sessions less as ephemeral chats and more as complex analytical documents that unfold over time. In particular, interfaces need to (i) surface and organize heterogeneous artifacts (code, tables, visualizations, insights) as first-class objects, (ii) recognize that different phases of the conversation emphasize different artifacts, and (iii) provide navigation and summarization support that respects this temporal structure rather than assuming a flat, turn-by-turn log.}}}

\paragraph{RQ2: Information needs when revisiting conversations.}
\rr{RQ2 investigates how data workers' information needs shape their approach to revisiting and making sense of prior analytical conversations. While real-time LLM use often supports sensemaking through processes such as interpretation~\cite{Gu2023HowDD}, validation~\cite{Gu2023HowDA, Xie2024WaitGPTMA}, and synthesis~\cite{Weng2024InsightLensDA}, our findings in \S\ref{sec:results_info_needs} reveal that revisitation introduces a distinct set of challenges. Participants consistently cycled through three recurring needs: \emph{orienting} themselves to the overall structure and progress of the conversation, \emph{recalling} specific artifacts or analytical threads (e.g., a particular chart or hypothesis), and \emph{prioritizing} segments most relevant to their current goals or audience. These needs often emerged after temporal gaps or when encountering unfamiliar conversations. This characterizes revisitation not as a one-off summarization activity, but as an evolving sensemaking process involving multiple granularities. These findings suggest concrete design directions for tools that support structured overviews, artifact-based retrieval, and audience-specific content selection.}

\aptLtoX{\begin{newimplications}
\rr{\textbf{Implication: Supporting multi-level navigation and sensemaking in analytical conversations.} Participants employed a range of navigation strategies when revisiting prior analyses, from top-down scanning of overviews to targeted retrieval based on memory. This suggests that tools should support coordinated multi-level navigation, where timelines, structured contents, and raw chat views stay in sync and allow fluid zooming in and out. Visual tags and icons can serve as powerful memory cues and should be designed as intentional, semantically meaningful wayfinding devices. Finally, filters based on speech acts, artifact types, or insight categories can help users carve out task-specific slices of long conversations, supporting both analytical foraging and downstream communication.}
\end{newimplications}}{
\implications{\rr{\textbf{Implication: Supporting multi-level navigation and sensemaking in analytical conversations.} Participants employed a range of navigation strategies when revisiting prior analyses, from top-down scanning of overviews to targeted retrieval based on memory. This suggests that tools should support coordinated multi-level navigation, where timelines, structured contents, and raw chat views stay in sync and allow fluid zooming in and out. Visual tags and icons can serve as powerful memory cues and should be designed as intentional, semantically meaningful wayfinding devices. Finally, filters based on speech acts, artifact types, or insight categories can help users carve out task-specific slices of long conversations, supporting both analytical foraging and downstream communication.}}}

\vspace{-4mm}

\paragraph{RQ3: Workflows, navigation strategies, and selection.}
\rr{RQ3 examines how data workers navigate and restructure analytical conversations to address these information needs, and what tensions and opportunities their workflows reveal for tool design. As detailed in \S\ref{sec:results_strategies_info_needs} and \S\ref{sec:results_strategies_summary_communication}, participants engaged in a repertoire of navigation strategies: \emph{sequential navigation} through the transcript, \emph{abstractive navigation} across multiple levels of detail, \emph{visual recall} using icons and visualizations as anchors, and \emph{filtering} based on speech acts or artifact types. In authoring summaries, they selectively assembled structured elements, injected process- and action-oriented context, adjusted length and granularity, and reorganized content to fit audience expectations. These behaviors expose a central tension between maintaining the narrative continuity of the conversation and efficiently extracting task‑specific slices for communication. RQ3 is addressed by identifying concrete workflows, both for navigation and for summary construction, that future tools should explicitly support through synchronized overview+detail views, semantically meaningful visual/structural cues, and interaction mechanisms that treat structured elements as first-class units for selection and recomposition. More broadly, these findings extend the principle of overview‑and‑detail~\cite{Shneiderman1996TheEH, Jiang2023GraphologueEL, Suh2023SensecapeEM} to analytical conversations, highlighting opportunities for conversation-specific affordances that strengthen visual recall and streamline sequential navigation to better support orienting, recalling, and prioritizing.}

\aptLtoX{\begin{newimplications}
\rr{\textbf{Implication: Scaffolding human-AI co-authorship in summary composition.} Participants used LLM-generated summaries as starting points but invested significant effort in curating, restructuring, and annotating them. Tools should therefore support co-authored summaries that (i) surface candidate content from the conversation, (ii) make it easy to inject process notes and domain context, (iii) provide controls over length and structure, and (iv) support lightweight formatting and reordering. Rather than aiming for fully automatic reporting, tools can create better value by scaffolding human narrative work while preserving traceability back to the underlying analytical conversation.}
\end{newimplications}}{
\implications{\rr{\textbf{Implication: Scaffolding human-AI co-authorship in summary composition.} Participants used LLM-generated summaries as starting points but invested significant effort in curating, restructuring, and annotating them. Tools should therefore support co-authored summaries that (i) surface candidate content from the conversation, (ii) make it easy to inject process notes and domain context, (iii) provide controls over length and structure, and (iv) support lightweight formatting and reordering. Rather than aiming for fully automatic reporting, tools can create better value by scaffolding human narrative work while preserving traceability back to the underlying analytical conversation.}}}

\paragraph{RQ4: Role of AI in structuring and communicating analyses.}
\rr{RQ4 explores how data workers perceive the role of AI in structuring and communicating analytical conversations, and what forms of human–AI collaboration are preferred. As detailed in \S\ref{sec:results_role_ai} and \S\ref{sec:results_strategies_summary_communication}, participants consistently valued AI for its ability to scaffold and accelerate their workflows, such as by generating initial summaries, suggesting structure, and proposing language. However, they expressed a strong preference for maintaining editorial control, particularly in high-stakes contexts. Participants viewed AI as a collaborator that could offer suggestions and surface relevant content, but not as an autonomous agent responsible for shaping tone, emphasis, or final framing. These findings address RQ4 by showing that the preferred model of AI integration is one of mixed-initiative collaboration, where AI is proactive enough to support and guide, yet remains responsive to human direction, domain expertise, and communicative intent. The observations echo broader concerns in co-writing and generative systems literature about preserving authorial agency and accountability~\cite{Hwang2025, Hoque2024}, and reflect the limitations of LLMs in adapting to domain-specific or audience-specific rhetorical norms~\cite{Gu2023HowDD}. Unlike program synthesis tasks, where correctness is paramount~\cite{Vaithilingam2022ExpectationVE}, communication in data analysis involves interpretive judgment, audience framing, and rhetorical nuance~\cite{Hullman2011Rhetoric, Mao2019, Piorkowski2021}. Thus, fully automatic summary generation may be ill-suited for such tasks; rather, tools should emphasize controllable, contextual AI assistance that supports human authorship.}

\aptLtoX{\begin{newimplications}
\rr{\textbf{Implication: Reframing the role of AI toward assistive and accountable collaboration.} Participants preferred AI as a collaborator that structures, suggests, and accelerates, rather than an autonomous agent that replaces their judgment. Tool designers should prioritize human-in-the-loop controls: mechanisms for pinning, bookmarking, and selectively including content; transparency about where summary statements come from; and workflows that support iterative refinement. In high-stakes settings, systems should foreground human authorship and accountability, positioning AI outputs as drafts or alternatives rather than final products.}
\end{newimplications}}
{\implications{\rr{\textbf{Implication: Reframing the role of AI toward assistive and accountable collaboration.} Participants preferred AI as a collaborator that structures, suggests, and accelerates, rather than an autonomous agent that replaces their judgment. Tool designers should prioritize human-in-the-loop controls: mechanisms for pinning, bookmarking, and selectively including content; transparency about where summary statements come from; and workflows that support iterative refinement. In high-stakes settings, systems should foreground human authorship and accountability, positioning AI outputs as drafts or alternatives rather than final products.}}}

\subsection{Design Implications for Future Conversational Interfaces for Analyses}
\label{sec:implications}
\subsubsection{Tools should support structured navigation and selection}
Our findings show that data workers relied on structured views to move between overviews and details and to prioritize elements such as charts, insights, or analysis steps (\S\ref{sec:results_strategies_info_needs}). Future tools should make this navigation and selection central, treating the structured elements as first-class units for interaction. Incorporating such structure can also enhance AI assistance by grounding generative outputs in relevant context, a capability increasingly adopted in commercial tools~\cite{CursorAtSymbolsOverview, NotebookLM}. Establishing a shared semantic vocabulary over this structure, e.g., ``the chart comparing preferences by age,'' could bridge natural language queries with underlying abstractions~\cite{Liu2023WhatIW}. In parallel, developing a visual language for structured elements could support visual recall and information scent~\cite{Pirolli2003InformationScent}. 

\subsubsection{Tools should treat communication as an ongoing, co-evolving activity}
While the focus of the study was post-hoc navigation and communication, participants emphasized the need to structure, navigate, and communicate their findings during analysis, not just after the process (\S\ref{sec:results_role_ai}). This need for in-the-moment structuring and articulation of insights mirrors prior observations from computation notebook use, where data workers interleave code, results, and narrative to externalize insights as they emerge~\cite{Kery2018TheSI, Wang2020CallistoCT}. To better support this evolving process, future tools should treat communication as an \textit{integrated}, iterative part of analysis, not simply a final output stage. Real-time structuring, tagging, and pruning could help data workers surface key insights as they explore their data. Systems like \textit{InsightLens}~\cite{Weng2024InsightLensDA} begin to address this by enabling real-time insight extraction and navigation, but broader affordances are needed. Tools could support features such as tagging key turns, flagging important artifacts, pruning less relevant discussion, and offering real-time sharing. These affordances might include integrations with communication and presentation tools (e.g., Slack, PowerPoint) and adjacent editors or side-pane workspaces (akin to \syncsense) for drafting summaries during analysis.

\subsubsection{Tools should provide controllable and contextual AI assistance}
In our probe, AI was scoped to specific tasks, such as structuring the conversation or rewriting user-authored summaries. However, participants' reflections on AI involvement point to a design opportunity: enabling users to adjust \textit{how much} and in \textit{what ways} AI contributes to communication tasks. This need aligns with prior findings on users' varied communication goals~\cite{Pang2022HowDD, Zhang2020HowDD} and levels of comfort and trust in AI systems~\cite{Gu2023HowDA, Gu2023HowDD, Vaithilingam2022ExpectationVE, Liang2023UnderstandingTU, Mcnutt2023OnTD}. As agentic systems increasingly automate analytical workflows~\cite{Fine2025DataScienceAgent, CursorAgentMode, ManusAI}, it remains critical to ensure that users can stay meaningfully involved, especially in communicating data analyses where domain knowledge and rhetorical nuance matter. To better support user agency and accommodate diverse communication needs, future tools should offer \textit{controllable and contextual AI assistance} in data analysis workflows. For example, systems could allow users to toggle between different levels of AI involvement, such as selecting relevant summary content, generating initial drafts and data stories~\cite{Sultanum2023DATATALESIT, Wang2025JupybaraOA}, or refining language~\cite{Lisnic2025PlumeST}. These systems could also learn and adapt~\cite{Gao2024AligningLA, zhao2025do} to individual preferences and communication styles over time.

\subsubsection{The Value of Integrated Navigation and Communication}
\rr{Studying navigation and communication together proved both productive and revealing. Our findings revealed substantial overlap in the underlying structures and affordances that support both tasks. Features such as filtering, multi-level abstraction, and visual recall facilitated both personal reorientation and summary authoring. This convergence validates our theoretical premise that navigation and communication share infrastructural foundations. Our findings surfaced a directional interplay between the two: communicative intent frequently drove navigational behavior. Several participants (e.g., P2, P10) deliberately structured their conversational interactions with future communication in mind by organizing threads, tagging insights, or calling out key artifacts to facilitate later summarization. These practices reflect Pirolli and Card's model of the sensemaking loop~\cite{Pirolli2007TheSP}, wherein synthesis (e.g., preparing a summary) reveals gaps in understanding, triggering renewed foraging and investigation. These observations suggest that navigation and communication are not sequential stages but interdependent activities. Tools should therefore support this bidirectional sensemaking loop, helping users fluidly move between exploration and explanation, using shared structural cues to guide both inquiry and expression.}

\subsection{Limitations and Future Work}
\label{sec:limitations}
Our study has several limitations that inform directions for future research. \rr{We conducted the study using a single dataset and a specific analysis scenario. While this controlled setup enabled consistent comparisons across participants and helped surface diverse strategies and feedback~\cite{Borlund2003TheIE}, the study design inherently narrows the domain of inquiry and may limit the generalizability of our findings to other datasets, analytical tasks, or organizational settings. Our scenario centered on consumer preference data and a single high-level business question; analysts in scientific, civic, or operational domains may face different constraints, stakes, and collaboration patterns. To help mitigate this limitation, we selected a real-world dataset comprising heterogeneous attribute types and data irregularities (e.g., missing values). Nonetheless, future research should extend this work across a wider range of datasets, task types, and domains to validate, challenge, and refine the findings presented here.}


We employed LLMs to extract conversational elements such as insights, visualizations, and speech acts; techniques that are susceptible to known limitations of stochasticity and hallucinations. In addition, our user study primarily focused on using our design probe to elicit observations of information needs, strategies, and grounded discussion, leaving the validation of the extraction quality unaddressed. While we did not observe any adverse effects on participant comprehension or interaction with the tool, the reliability of these LLM-generated structures remains an open question. Future research should systematically evaluate the accuracy and robustness of LLM-based extraction techniques, especially for high-stakes analytical contexts.

\rr{Our study involved a relatively small sample drawn exclusively from a single organization, which may constrain the diversity of perspectives and limit the broader applicability of our findings. The participants' shared institutional context could reflect organizational norms around tools, documentation practices, and collaboration styles that are not representative of other settings. Furthermore, our participants were experienced data workers, which likely influenced their expectations around provenance, editorial control, and communication fidelity. These characteristics may not generalize to novice analysts, cross-functional stakeholders, or teams operating under different constraints or technical infrastructures. That said, the study design incorporated varied communication tasks (e.g., summarizing for an investor vs. an analyst business partner, via direct message or formal data report), allowing us to gather observations across a range of contexts. To validate and refine our design implications, future work should recruit a larger and more heterogeneous participant pool spanning diverse professional roles, experience levels, accessibility needs, and organizational environments.}


The structured elements we introduced were evaluated only within the scope of our study dataset and scenario. Participants engaged with nearly all categories (excluding one insight type), but it remains unclear how comprehensive these definitions are. Additional elements, such as key decisions or action-oriented context, may also be relevant to capture. Future research should refine and validate this schema across more complex or domain-specific environments.


\rr{Finally, although our analysis identifies descriptive patterns linking navigation strategies (e.g., filtering, abstractive navigation, visual recall) with the structure of participants' summaries, the study was not designed to support causal inferences regarding the impact of navigation on summary quality. Establishing such causal relationships would require a more controlled experimental design. A natural next step is to conduct a larger-scale, systematically varied study in which navigation supports are manipulated, such as the availability of filtering tools or visual cues, and downstream outcomes are evaluated. These outcomes could include summary usefulness, perceived quality, efficiency of composition, and communicative effectiveness across different stakeholder audiences.}

\section{Conclusion}
In this paper, we present an exploratory study on how data workers revisit, summarize, and communicate their analytical conversations with LLMs. To elicit these behaviors, we deployed a design probe that introduced affordances for on-demand overview and detail, alongside AI assistance features designed to preserve human agency during summarization. In a user study, participants used the probe to summarize their conversations for different audiences and communication media. We observed recurring information needs (i.e., orienting, recalling, and prioritizing) alongside the strategies data workers employed to navigate, distill, and communicate effectively. Our findings highlight opportunities for structured navigation and selective content composition, supporting communication as an ongoing, co-evolving activity. These observations also suggest directions for AI assistance that is controllable, communicatively aware, and better aligned with users' goals.



\bibliographystyle{ACM-Reference-Format}
\bibliography{main}


\clearpage

\appendix
\section{Supplementary Material}
\begin{figure}[h]
    \centering
  \includegraphics[width=0.9\linewidth]{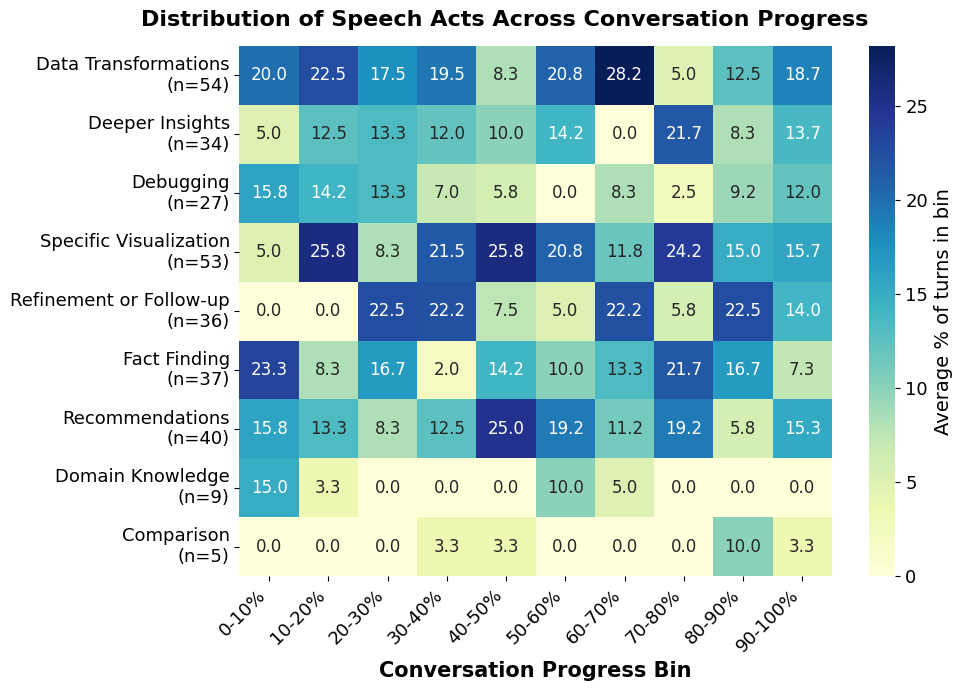}
    \caption{Frequency of each speech act at each turn decile. Each conversation is normalized by its number of turns and at each turn decile, the frequency is normalized by speech act types (i.e., columns sum to 100\%). This is averaged across all participant conversations. Participants focus on data transformations, visualizations and refinement throughout the conversation, indicating the iterative nature of analytical conversations
    }
  \label{fig:results_speechact_heatmap}
\end{figure}

\begin{figure}[h]
    \centering
  \includegraphics[width=0.9\linewidth]{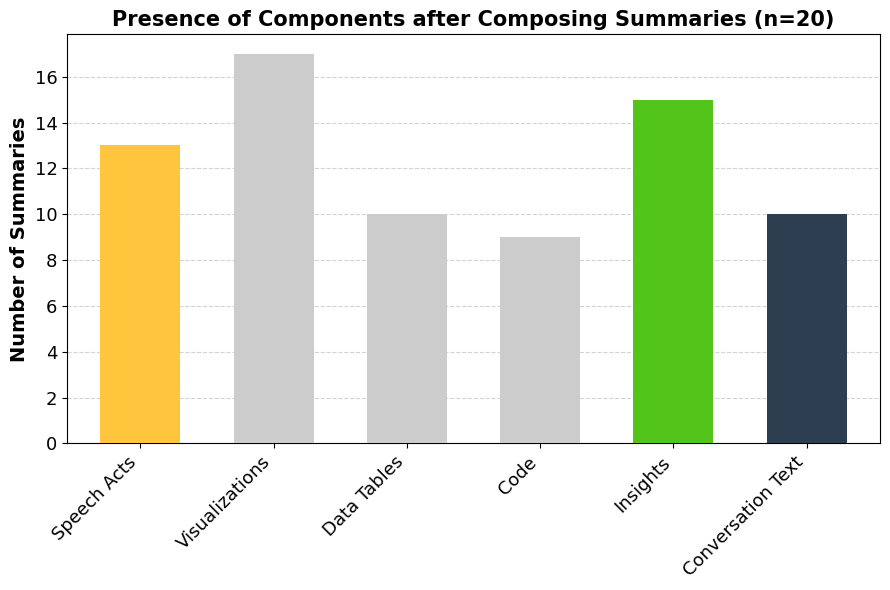}
    \caption{Presence of conversation elements in the drop container just before participants serialized their final summaries. Speech acts encompass~\syncsense summary of the given turn. Bars reflect binary presence, whether a element appeared in a summary, not frequency of use. Visualizations and insights appeared in over 75\% of summaries, with conversation text also commonly included for added context.}
  \label{fig:results_summary_composition}
\end{figure}

\begin{figure*}[t!]
    \centering
  \includegraphics[width=1.0\linewidth]{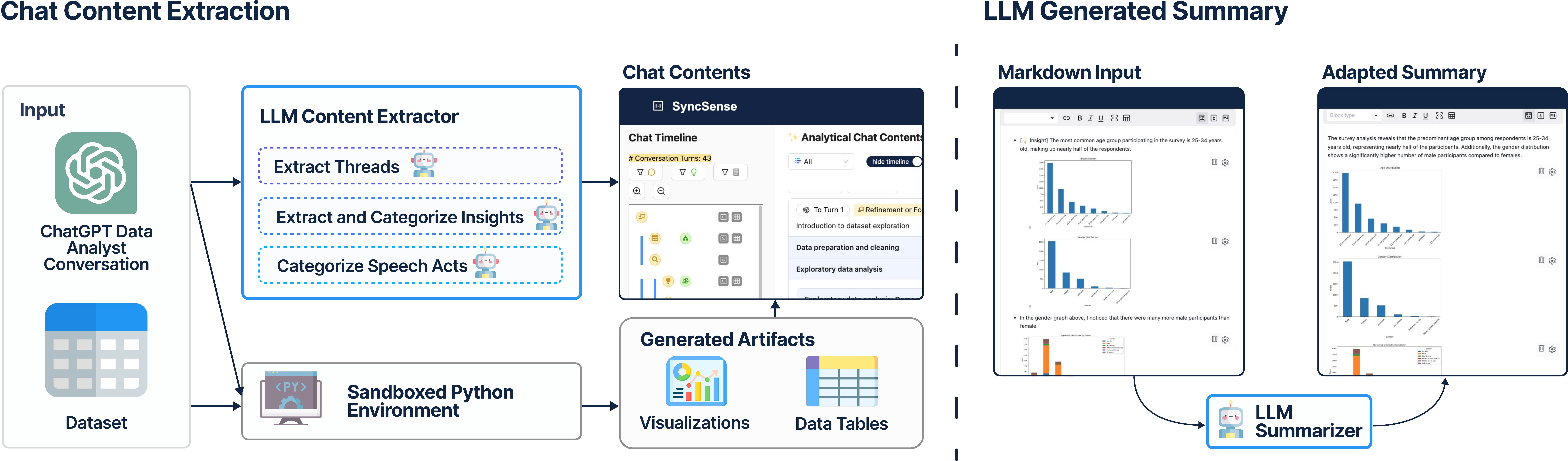}
  \caption{{\syncsense's Backend and LLM modules. The chat element extraction process takes the original ChatGPT analytical conversation and dataset. The code from the conversation is executed to generate the artifacts, while three LLM modules (three separate LLM API calls) are used to extract threads, insights, and speech acts, respectively (left). The LLM-generated summary in the \authoring panel takes the markdown and user configurations in the post-drop editor as input to adapt and generates (via an API call) the final summary (right).}}
  \label{fig:system}
\end{figure*}

The supplementary material includes the following:
\begin{enumerate}
    \item (\textbf{\S~\ref{appendix:additional_figures}}). Additional figures that show the backend design of \syncsense~(Fig.~\ref{fig:system}), how speech acts are distributed throughout an analytical conversation (Fig.~\ref{fig:results_speechact_heatmap}), and the composition of authored summaries (Fig.~\ref{fig:results_summary_composition}).
    \item(\textbf{\S~\ref{appendix:result_codes}}) A table of the codebook for the qualitative findings of our study (Table~\ref{tab:codes_summary}).
    \item (\textbf{\S~\ref{appendix:prompts}}) Prompt templates used in the backend of \syncsense.
    \item {\texttt{\textbf{study\_analytical\_conversations.csv}}} containing the links to the original ChatGPT Data Analyst conversations along with extracted statistics for all analytical conversations in our user study.
    \item {\texttt{\textbf{syncsense\_tutorial\_walkthrough.csv}}} containing screenshots of the tutorial walkthrough used to introduce participants to \syncsense.
    
\end{enumerate}

\subsection{Additional Figures}
\label{appendix:additional_figures}


\subsection{Study Result Codes}
\label{appendix:result_codes}
We include the qualitative codes, definitions, and occurrences in study sessions in Table~\ref{tab:codes_summary}.


\begin{table*}[ht]
\centering
\small
\begin{tabular}{|m{2.3cm}|p{4cm}|p{6cm}|p{3cm}|}
\hline
\textbf{Section} & \textbf{Code} & \textbf{Definition} & \textbf{Participants} \\
\hline
\multirow{3}{=}{Information Needs \newline(\textbf{RQ2}, Sec.~\ref{sec:results_info_needs})} 
& Orienting & Needing to situate oneself in the conversation to understand its structure and content. & P1–P10 \\
\cline{2-4}
& Recalling & Retrieving specific information or insights from memory. & P1, P2, P3, P10 \\
\cline{2-4}
& Prioritization & Determining which parts of the conversation are most relevant for summarization. & P1–P10 \\
\hline

\multirow{4}{=}{Strategies for \newline Information Needs  \newline  (\textbf{RQ3}, Sec.~\ref{sec:results_strategies_info_needs})}
& Sequential Navigation & Exploring contents in order, either through high-level or detailed views. & P1–P10 \\
\cline{2-4}
& Abstractive Navigation & Moving between different abstraction levels (e.g., summary to detail). & P1, P2, P3, P4, P7, P10 \\
\cline{2-4}
& Visual Recall & Using visuals, tags, or icons to trigger memory of key content. & P1, P2, P3, P4, P5, P8 \\
\cline{2-4}
& Filtering & Narrowing the view based on specific criteria. & P2, P3, P4, P6, P7 \\
\hline

\multirow{4}{=}{Strategies for Summary Communication \newline(\textbf{RQ3}, Sec.~\ref{sec:results_strategies_summary_communication})}
& Add Process Details & Including information about analytical steps or methodology. & P1, P2, P3, P4, P5, P9, P10 \\
\cline{2-4}
& Add Action-Oriented Context & Framing summaries in terms of goals, implications, and next steps. & P2, P3, P4, P5, P6, P7, P9 \\
\cline{2-4}
& Adjust Overall Length & Editing summaries for conciseness or comprehensiveness. & P4, P5, P6, P7, P8, P10 \\
\cline{2-4}
& Reorder/Format Content & Changing structure or visual style to improve readability and focus. & P2, P3, P6, P7, P10 \\
\hline

\multirow{4}{=}{Preferred Role of AI  \newline(\textbf{RQ4}, Sec.~\ref{sec:results_role_ai})}
& Retain Editorial Control & Wanting final say over summary content, especially in high-stakes settings. & P1, P2, P5, P7 \\
\cline{2-4}
& Selective + Structured Authoring & Leveraging AI for structure while choosing content manually. & P2, P3, P4, P6, P10 \\
\cline{2-4}
& Acceleration, Not Automation & Using AI to save time or surface drafts without ceding control. & P2, P5, P6, P10 \\
\cline{2-4}
& Scaffolding analyses & Using AI to help structure the analysis process & P2, P3, P10 \\
\hline
\end{tabular}
\caption{Codes organized by analytic focus area, with definitions and participants.}
\label{tab:codes_summary}
\end{table*}

\clearpage
\onecolumn

\subsection{Prompt Templates}
\label{appendix:prompts}

\subsubsection{Prompts for summarizing turns and extracting threads}

\cprotect[mm]\aptLtoX{\begin{tabular}{p{500pt}}
\cellcolor{gray!40}\textcolor{white}{Summarize Turns. We first summarize each turn before extracting the threads.}\\
\textbf{System prompt}\\
You are an AI Conversational Data Analysis Assistant who is an expert at understanding a data conversation. In a conversation, because data analysis is non-linear and iterative, there can be multiple threads of analysis being explored. Analysis threads are distinct, focused lines of investigation or exploration within a larger analytical conversation. Each thread can address a specific question, hypothesis, or aspect of the overall topic, often requiring its own set of methods and analyses. There can also be sub-threads within a thread that go into more detail on a specific aspect of the thread.\\
\hline
\includegraphics[width=\iconwidth]{emojis/user.png} \textbf{User:}\\
<Instruction> We want to capture the structure of analysis threads in a conversation.\\
First, we need to understand what the each turn in the conversation is exploring.
Given a conversation history which is a list of ordered messages between
the user and the assistant with each message in the following format:\\
\begin{imageonly}
\begin{jsonblock}
```json
{
  "turn": 0,  // id for the turn in the conversation PAY ATTENTION TO THIS 
  "role": "user",  // "user" or "assistant" or "tool"
  "content": "Hello World",  // the content of this message, in markdown format string
}
```
\end{jsonblock}\end{imageonly}\\
return a short half-sentence summary for each turn. 
A single turn encompasses both the user query and the assistant messages. 
Therefore, there are more than one message (JSON) with the same turn ID.\\
**Note**: \\
- The number of threads in the conversation should match the number of turns in the conversation (as specified by the "turn" field).\\
- This means that the number of threads should be the same as the number of unique turn IDs in the conversation history.\\
- Be sure to include the turn number that the summary references
Please return the result in the format specified in **Format Instructions** enclosed in triple backticks.\\
</Instruction>\\
<Format Instructions>\\
\begin{imageonly}
\begin{jsonblock}
// Format instructions based on LangChain's Pydantic JSON Parser (hidden for brevity)
\end{jsonblock}
\end{imageonly}\\
<Format Instructions/>\\
<Example>\\
Conversation History:\\
\begin{imageonly}
\begin{jsonblock}
// Real example of an analytical conversation analyzing airlines (hidden for brevity)
\end{jsonblock}
\end{imageonly}\\
Analysis Turn Summaries:\\
\begin{imageonly}
\begin{jsonblock}
```json
{
  "threads": [
    { "tid": 1, "title": "Any source airports? no", "turn": 0 },
    { "tid": 2, "title": "Any destination airports? no", "turn": 1 },
    { "tid": 3, "title": "What is the most traveled route", "turn": 2 },
    { "tid": 4, "title": "Pricing Per Airline", "turn": 3 },
    { "tid": 5, "title": "Airline Class Differences", "turn": 4 },
    ... // (hidden for brevity)
    { "tid": 15, "title": "Plot Improvement 1: breakdown by airline", "turn": 14 },
    { "tid": 16, "title": "Plot Improvement 2: average price per airline", "turn": 15 },
    { "tid": 17, "title": "Plot Alternative 3: merge all airlines", "turn": 16 },
    { "tid": 18, "title": "Conversation Summary", "turn": 17 }
  ]
}
```
\end{jsonblock}\end{imageonly}\\
</Example>\\

Begin.\\
Conversation History:\\
\\
\cellcolor{gray!40}{EXAMPLE INPUT}\\
\begin{imageonly}
\begin{jsonblock}
```json
[
  {
    "role": "user",
    "content": "Dear Chatty, I have a dataset to analyse. Let`s imagine that me and my business partner are planning to open a coffee shop in our city. To make informed decisions, we will analyze data to guide key strategies such as which coffee to source, pricing, and target customer demographics. Our goal is to uncover actionable insights to shape your business plan. The data set that we have at our disposal is called \"The Great American Coffee Taste Test\".  The link to the dataset is here : [csv_link]\n\n Dataset Description: In October 2023, \"world champion barista\" James Hoffmann and coffee company Cometeer held the \"Great American Coffee Taste Test\" on YouTube, during which viewers were asked to fill out a survey about 4 coffees they ordered from Cometeer for the tasting.",
    "turn": 0
  },
  {
    "role": "assistant",
    "content": "Got it! You`ve uploaded the dataset, so let`s start by exploring its structure. I`ll load the data and give you a quick summary, including column names, data types, and a few sample rows.",
    "turn": 0
  },
  ... // (hidden for brevity)
]
```
\end{jsonblock}
\end{imageonly}\\
Analysis Turn Summaries:\\
\hline
\includegraphics[width=\iconwidth]{emojis/assistant.png} \textbf{Assistant:}\\
\cellcolor{gray!40}{EXAMPLE INPUT}\\
\begin{imageonly}
\begin{jsonblock}
```json
{
  "threads": [
    {"tid": 1, "title": "Initial setup and dataset overview", "turn": 0 },
    {"tid": 2, "title": "List of column names and their top categories", "turn": 1 },
    ... // (hidden to save space)
    {"tid": 7, "title": "Re-upload dataset due to reset", "turn": 6 },
    {"tid": 8, "title": "Summary of detailed coffee preferences", "turn": 7 },
    {"tid": 9, "title": "Number of people matching each category", "turn": 8 },
    {"tid": 10, "title": "Clarifying categorization criteria for accuracy", "turn": 9 },
    {"tid": 11, "title": "Confirming distinct elements of where_drink column", "turn": 10 },
    {"tid": 12, "title": "Calculating count distinct for customer categories", "turn": 11 }
  ]
}
```
\end{jsonblock}
\end{imageonly}\\
\end{tabular}}{
\begin{prompt}[Summarize Turns. We first summarize each turn before extracting the threads.]

\begin{systemmessage}
You are an AI Conversational Data Analysis Assistant who is an expert at understanding a data conversation. In a conversation, because data analysis is non-linear and iterative, there can be multiple threads of analysis being explored. Analysis threads are distinct, focused lines of investigation or exploration within a larger analytical conversation. Each thread can address a specific question, hypothesis, or aspect of the overall topic, often requiring its own set of methods and analyses. There can also be sub-threads within a thread that go into more detail on a specific aspect of the thread.
\end{systemmessage}

\begin{usermessage}
<Instruction> We want to capture the structure of analysis threads in a conversation.

First, we need to understand what the each turn in the conversation is exploring.
Given a conversation history which is a list of ordered messages between
the user and the assistant with each message in the following format:

\begin{jsonblock}
```json
{
  "turn": 0,  // id for the turn in the conversation PAY ATTENTION TO THIS 
  "role": "user",  // "user" or "assistant" or "tool"
  "content": "Hello World",  // the content of this message, in markdown format string
}
```
\end{jsonblock}
return a short half-sentence summary for each turn. 
A single turn encompasses both the user query and the assistant messages. 
Therefore, there are more than one message (JSON) with the same turn ID.

**Note**: 

- The number of threads in the conversation should match the number of turns in the conversation (as specified by the "turn" field).

- This means that the number of threads should be the same as the number of unique turn IDs in the conversation history.

- Be sure to include the turn number that the summary references
Please return the result in the format specified in **Format Instructions** enclosed in triple backticks.

</Instruction>

<Format Instructions>
\begin{jsonblock}
// Format instructions based on LangChain's Pydantic JSON Parser (hidden for brevity)
\end{jsonblock}
<Format Instructions/>

<Example>

Conversation History:
\begin{jsonblock}
// Real example of an analytical conversation analyzing airlines (hidden for brevity)
\end{jsonblock}

Analysis Turn Summaries:

\begin{jsonblock}
```json
{
  "threads": [
    { "tid": 1, "title": "Any source airports? no", "turn": 0 },
    { "tid": 2, "title": "Any destination airports? no", "turn": 1 },
    { "tid": 3, "title": "What is the most traveled route", "turn": 2 },
    { "tid": 4, "title": "Pricing Per Airline", "turn": 3 },
    { "tid": 5, "title": "Airline Class Differences", "turn": 4 },
    ... // (hidden for brevity)
    { "tid": 15, "title": "Plot Improvement 1: breakdown by airline", "turn": 14 },
    { "tid": 16, "title": "Plot Improvement 2: average price per airline", "turn": 15 },
    { "tid": 17, "title": "Plot Alternative 3: merge all airlines", "turn": 16 },
    { "tid": 18, "title": "Conversation Summary", "turn": 17 }
  ]
}
```
\end{jsonblock}
</Example>

Begin.

Conversation History:
\begin{describedexample}{}{Example input}
\begin{jsonblock}
```json
[
  {
    "role": "user",
    "content": "Dear Chatty, I have a dataset to analyse. Let`s imagine that me and my business partner are planning to open a coffee shop in our city. To make informed decisions, we will analyze data to guide key strategies such as which coffee to source, pricing, and target customer demographics. Our goal is to uncover actionable insights to shape your business plan. The data set that we have at our disposal is called \"The Great American Coffee Taste Test\".  The link to the dataset is here : [csv_link]\n\n Dataset Description: In October 2023, \"world champion barista\" James Hoffmann and coffee company Cometeer held the \"Great American Coffee Taste Test\" on YouTube, during which viewers were asked to fill out a survey about 4 coffees they ordered from Cometeer for the tasting.",
    "turn": 0
  },
  {
    "role": "assistant",
    "content": "Got it! You`ve uploaded the dataset, so let`s start by exploring its structure. I`ll load the data and give you a quick summary, including column names, data types, and a few sample rows.",
    "turn": 0
  },
  ... // (hidden for brevity)
]
```
\end{jsonblock}
\end{describedexample}

Analysis Turn Summaries:
\end{usermessage}

\begin{assistantmessage}
\begin{describedexample}{}{Example Output}
\begin{jsonblock}
```json
{
  "threads": [
    {"tid": 1, "title": "Initial setup and dataset overview", "turn": 0 },
    {"tid": 2, "title": "List of column names and their top categories", "turn": 1 },
    ... // (hidden to save space)
    {"tid": 7, "title": "Re-upload dataset due to reset", "turn": 6 },
    {"tid": 8, "title": "Summary of detailed coffee preferences", "turn": 7 },
    {"tid": 9, "title": "Number of people matching each category", "turn": 8 },
    {"tid": 10, "title": "Clarifying categorization criteria for accuracy", "turn": 9 },
    {"tid": 11, "title": "Confirming distinct elements of where_drink column", "turn": 10 },
    {"tid": 12, "title": "Calculating count distinct for customer categories", "turn": 11 }
  ]
}
```
\end{jsonblock}
\end{describedexample}
\end{assistantmessage}
\end{prompt}}

\cprotect[mm]\aptLtoX{\begin{tabular}{p{500pt}}
\cellcolor{gray!40}\textcolor{white}{Consolidate Threads. We group similar threads level by level starting with the turn summaries.}\\
\textbf{System prompt:}\\
You are an AI Conversational Data Analysis Assistant who is an expert at understanding a data conversation. In a conversation, because data analysis is non-linear and iterative, there can be multiple threads of analysis being explored. Analysis threads are distinct, focused lines of investigation or exploration within a larger analytical conversation. Each thread can address a specific question, hypothesis, or aspect of the overall topic, often requiring its own set of methods and analyses. There can also be sub-threads within a thread that go into more detail on a specific aspect of the thread.\\
\hline
\includegraphics[width=\iconwidth]{emojis/user.png} \textbf{User:}\\
 <Instruction> We want to iteratively capture the structure of analysis threads in a conversation.\\
Given the analysis threads summary and how the conversation is structured, determine if any of the threads can be grouped together to represent a larger thread. If so, group the threads together and provide a new summary for the grouped thread.\\
Please return the grouped analysis thread summaries based on the format specified in **Format Instructions** enclosed in triple backticks.
For any new thread, please give a new unique thread ID. In addition, be sure to reference the input thread IDs that were grouped together.
If no threads can be grouped together, return an empty YAML.\\
</Instruction>\\
<Format Instructions>\\
\begin{imageonly}
\begin{jsonblock}
// Format instructions based on LangChain's Pydantic JSON Parser (hidden to save space)
\end{jsonblock}
\end{imageonly}\\
<Format Instructions/>\\
Begin.\\
Analysis Threads Summary:\\
\\
\cellcolor{gray!40}{EXAMPLE INPUT}\\
\begin{imageonly}
\begin{jsonblock}
```json
{
  "threads": [
    {"tid": 1, "title": "Initial setup and dataset overview", "turn": 0 },
    {"tid": 2, "title": "List of column names and their top categories", "turn": 1 },
    ... // (hidden to save space)
    {"tid": 9, "title": "Number of people matching each category", "turn": 8 },
    {"tid": 10, "title": "Clarifying categorization criteria for accuracy", "turn": 9 },
    {"tid": 11, "title": "Confirming distinct elements of where_drink column", "turn": 10 },
    {"tid": 12, "title": "Calculating count distinct for customer categories", "turn": 11 }
  ]
}
```
\end{jsonblock}
\end{imageonly}\\
Grouped Analysis Threads Summary: \\
\hline
\includegraphics[width=\iconwidth]{emojis/assistant.png} \textbf{Assistant:}\\
\cellcolor{gray!40}{EXAMPLE OUPUT}\\
\begin{imageonly}
\begin{jsonblock}
```json
{
  "threads": [{
      "tid": 13,
      "title": "Coffee drinker categorization and detailed analysis",
      "child_ids": [3, 4, 5, 6, 8, 9, 10, 12]
    }]
}
```
\end{jsonblock}
\end{imageonly}\\
\end{tabular}}{
\begin{prompt}[Consolidate Threads. We group similar threads level by level starting with the turn summaries.]
\begin{systemmessage}
You are an AI Conversational Data Analysis Assistant who is an expert at understanding a data conversation. In a conversation, because data analysis is non-linear and iterative, there can be multiple threads of analysis being explored. Analysis threads are distinct, focused lines of investigation or exploration within a larger analytical conversation. Each thread can address a specific question, hypothesis, or aspect of the overall topic, often requiring its own set of methods and analyses. There can also be sub-threads within a thread that go into more detail on a specific aspect of the thread.
\end{systemmessage}

\begin{usermessage}
<Instruction> We want to iteratively capture the structure of analysis threads in a conversation.

Given the analysis threads summary and how the conversation is structured, determine if any of the threads can be grouped together to represent a larger thread. If so, group the threads together and provide a new summary for the grouped thread.
    
Please return the grouped analysis thread summaries based on the format specified in **Format Instructions** enclosed in triple backticks.
For any new thread, please give a new unique thread ID. In addition, be sure to reference the input thread IDs that were grouped together.
If no threads can be grouped together, return an empty YAML.

</Instruction>

<Format Instructions>
\begin{jsonblock}
// Format instructions based on LangChain's Pydantic JSON Parser (hidden to save space)
\end{jsonblock}
<Format Instructions/>

Begin.

Analysis Threads Summary:
\begin{describedexample}{}{Example Input}
\begin{jsonblock}
```json
{
  "threads": [
    {"tid": 1, "title": "Initial setup and dataset overview", "turn": 0 },
    {"tid": 2, "title": "List of column names and their top categories", "turn": 1 },
    ... // (hidden to save space)
    {"tid": 9, "title": "Number of people matching each category", "turn": 8 },
    {"tid": 10, "title": "Clarifying categorization criteria for accuracy", "turn": 9 },
    {"tid": 11, "title": "Confirming distinct elements of where_drink column", "turn": 10 },
    {"tid": 12, "title": "Calculating count distinct for customer categories", "turn": 11 }
  ]
}
```
\end{jsonblock}
\end{describedexample}

Grouped Analysis Threads Summary: 
\end{usermessage}

\begin{assistantmessage}
\begin{describedexample}{}{Example Output}
\begin{jsonblock}
```json
{
  "threads": [{
      "tid": 13,
      "title": "Coffee drinker categorization and detailed analysis",
      "child_ids": [3, 4, 5, 6, 8, 9, 10, 12]
    }]
}
```
\end{jsonblock}
\end{describedexample}
\end{assistantmessage}
\end{prompt}}

\clearpage
\subsubsection{Prompt for extracting speech acts}

\cprotect[mm]\aptLtoX{\begin{tabular}{p{500pt}}
\cellcolor{gray!40}\textcolor{white}{Speech Acts Extraction Prompt}\\
\textbf{System prompt:}\\
You are an AI Conversational Data Analysis Assistant who
is an expert at understanding a analytical conversation.
The conversation consists of user queries and assistant responses.
We want to understand the speech acts in the conversation.
Speech acts is a kind of action being performed by the speaker.
The analytical conversation will consist of the following speech acts from the user: 
\\
- Fact Finding: expecting a single response such as a number, or table, or “yes/no” (e.g., “did any passengers survive on Titanic?”)\\
- Specific Visualization: requesting a specific type of visualization (e.g., “can you plot a bar chart of the number of passengers by class?”)\\
- Domain Knowledge: asking about domain knowledge that is not directly answered by the data (ONLY include when the assistant does not have code in the response) \\
- Deeper Insights: e.g., "have there been outliers per class?”, “How much more likely was a passenger to survive if they were in first class?”\\
- Data Transformations: data manipulations such as filtering, sorting, or aggregating (e.g., “can you filter the data to show only passengers in first class?”)\\
- Recommendations: asking for recommendations or predictions based on the data (e.g., “what is the best course of action for the airline?”)\\
- Refinement or Follow-up: refining a previous query or asking for more details (e.g., “can you show the survival rates by age group?”)\\
- Debugging: asking for clarification or help with the analysis (e.g., “I don’t understand the visualization”)\\
\hline

\includegraphics[width=\iconwidth]{emojis/user.png} \textbf{User:}\\
<Instruction> Given the user queries used in the conversation, which is a list of \
user messages with each message in the following format:\\
\begin{imageonly}
\begin{jsonblock}    
```json
{
  "turn": 0,  // id for the turn in the conversation
  "role": "USER",  // "USER
  "content": "Hello World",  // the content of this message, in markdown format string
}
```
\end{jsonblock}
\end{imageonly}\\
please return the speech act for each user query.
Please return the result based on the format specified in **Format Instructions**.\\
</Instruction>\\
<Format Instructions>\\
\begin{imageonly}
\begin{jsonblock}
// Format instructions based on LangChain's Pydantic JSON Parser (hidden to save space)
\end{jsonblock}
\end{imageonly}\\
</Format Instructions>\\
Begin.\\
User Queries:\\
\cellcolor{gray!40}{EXAMPLE INPUT}\\
\begin{imageonly}
\begin{jsonblock}
```json
[
  {
    "role": "user",
    "content": "Dear Chatty, I have a dataset to analyse...", // (shortened for brevity)
    "turn": 0
  },
  {
    "role": "user",
    "content": "Can you list all the columns names and their 3 main attributes, so I have a sense of the data ? For the top 3 categories for each columns names give me the frequency in number and %",
    "turn": 1,
  }
  ... // (hidden for brevity)
]
```
\end{jsonblock}
\end{imageonly}\\
Speech Acts:\\
\includegraphics[width=\iconwidth]{emojis/assistant.png} \textbf{Assistant:}\\
\cellcolor{gray!40}{EXAMPLE OUPUT}\\
\begin{imageonly}
        \begin{jsonblock}
```json
{
  "speech_acts": [
    { "msg_turn_id": 0, "speech_act": "Fact Finding" },
    { "msg_turn_id": 1, "speech_act": "Fact Finding" },
    { "msg_turn_id": 2, "speech_act": "Domain Knowledge" },
    { "msg_turn_id": 3, "speech_act": "Data Transformations" },
    { "msg_turn_id": 4, "speech_act": "Refinement or Follow-up" },
    { "msg_turn_id": 5, "speech_act": "Refinement or Follow-up" },
    { "msg_turn_id": 6, "speech_act": "Fact Finding" },
    { "msg_turn_id": 7, "speech_act": "Refinement or Follow-up" },
    { "msg_turn_id": 8, "speech_act": "Fact Finding" },
    { "msg_turn_id": 9, "speech_act": "Fact Finding" },
    { "msg_turn_id": 10, "speech_act": "Refinement or Follow-up" },
    { "msg_turn_id": 11, "speech_act": "Fact Finding" },
    { "msg_turn_id": 12, "speech_act": "Fact Finding" }
  ]
}
```
        \end{jsonblock}
        \end{imageonly}\\
\end{tabular}}{
\begin{prompt}[Speech Acts Extraction Prompt]
\begin{systemmessage}
You are an AI Conversational Data Analysis Assistant who
is an expert at understanding a analytical conversation.
The conversation consists of user queries and assistant responses.
We want to understand the speech acts in the conversation.
Speech acts is a kind of action being performed by the speaker.
The analytical conversation will consist of the following speech acts from the user: 
\\
- Fact Finding: expecting a single response such as a number, or table, or “yes/no” (e.g., “did any passengers survive on Titanic?”)

- Specific Visualization: requesting a specific type of visualization (e.g., “can you plot a bar chart of the number of passengers by class?”)

- Domain Knowledge: asking about domain knowledge that is not directly answered by the data (ONLY include when the assistant does not have code in the response) 

- Deeper Insights: e.g., "have there been outliers per class?”, “How much more likely was a passenger to survive if they were in first class?”

- Data Transformations: data manipulations such as filtering, sorting, or aggregating (e.g., “can you filter the data to show only passengers in first class?”)

- Recommendations: asking for recommendations or predictions based on the data (e.g., “what is the best course of action for the airline?”)

- Refinement or Follow-up: refining a previous query or asking for more details (e.g., “can you show the survival rates by age group?”)

- Debugging: asking for clarification or help with the analysis (e.g., “I don’t understand the visualization”)
\end{systemmessage}

\begin{usermessage}
<Instruction> Given the user queries used in the conversation, which is a list of \
user messages with each message in the following format:
\begin{jsonblock}    
```json
{
  "turn": 0,  // id for the turn in the conversation
  "role": "USER",  // "USER
  "content": "Hello World",  // the content of this message, in markdown format string
}
```
\end{jsonblock}

please return the speech act for each user query.
Please return the result based on the format specified in **Format Instructions**.

</Instruction>

<Format Instructions>
\begin{jsonblock}
// Format instructions based on LangChain's Pydantic JSON Parser (hidden to save space)
\end{jsonblock}
</Format Instructions>

Begin.

User Queries:
\begin{describedexample}{}{Example Input}
\begin{jsonblock}
```json
[
  {
    "role": "user",
    "content": "Dear Chatty, I have a dataset to analyse...", // (shortened for brevity)
    "turn": 0
  },
  {
    "role": "user",
    "content": "Can you list all the columns names and their 3 main attributes, so I have a sense of the data ? For the top 3 categories for each columns names give me the frequency in number and %",
    "turn": 1,
  }
  ... // (hidden for brevity)
]
```
\end{jsonblock}
\end{describedexample}
Speech Acts:
\end{usermessage}
\begin{assistantmessage}
    \begin{describedexample}{}{Example Output}
        \begin{jsonblock}
```json
{
  "speech_acts": [
    { "msg_turn_id": 0, "speech_act": "Fact Finding" },
    { "msg_turn_id": 1, "speech_act": "Fact Finding" },
    { "msg_turn_id": 2, "speech_act": "Domain Knowledge" },
    { "msg_turn_id": 3, "speech_act": "Data Transformations" },
    { "msg_turn_id": 4, "speech_act": "Refinement or Follow-up" },
    { "msg_turn_id": 5, "speech_act": "Refinement or Follow-up" },
    { "msg_turn_id": 6, "speech_act": "Fact Finding" },
    { "msg_turn_id": 7, "speech_act": "Refinement or Follow-up" },
    { "msg_turn_id": 8, "speech_act": "Fact Finding" },
    { "msg_turn_id": 9, "speech_act": "Fact Finding" },
    { "msg_turn_id": 10, "speech_act": "Refinement or Follow-up" },
    { "msg_turn_id": 11, "speech_act": "Fact Finding" },
    { "msg_turn_id": 12, "speech_act": "Fact Finding" }
  ]
}
```
        \end{jsonblock}
    \end{describedexample}
\end{assistantmessage}
\end{prompt}}

\clearpage
\subsubsection{Prompt for extracting insights.}

\cprotect[mm]\aptLtoX{\begin{tabular}{p{500pt}}
\cellcolor{gray!40}\textcolor{white}{Insight Extraction Prompt}\\
\textbf{System prompt:}\\
You are an AI Data Analysis Assistant who is an expert at extracting insights from a data conversation. Here is some background information about data insights:\\
*Data Insight**\\
Data insights are numerical or statistical results derived from data.\\
**Insight Category**\\
There are 11 types of data insights: Value, Proportion, Difference, Distribution, Trend, Rank, Aggregation, Association, Extreme, Categorization and Outlier. The explanations of these types are listed below: \\
- Value: Retrieve the exact value of data attribute(s) under a set of specific criteria.\\
- Proportion: Measure the proportion of selected data attribute(s) within a specified set.\\
- Difference: Compare any two/more data attributes or compare the target object with previous values along with the time series.\\
- Distribution: Demonstrate the amount of value shared across the selected data attribute or show the breakdown of all data attributes.\\
- Trend: Present a general tendency over a time segment.\\
- Rank: Sort the data attributes based on their values and show the breakdown of selected data attributes.\\
- Aggregation: Calculate the descriptive statistical indicators (e.g., average, sum, count, etc. ) based on the data attributes.\\
- Association: Identify the useful relationship between two data attributes or among multiple attributes.\\
- Extreme: Find the extreme data cases along with the data attributes or within a certain range.\\
- Categorization: Select the data attribute(s) that satisfy certain conditions.\\
- Outlier: Explore the unexpected data attribute(s) or statistical outlier(s) from a given set.\\
**Insight Evidence**\\
Insight evidence are intermediate outputs in the conversation that **DIRECTLY** support each insight. Here are the types of insight evidence and their explanations:\\
* Code: The **EXACT** piece of code that **DIRECTLY** derives the insight. Dependency import and visualization code are **EXCLUDED** from consideration. For example, if the insight is "The positive rate for females is more than double that of males", the piece of code we focus on is \\
\begin{imageonly}
\begin{jsonblock}
  ```python
  \# Filter only the positive rates
  male_positive_rate = gender_positive_rate.loc['Male', 'Positive']
  female_positive_rate = gender_positive_rate.loc['Female', 'Positive']
  ```
\end{jsonblock}
\end{imageonly}\\
* Code Console Output: The **EXACT** code output that **DIRECTLY** derives the insight. For example, if the insight is "There is a particular peak of worldwide gross in 1999", the code output we focus on is\\
\begin{imageonly}
\begin{jsonblock}
  ```console
  Release Year  Worldwide Gross
  ...           ...
  14            1999
  ...           ...
  ```
\end{jsonblock}
\end{imageonly}\\
* Source Text: The **quote** of conversation where the insight is summarized. You must identify the **minimal but critical** part of the conversation that **DIRECTLY** derives the insight.\\
\hline

\includegraphics[width=\iconwidth]{emojis/user.png} \textbf{User:}\\
<Instruction> Given an analysis conversation which is a list of ordered messages and 
each message is in the following format: \\
\begin{imageonly}
\begin{jsonblock}    
```json
{{
  "id": 0,  // unique id of every message
  "role": "USER",  // "USER" or "ASSISTANT"
  "content": "Hello World",  // the content of this message, in markdown format string
}}
```
\end{jsonblock}
\end{imageonly}\\
extract all insights from the conversation. Please return the insights based on the format specified in **Format Instructions**.\\
**IMPORTANT**: \\
- Find as **MANY** insights as possible from the conversation. But remember, **QUALITY** is more important than **QUANTITY**. **NEVER EVER** make up insights. **ONLY** derive insights from the conversation and the dataset.\\
- You should only extract insights that are directly related to **Data Analysis**. Irrelevant insights **MUST NOT** be included.\\
- **NEVER EVER** make up insight types. **ONLY** classify the insights into the 11 categories: [Value, Proportion, Difference, Distribution, Trend, Rank, Aggregation, Association, Extreme, Categorization, Outlier].\\
- Insight evidence, especially **CODE** and **CODE OUTPUT**, must be **DIRECTLY** releated to the corresponding insight, following the instructions above. Make them **ACCURATE** and **CONCISE**.\\
- The insights should reference the original messages (via the `id` field) that are part of a given thread. These messages do not have to be continuous and may contain disjoint parts of the conversation.\\
</Instruction>\\
<Format Instructions>\\
\begin{imageonly}
\begin{jsonblock}
// Format instructions based on LangChain's Pydantic JSON Parser (hidden to save space)
\end{jsonblock}
\end{imageonly}\\
</Format Instructions>\\
Begin.\\

Convsersation History: \\
\cellcolor{gray!40}{EXAMPLE INPUT}\\
\begin{imageonly}
\begin{jsonblock}
```json
[
  {
    "role": "user",
    "content": "Dear Chatty, I have a dataset to analyse. Let`s imagine that me and my business partner are planning to open a coffee shop in our city. To make informed decisions, we will analyze data to guide key strategies such as which coffee to source, pricing, and target customer demographics. Our goal is to uncover actionable insights to shape your business plan. The data set that we have at our disposal is called \"The Great American Coffee Taste Test\".  The link to the dataset is here : [csv_link]\n\n Dataset Description: In October 2023, \"world champion barista\" James Hoffmann and coffee company Cometeer held the \"Great American Coffee Taste Test\" on YouTube, during which viewers were asked to fill out a survey about 4 coffees they ordered from Cometeer for the tasting.",
    "turn": 0
  },
  {
    "role": "assistant",
    "content": "Got it! You`ve uploaded the dataset, so let`s start by exploring its structure. I`ll load the data and give you a quick summary, including column names, data types, and a few sample rows.",
    "turn": 0
  },
  ... // (hidden for brevity)
]
```
\end{jsonblock}
\end{imageonly}
Insights:\\
\hline

\includegraphics[width=\iconwidth]{emojis/assistant.png} \textbf{Assistant:}\\
\cellcolor{gray!40}{EXAMPLE OUPUT}\\
\begin{imageonly}
        \begin{jsonblock}
```json
{
  "insights": [
    {
      "insight": "The dataset contains 4,042 entries and 57 columns, covering various aspects of coffee consumption, preferences, and demographics.",
      "keywords": ["dataset", "entries", "columns", "structure"],
      "sourceMessageIds": [3],
      "types": ["Value"],
      "evidence": {
        "code_console": ... // (hidden for brevity)
    }},
    {
      "insight": "The top 3 values for the 'age' column are '25-34 years old' (1986, 49.51%), '35-44 years old' (960, 23.93%), and '18-24 years old' (461, 11.49%).",
      "keywords": ["age", "values", "frequency"],
      "sourceMessageIds": [8],
      "types": ["Value", "Rank"],
      "evidence": {
        "code_console": "{'age': [('25-34 years old', 1986, 49.51),\n  ('35-44 years old', 960, 23.93),\n  ('18-24 years old', 461, 11.49)]}"
      }
    },
    {
      "insight": "The top 3 values for the 'gender' column are 'Male' (2524, 71.64%), 'Female' (853, 24.21%), and 'Non-binary' (103, 2.92%).",
      "keywords": ["gender", "values", "frequency"],
      "sourceMessageIds": [8],
      "types": ["Value", "Rank"],
      "evidence": {
        "code_console": "{'gender': [('Male', 2524, 71.64),\n  ('Female', 853, 24.21),\n  ('Non-binary', 103, 2.92)]}"
      }
    }
    ... // (hidden for brevity)
    ]
}
```
        \end{jsonblock}
        \end{imageonly}
\end{tabular}}{
\begin{prompt}[Insight Extraction Prompt]
\begin{systemmessage}
You are an AI Data Analysis Assistant who is an expert at extracting insights from a data conversation. Here is some background information about data insights:

*Data Insight**

Data insights are numerical or statistical results derived from data.

**Insight Category**

There are 11 types of data insights: Value, Proportion, Difference, Distribution, Trend, Rank, Aggregation, Association, Extreme, Categorization and Outlier. The explanations of these types are listed below: 

- Value: Retrieve the exact value of data attribute(s) under a set of specific criteria.

- Proportion: Measure the proportion of selected data attribute(s) within a specified set.

- Difference: Compare any two/more data attributes or compare the target object with previous values along with the time series.

- Distribution: Demonstrate the amount of value shared across the selected data attribute or show the breakdown of all data attributes.

- Trend: Present a general tendency over a time segment.

- Rank: Sort the data attributes based on their values and show the breakdown of selected data attributes.

- Aggregation: Calculate the descriptive statistical indicators (e.g., average, sum, count, etc. ) based on the data attributes.

- Association: Identify the useful relationship between two data attributes or among multiple attributes.

- Extreme: Find the extreme data cases along with the data attributes or within a certain range.

- Categorization: Select the data attribute(s) that satisfy certain conditions.

- Outlier: Explore the unexpected data attribute(s) or statistical outlier(s) from a given set.

**Insight Evidence**

Insight evidence are intermediate outputs in the conversation that **DIRECTLY** support each insight. Here are the types of insight evidence and their explanations:

* Code: The **EXACT** piece of code that **DIRECTLY** derives the insight. Dependency import and visualization code are **EXCLUDED** from consideration. For example, if the insight is "The positive rate for females is more than double that of males", the piece of code we focus on is 
\begin{jsonblock}
  ```python
  \# Filter only the positive rates
  male_positive_rate = gender_positive_rate.loc['Male', 'Positive']
  female_positive_rate = gender_positive_rate.loc['Female', 'Positive']
  ```
\end{jsonblock}
* Code Console Output: The **EXACT** code output that **DIRECTLY** derives the insight. For example, if the insight is "There is a particular peak of worldwide gross in 1999", the code output we focus on is
\begin{jsonblock}
  ```console
  Release Year  Worldwide Gross
  ...           ...
  14            1999
  ...           ...
  ```
\end{jsonblock}

* Source Text: The **quote** of conversation where the insight is summarized. You must identify the **minimal but critical** part of the conversation that **DIRECTLY** derives the insight.
\end{systemmessage}
\begin{usermessage}
<Instruction> Given an analysis conversation which is a list of ordered messages and 
each message is in the following format: 
\begin{jsonblock}    
```json
{{
  "id": 0,  // unique id of every message
  "role": "USER",  // "USER" or "ASSISTANT"
  "content": "Hello World",  // the content of this message, in markdown format string
}}
```
\end{jsonblock}
extract all insights from the conversation. Please return the insights based on the format specified in **Format Instructions**.

**IMPORTANT**: 

- Find as **MANY** insights as possible from the conversation. But remember, **QUALITY** is more important than **QUANTITY**. **NEVER EVER** make up insights. **ONLY** derive insights from the conversation and the dataset.

- You should only extract insights that are directly related to **Data Analysis**. Irrelevant insights **MUST NOT** be included.

- **NEVER EVER** make up insight types. **ONLY** classify the insights into the 11 categories: [Value, Proportion, Difference, Distribution, Trend, Rank, Aggregation, Association, Extreme, Categorization, Outlier].

- Insight evidence, especially **CODE** and **CODE OUTPUT**, must be **DIRECTLY** releated to the corresponding insight, following the instructions above. Make them **ACCURATE** and **CONCISE**.

- The insights should reference the original messages (via the `id` field) that are part of a given thread. These messages do not have to be continuous and may contain disjoint parts of the conversation.

</Instruction>

<Format Instructions>
\begin{jsonblock}
// Format instructions based on LangChain's Pydantic JSON Parser (hidden to save space)
\end{jsonblock}
</Format Instructions>

Begin.

Convsersation History: 

\begin{describedexample}{}{Example input}
\begin{jsonblock}
```json
[
  {
    "role": "user",
    "content": "Dear Chatty, I have a dataset to analyse. Let`s imagine that me and my business partner are planning to open a coffee shop in our city. To make informed decisions, we will analyze data to guide key strategies such as which coffee to source, pricing, and target customer demographics. Our goal is to uncover actionable insights to shape your business plan. The data set that we have at our disposal is called \"The Great American Coffee Taste Test\".  The link to the dataset is here : [csv_link]\n\n Dataset Description: In October 2023, \"world champion barista\" James Hoffmann and coffee company Cometeer held the \"Great American Coffee Taste Test\" on YouTube, during which viewers were asked to fill out a survey about 4 coffees they ordered from Cometeer for the tasting.",
    "turn": 0
  },
  {
    "role": "assistant",
    "content": "Got it! You`ve uploaded the dataset, so let`s start by exploring its structure. I`ll load the data and give you a quick summary, including column names, data types, and a few sample rows.",
    "turn": 0
  },
  ... // (hidden for brevity)
]
```
\end{jsonblock}
\end{describedexample}

Insights:
\end{usermessage}
\begin{assistantmessage}
    \begin{describedexample}{}{Example Output}
        \begin{jsonblock}
```json
{
  "insights": [
    {
      "insight": "The dataset contains 4,042 entries and 57 columns, covering various aspects of coffee consumption, preferences, and demographics.",
      "keywords": ["dataset", "entries", "columns", "structure"],
      "sourceMessageIds": [3],
      "types": ["Value"],
      "evidence": {
        "code_console": ... // (hidden for brevity)
        }
    },
    {
      "insight": "The top 3 values for the 'age' column are '25-34 years old' (1986, 49.51%), '35-44 years old' (960, 23.93%), and '18-24 years old' (461, 11.49%).",
      "keywords": ["age", "values", "frequency"],
      "sourceMessageIds": [8],
      "types": ["Value", "Rank"],
      "evidence": {
        "code_console": "{'age': [('25-34 years old', 1986, 49.51),\n  ('35-44 years old', 960, 23.93),\n  ('18-24 years old', 461, 11.49)]}"
      }
    },
    {
      "insight": "The top 3 values for the 'gender' column are 'Male' (2524, 71.64%), 'Female' (853, 24.21%), and 'Non-binary' (103, 2.92%).",
      "keywords": ["gender", "values", "frequency"],
      "sourceMessageIds": [8],
      "types": ["Value", "Rank"],
      "evidence": {
        "code_console": "{'gender': [('Male', 2524, 71.64),\n  ('Female', 853, 24.21),\n  ('Non-binary', 103, 2.92)]}"
      }
    }
    ... // (hidden for brevity)
    ]
}
```
        \end{jsonblock}
    \end{describedexample}
\end{assistantmessage}
\end{prompt}}

\clearpage
\subsubsection{Prompt for rewriting from serialized markdown}
For this prompt, we have preset definitions for the length, technical detail, and formality (see Table~\ref{tab:prompt_sliders}). These correspond to the sliders in \syncsense's \authoring panel.

\begin{table}[h]
\centering
\begin{tabular}{|l|c|p{12cm}|}
\hline
\textbf{Slider} & \textbf{Level} & \textbf{Description} \\
\hline
\multirow{3}{*}{Length} 
& 1 & The summary length should be very short and concise. It should be read in 30 seconds or less (100 words or less). \\
& 2 & The summary length should be moderate. It should be read in 2 minutes or less (around 300–400 words). \\
& 3 & The summary length can be longer. It should be read in 5 minutes or less (around 800–1,000 words). \\
\hline
\multirow{3}{*}{Technical Detail} 
& 1 & The summary should focus on key insights with minimal technical details for a non-technical audience. \\
& 2 & The summary should balance high-level insights with details that include the concrete data analysis and statistical methods for a semi-technical audience. \\
& 3 & The summary should focus on the concrete data analysis and statistical methods for a technical audience. \\
\hline
\multirow{3}{*}{Formality} 
& 1 & The summary could be informal. \\
& 2 & The summary should be semi-formal. \\
& 3 & The summary should be formal. \\
\hline
\end{tabular}
\caption{Slider definitions for summary length, technical detail, and formality.}
\label{tab:prompt_sliders}
\end{table}

\cprotect[mm]\aptLtoX{\begin{tabular}{p{500pt}}
\cellcolor{gray!40}\textcolor{white}{Summary from Markdown Prompt}\\
\textbf{System prompt:}\\
You are an AI Conversational Data Analysis Assistant, \
specialized in summarizing analytical conversations. \
Before proceeding with any analysis, it’s important to understand the key \
elements of an analytical conversation.\\
The conversation is structured as follows:\\
1. **Analysis Threads**: These represent distinct lines of investigation or exploration within \
the conversation. Each thread focuses on a particular question, hypothesis, or aspect of the overall topic, requiring specific methods and analyses. A thread is a collection of related turns, which may not always be consecutive. Sub-threads may also exist, diving deeper into particular aspects of the main thread.\\
2. **Insights**: These are key findings or observations that emerge from the data. Insights come from certain turns in the conversation and are supported by evidence in the form of code, code console output, or source text in one or more turns.\\
3. **Artifacts**: These are the artifacts generated during the analysis process, such as visualizations, code, or data tables. \\
\hline
\includegraphics[width=\iconwidth]{emojis/user.png} \textbf{User:}\\
<Instruction> \\
**Background** \\
We have distilled the conversation into the elements that we want to be summarized written \
in markdown format. The elements can include analysis threads, conversation turns, insights, and artifacts (visualizations, code, and data tables). \\
    Now, we want to reformat the extracted content into a coherent and concise summary that is tailored to \
specific user specitifcations. These specifications center around the length of the summary, the level of technical detail, \
and the formality of the language used. \\
    For instance, the level of technical detail can range from a high-level overview of the analysis to a detailed explanation of the methods and results. \\
    **Task**\\
Given the the extracted summarized content and user specified length, technical detail, and language formality requirements please reformat the content that is tailored to the specification. \\
    **Important**:\\
- DO NOT add any new information that is not present in the extracted content such as any personal opinions, interpretations, insights, or additional analysis.\\
- Do your best to adhere to the user specified requirements for length, technical detail, and language formality but DO NOT add extra content that is not present in the extracted content.\\
- IF there are any attached code, and data tables, they will be links in the extracted content and should be referenced in the summary.\\
- IF there are any images, they will also be links in the extracted content and should be included.\\
- DO NOT include any links that are not present in the extracted content.\\
</Instruction>\\
Begin. \\
Contents to be reformatted:\\
\cellcolor{gray!40}{EXAMPLE INPUT}\\
\begin{imageonly}
\begin{jsonblock}
```
* [Chat Turn 5] Coffee shop preferences
  * [Insight] Specialty Coffee Shops are a dominant preference among respondents, chosen by 16.3% of respondents.
  * [Table for Coffee shop preferences | Turn 6](table.csv)
* ![Title - Favorite Coffee Types by Employment Status, Title - Brewing Methods by Employment Status... Turn 8](viz_0.png)
* [Chat Turn 10] Coffee preferences by gender
  * [Code for Coffee preferences by gender | Turn 11](code_snippet.py)
  * ![Title - Favorite Coffee Types by Gender, Title - Brewing Methods by Gender... Turn 11](/viz_1.png)
* [Insight] The dataset contains 4011 responses with 7 unique age groups, where the most common age group is "25-34 years old" (1986 respondents).
```
\end{jsonblock}
\end{imageonly}\\
\# Length specification:\\
\cellcolor{gray!40}{EXAMPLE INPUT}\\
\cellcolor{gray!40}The summary length should be very short and concise. It should be read in 30 seconds or less (100 words or less).\\
\# Technical detail specification:\\
When thinking about the level of technical detail, consider technical and non-technical audiences.\\
Non-technical audience focuses more on the big-picture goals of the work and interpretation
of the findings rather than specific details.\\
Technical audience, however, is able to interpret the concrete data analysis methods.\\
Based on the above definition, the technical detail should be the following:\\

\cellcolor{gray!40}{EXAMPLE INPUT}\\
\cellcolor{gray!40}The summary should focus on key insights with minimal technical details for a non-technical audience.\\
\#Formality specification:\\
Formal and informal writing differ in tone, purpose, and audience.
Here’s a comprehensive overview of each style.\\
**Formal Writing**\\
Used for serious, professional contexts, formal writing is clear, respectful, and direct, often reserved for business, academic, or journalistic settings. This tone is suited for addressing unfamiliar audiences, such as potential employers or formal reports.\\
Key Characteristics:
\begin{itemize}
\item Word Choice: Uses precise language and avoids slang. Example: “Please eliminate that slide from the presentation.”    
\item Punctuation: No exclamatory punctuation. Example: “I appreciate your help.”    
\item Structure: Full sentences, often with complex syntax. Example: “Have you completed the monthly report?”
\item Clarity: Avoids ambiguity, slang, or emoji. Example: “I’m running late for our meeting. I’ll be there soon.”
\end{itemize}\\
Formal Writing Techniques:
\begin{itemize}
\item Avoids contractions and personal pronouns.    
\item Includes technical terminology relevant to the topic.    
\item Uses respectful, elaborate descriptions for clarity.    
\item Minimizes personal opinions and feelings.
\end{itemize}\\
**Informal Writing**\\
Informal writing has a conversational, relaxed tone and is commonly used in casual communication, such as chatting with friends or writing blog posts.\\
Key Characteristics:
\begin{itemize}
\item Word Choice: Includes slang, idioms, and casual expressions. Example: “Let’s lose slide \#23.”    
\item Punctuation: Exclamation points and emoji can be used. Example: “I appreciate your help!”    
\item Structure: Often shorter, with direct or even fragmented sentences. Example: “Monthly report ready?”    
\item Tone: Personal opinions and creativity are encouraged.
\end{itemize}\\
Informal Writing Techniques:
\begin{itemize}
\item Uses contractions and pronouns freely. 
\item Can meander off-topic, mimicking conversational flow.
\item Prioritizes tone over strict grammar and spelling.
\end{itemize}\\
Based on the above definition, the formality should be the following:\\
\cellcolor{gray!40}{EXAMPLE INPUT}\\
\cellcolor{gray!40}The summary should be semi-formal.\\
Final summary:\\
\hline
\includegraphics[width=\iconwidth]{emojis/assistant.png} \textbf{Assistant:}\\
\cellcolor{gray!40}{EXAMPLE OUPUT}\\
    \begin{imageonly}
    \begin{jsonblock}
```        
The analysis reveals key insights regarding coffee preferences. Specialty Coffee Shops are a top choice, preferred by 16.3% of respondents (refer to Table). The dataset comprises 4011 responses across 7 age groups, with the "25-34 years old" group being the most represented (1986 respondents). Additionally, coffee preferences were examined by gender (refer to attached Code and Image). The study also includes visualizations on favorite coffee types and brewing methods by employment status (see Image).

![Coffee Preferences Table](table.csv)
![Coffee Types and Brewing Methods by Employment Status](viz_0.png)
![Coffee Types and Brewing Methods by Gender](viz_1.png)
```
    \end{jsonblock}
    \end{imageonly}
\end{tabular}}{
\begin{prompt}[Summary from Markdown Prompt]
\begin{systemmessage}
You are an AI Conversational Data Analysis Assistant, \
specialized in summarizing analytical conversations. \
Before proceeding with any analysis, it’s important to understand the key \
elements of an analytical conversation.

The conversation is structured as follows:

1. **Analysis Threads**: These represent distinct lines of investigation or exploration within \
the conversation. Each thread focuses on a particular question, hypothesis, or aspect of the overall topic, requiring specific methods and analyses. A thread is a collection of related turns, which may not always be consecutive. Sub-threads may also exist, diving deeper into particular aspects of the main thread.

2. **Insights**: These are key findings or observations that emerge from the data. Insights come from certain turns in the conversation and are supported by evidence in the form of code, code console output, or source text in one or more turns.

3. **Artifacts**: These are the artifacts generated during the analysis process, such as visualizations, code, or data tables. 
\end{systemmessage}

\begin{usermessage}
<Instruction> 

**Background** 
We have distilled the conversation into the elements that we want to be summarized written \
in markdown format. The elements can include analysis threads, conversation turns, insights, and artifacts (visualizations, code, and data tables). \
    
Now, we want to reformat the extracted content into a coherent and concise summary that is tailored to \
specific user specitifcations. These specifications center around the length of the summary, the level of technical detail, \
and the formality of the language used. \
    
For instance, the level of technical detail can range from a high-level overview of the analysis to a detailed explanation of the methods and results. \
    
**Task**
Given the the extracted summarized content and user specified length, technical detail, and language formality requirements please reformat the content that is tailored to the specification. \
    
**Important**:

- DO NOT add any new information that is not present in the extracted content such as any personal opinions, interpretations, insights, or additional analysis.

- Do your best to adhere to the user specified requirements for length, technical detail, and language formality but DO NOT add extra content that is not present in the extracted content.

- IF there are any attached code, and data tables, they will be links in the extracted content and should be referenced in the summary.

- IF there are any images, they will also be links in the extracted content and should be included.

- DO NOT include any links that are not present in the extracted content.

</Instruction>

Begin. 

Contents to be reformatted:
\begin{describedexample}{}{Example input}
\begin{jsonblock}
```
* [Chat Turn 5] Coffee shop preferences
  * [Insight] Specialty Coffee Shops are a dominant preference among respondents, chosen by 16.3% of respondents.
  * [Table for Coffee shop preferences | Turn 6](table.csv)
* ![Title - Favorite Coffee Types by Employment Status, Title - Brewing Methods by Employment Status... Turn 8](viz_0.png)
* [Chat Turn 10] Coffee preferences by gender
  * [Code for Coffee preferences by gender | Turn 11](code_snippet.py)
  * ![Title - Favorite Coffee Types by Gender, Title - Brewing Methods by Gender... Turn 11](/viz_1.png)
* [Insight] The dataset contains 4011 responses with 7 unique age groups, where the most common age group is "25-34 years old" (1986 respondents).
```
\end{jsonblock}
\end{describedexample}
\# Length specification:
\begin{describedexample}{}{Example input}
The summary length should be very short and concise. It should be read in 30 seconds or less (100 words or less).
\end{describedexample}
\# Technical detail specification:

When thinking about the level of technical detail, consider technical and non-technical audiences.

Non-technical audience focuses more on the big-picture goals of the work and interpretation
of the findings rather than specific details.

Technical audience, however, is able to interpret the concrete data analysis methods.

Based on the above definition, the technical detail should be the following:

\begin{describedexample}{}{Example input}
The summary should focus on key insights with minimal technical details for a non-technical audience.
\end{describedexample}

\#Formality specification:

Formal and informal writing differ in tone, purpose, and audience.
Here’s a comprehensive overview of each style.

**Formal Writing**

Used for serious, professional contexts, formal writing is clear, respectful, and direct, often reserved for business, academic, or journalistic settings. This tone is suited for addressing unfamiliar audiences, such as potential employers or formal reports.

Key Characteristics:

	•	Word Choice: Uses precise language and avoids slang. Example: “Please eliminate that slide from the presentation.”
    
	•	Punctuation: No exclamatory punctuation. Example: “I appreciate your help.”
    
	•	Structure: Full sentences, often with complex syntax. Example: “Have you completed the monthly report?”

	•	Clarity: Avoids ambiguity, slang, or emoji. Example: “I’m running late for our meeting. I’ll be there soon.”

Formal Writing Techniques:

	•	Avoids contractions and personal pronouns.

	•	Includes technical terminology relevant to the topic.

	•	Uses respectful, elaborate descriptions for clarity.

	•	Minimizes personal opinions and feelings.

**Informal Writing**

Informal writing has a conversational, relaxed tone and is commonly used in casual communication, such as chatting with friends or writing blog posts.

Key Characteristics:

	•	Word Choice: Includes slang, idioms, and casual expressions. Example: “Let’s lose slide \#23.”
    
	•	Punctuation: Exclamation points and emoji can be used. Example: “I appreciate your help!”
    
	•	Structure: Often shorter, with direct or even fragmented sentences. Example: “Monthly report ready?”
    
	•	Tone: Personal opinions and creativity are encouraged.

Informal Writing Techniques:

	•	Uses contractions and pronouns freely.
    
	•	Can meander off-topic, mimicking conversational flow.
    
	•	Prioritizes tone over strict grammar and spelling.

Based on the above definition, the formality should be the following:
\begin{describedexample}{}{Example input}
The summary should be semi-formal.
\end{describedexample}
Final summary:

\end{usermessage}
\begin{assistantmessage}
    \begin{describedexample}{}{Example Output}
    \begin{jsonblock}
```        
The analysis reveals key insights regarding coffee preferences. Specialty Coffee Shops are a top choice, preferred by 16.3% of respondents (refer to Table). The dataset comprises 4011 responses across 7 age groups, with the "25-34 years old" group being the most represented (1986 respondents). Additionally, coffee preferences were examined by gender (refer to attached Code and Image). The study also includes visualizations on favorite coffee types and brewing methods by employment status (see Image).

![Coffee Preferences Table](table.csv)
![Coffee Types and Brewing Methods by Employment Status](viz_0.png)
![Coffee Types and Brewing Methods by Gender](viz_1.png)
```
    \end{jsonblock}

    \end{describedexample}
\end{assistantmessage}

\end{prompt}}


\end{document}